\documentclass[11pt]{article}

\usepackage{times}

\usepackage{epsfig}
\usepackage{graphicx}
\usepackage{colordvi}

\usepackage{natbib}

\usepackage{amsmath}%
\usepackage{amsfonts}%
\usepackage{amssymb}%

\usepackage[oneside]{fltpage}

\renewcommand\b{\begin{equation}}
\newcommand\e{\end{equation}}

\def\sign{\mathrm{sign}}
\def\erf{\mathrm{erf}}

\newenvironment{sciabstract}{%
\begin{quote} \bf}
{\end{quote}}

\newcommand{\nn}{\nonumber}
\newcommand{\al}{\alpha}
\newcommand{\ep}{\epsilon}
\newcommand{\la}{\lambda}
\newcommand{\ve}{\varepsilon}
\newcommand{\mc}{\mathcal}
\newcommand{\mr}{\mathrm}
\def\[{\begin{eqnarray}}
\def\]{\end{eqnarray}}

\newcounter{lastnote}

\title{Computational principles of biological memory}

\author
{Marcus K. Benna$^{1}$, Stefano Fusi$^{1\ast}$\\
\\
\normalsize{$^{1}$Center for Theoretical Neuroscience, Columbia University,}\\
\normalsize{College of Physicians and Surgeons, New York, NY 10032, USA}\\
\\
\normalsize{$^\ast$To whom correspondence should be addressed; E-mail:  sf2237@columbia.edu.}
}

\begin{document}


\maketitle


\begin{sciabstract}

Memories are stored, retained, and recollected through complex, coupled processes operating on multiple timescales. To understand the computational principles behind these intricate networks of interactions we construct a broad class of synaptic models that efficiently harnesses biological complexity to preserve numerous memories. The memory capacity scales almost linearly with the number of synapses, which is a substantial improvement over the square root scaling of previous models. This was achieved by combining multiple dynamical processes that initially store memories in fast variables and then progressively transfer them to slower variables. Importantly, the interactions between fast and slow variables are bidirectional. The proposed models are robust to parameter perturbations and can explain several properties of biological memory, including delayed expression of synaptic modifications, metaplasticity, and spacing effects.	

\end{sciabstract}

\section*{Introduction}

The complexity and diversity of the numerous biological mechanisms that underlie memory is both fascinating and disconcerting. The molecular machinery that is responsible for memory consolidation at the level of synaptic connections is believed to employ a complex network of highly diverse biochemical processes that operate on different timescales \citep[see e.g.][]{kandelbook,Bhalla2014}. Understanding how these processes are orchestrated to preserve memories over a lifetime requires guiding principles to interpret the complex organization of the observed synaptic molecular interactions and explain its computational advantage. Here we present a class of synaptic models that can efficiently harness biological complexity to store and preserve a huge number of memories on long timescales, vastly outperforming all previous synaptic models of memory.

The models that we construct solve a long-standing dilemma: in a memory system that is continually receiving and storing new information, synaptic strengths representing old memories must be protected from being overwritten during the storage of new information. Failure to provide such protection results in memory lifetimes that are catastrophically low \citep{AmitFusi1994,Fusi2002,FusiAbbott2007}.
On the other hand, protecting old memories too rigidly causes memory traces of new information to be extremely weak, being represented by a small number of synapses. This is one aspect of the plasticity-rigidity dilemma \citep[see also][]{McCloskey1989,cg91,mcclelland95,FusiDrewAbbott2005}. Synapses that are highly plastic are good at storing new memories but poor at retaining old ones. Less plastic synapses are good at preserving memories, but poor at storing new ones.

Previous theoretical works have estimated the consequences of the plasticity-rigidity dilemma on the memory performance for various synaptic models characterized by different degrees of complexity. For many years, long-term potentiation of synapses was represented, at least by the modeling community, as a simple switch-like change in synaptic state. Memory models studied in the 1980's \citep[see][]{Hopfield1982} suggested that networks of neurons connected by such synapses could preserve a number of memories that scales linearly with the size of the network. However, subsequent theoretical analyses \citep{AmitFusi1994,Fusi2002,FusiAbbott2007} revealed that what had appeared to be a harmless assumption in the theoretical calculations was actually a fatal flaw. The unfortunate approximation was ignoring the limits on synaptic strengths imposed on any real physical or biological system. When these limits are included, e.g.~in the extreme case of binary synapses in which the weight takes only two distinct values, the memory capacity grows only logarithmically with the number of synapses $N$ for highly plastic synapses, and like $\sqrt{N}$ for more rigid synapses that are able to store only a small amount of information per memory.

A possible resolution of this dilemma is to make each synapse complex enough to contain both highly plastic and rigid components. In many models the plastic components are represented by fast biochemical processes, which can change rapidly to acquire and store a large amount of information about new memories. This initial memory trace is strong but labile; it decays quickly when other memories are stored. Memories can be consolidated if the information about each new memory is progressively transferred to the slow components, which can preserve memories on longer timescales. This mechanism is widely used in artificial devices (e.g.~computer memories, which include fast RAM and hard drives), it was proposed to explain memory consolidation at the systems level \citep{mcclelland95,Roxin2013}, and it was incorporated into a model of synaptic memory based on a cascade of biochemical processes that operate on different timescales~\citep{FusiDrewAbbott2005}. This form of synaptic complexity can greatly extend memory lifetimes without sacrificing the initial memory strength, accounting for our remarkable ability to remember for long times a large number of details even when memories are learned in one shot~\citep{Brady2008}. The two quantities that characterize memory performance, memory lifetime and the strength of the initial memory trace, scale like the square root of the number of synapses ($\sqrt{N}$) in the cascade model \citep{FusiDrewAbbott2005}.

Here we show that these models can be significantly improved when the network of interactions between the multiple biochemical processes that control the synaptic dynamics is bidirectional and appropriately tuned. In this case, the decay of the memory trace is substantially slower than in all previous models, leading to a memory lifetime that scales almost linearly with the number of synapses. Importantly, in our model, improved memory lifetime does not require a systematic reduction in the initial memory strength, which also scales approximately like the square root of the number of synapses. Although the proposed synaptic model requires tuning, it is robust to noise and variations in its parameters. Finally, we construct a broad class of synaptic models that are equivalent in terms of memory performance. These different models capture the complexity and diversity of biochemical processes believed to be involved in memory consolidation. Thanks to their complexity, they can also reproduce the rich phenomenology of a plethora of biology and psychology experiments, including power-law memory decay \citep{WixtedEbbesen1991,WixtedEbbesen1997}, synaptic metaplasticity \citep{Abraham2008}, delayed expression of synaptic potentiation and depression, and spacing effects \citep[see e.g.][]{Anderson2000}.

\section*{The memory benchmark}

To study the process of storing multiple memories and to benchmark memory models we need to make assumptions about the nature of memories. Storage of new memories is likely to exploit similarities with previously stored information (see e.g.~semantic memories). In what follows, we focus on mechanisms responsible for storing new information that has already been preprocessed in this way and is thus incompressible. For this reason, we consider memories that are unstructured (random) and do not have any correlations with previously stored information (uncorrelated). Although this may appear to be a strong and limiting assumption, it is widely considered as the standard benchmark for synaptic models, mainly because theoretical studies on random and uncorrelated memories are often predictive of the scaling properties of the memory performance in more general cases \citep[see e.g.~the case of the perceptron,][]{Rosenblatt1958}.

Consider an ensemble of $N$ synapses which is exposed to a continuous stream of modifications, each leading to the storage of a new memory. We express the assumption that the stored memories are unstructured by hypothesizing that the synaptic modifications are random and uncorrelated. Each synapse thus experiences a random sequence of potentiations and depressions, and the sequences are different and uncorrelated for different synapses. The memory of an event is defined by the pattern of $N$ synaptic modifications potentially induced by it. We will select arbitrarily one of these memories and track it over time. The selected memory is not different or special in any way, so that the results for this particular memory apply equally to all the memories being stored.

To track the selected memory we take the point of view of an ideal observer that knows the strengths of all the synapses relevant to a particular memory trace \citep[see e.g.][]{Fusi2002,FusiDrewAbbott2005}. Of course in the brain the readout is implemented by complex neural circuitry, and estimates of the strength of the memory trace based on the ideal observer approach may be significantly larger than the memory trace that is actually usable by the neural circuits. However, given the remarkable memory capacity of biological systems, it is not unreasonable to assume that the readout circuits perform almost optimally. Moreover, we will show that the ideal observer approach predicts the correct scaling properties of the memory capacity of simple neural circuits that actually perform memory retrieval (see the Discussion and Suppl. Info. \ref{memretrievalsupp}).

More quantitatively, we define the memory signal as the correlation between the state of the synaptic ensemble and the pattern of synaptic modifications originally imposed by the event being remembered. Previously stored memories, which are assumed to be random and uncorrelated, make the memory trace noisy. Memories that are stored after the tracked one continuously degrade the memory signal and also contribute to its fluctuations. We will monitor the signal to noise ratio (SNR) of a memory, which is defined as the ratio between the memory signal and its standard deviation (see Methods \ref{snrdef} for a more formal definition). One measure of memory performance is the memory lifetime, the maximal time since storage over which a memory can be detected, i.e.~for which the SNR is larger than one. The scaling properties of the memory performance that we will derive do not depend on the specific choice of the critical SNR value, as long as it is of order one.
The memory lifetime is also a measure of the memory capacity because all memories that have been stored more recently than the tracked one will have a larger SNR, and hence if the tracked memory is retrievable, then all the more recent memories will be retrievable a fortiori.

\section*{Constructing the synaptic model}

The value of a synaptic weight $w$ at any given time is typically the result of multiple synaptic modifications. To build an efficient synaptic model, it is instructive to start from an abstract memory model in which the present weight is expressed as a sum of synaptic modifications $\Delta w$, weighted by a factor $r$ that decreases with the age of the modification $t-t_l$, where $t$ is the current time and $t_l$ is the time of the $l$th modification. In this case the signal of the corresponding memory would decay as $r(t-t_l)$. The noise would be approximately proportional to square root of the the variance of $w$ at time $t$, 
\begin{equation}
\mathrm{Var}(w(t)) = \sum_{l\, :\, t_l<t}\left[\Delta w(t_l)\, r(t-t_l)\right]^2 \ ,
\label{varw}
\end{equation}

where we have assumed that the average of $\Delta w(t_l)$ is zero, which is equivalent to hypothesizing that synaptic potentiation and depression are balanced.
A slowly decaying $r$ would enable the synaptic weight to depend on a large number of modifications, but it would also induce a large variance for $w(t)$, potentially arbitrarily large if the sum extends over an arbitrary number of modifications. On the other hand, fast decays would limit the memory capacity. From eqn.~(\ref{varw}) it is apparent that in the case of random and uncorrelated modifications, the slowest power-law decay one can afford while keeping $w$ finite is approximately $r(t)\sim t^{-1/2}$ (see also Methods~\ref{optimaldecays}). In Suppl.~Info.~\ref{optimization}, we show that under some conditions this is approximately the optimal solution among all possible functional decays (see also the Discussion).

This abstract model reveals what kind of decay of the memory signal is desirable, but it does not explain how this behavior is achievable by synaptic dynamics. The next step is to construct a model that implements the desired power-law decay. One simple way would be to endow each synapse with a timer and introduce a mechanism to decrease the relative weight of each synaptic modification on the basis of the age of the modification \cite[see e.g.][]{Wu2009}, but this would just move the problem to the encoding and preservation of the age of a memory, which is potentially as difficult as the original memory problem we intend to solve. As we will show, there is no need for a timer, as there are synaptic models in which the $1/\sqrt{t}$ decay emerges naturally from the interaction of multiple processes.

We will start with the construction of a simple chain model that captures and illustrates all the relevant scaling properties of more complex models. Then we will show how to generalize the model to incorporate less orderly interactions that are more similar to those observed in biological synapses. The simple chain model is described in Fig.~\ref{model}A and is characterized by multiple dynamical variables, each representing a different biochemical process. The first variable, which is the most plastic one, represents the strength of the synaptic weight. It is rapidly modified every time the conditions for synaptic potentiation or depression are met. For example, in the case of STDP, the synapse is potentiated when there is a pre-synaptic spike that precedes a post-synaptic action potential. The other dynamical variables are hidden (i.e.~not directly coupled to neural activity) and represent other biochemical processes that are affected by changes in the first variable. In the simplest configuration, these variables are arranged in a linear chain, and each variable interacts with its two nearest neighbors. These hidden variables tend to equilibrate around the weighted average of the neighboring variables. When the first variable is modified, the second variable tends to follow it. In this way a potentiation/depression is propagated downstream, through the chain of all variables. Importantly, the downstream variables also affect the upstream variables as the interactions are bidirectional.

To gain insight into the way this type of synapse works, it is useful to resort to an analogy with a set of communicating vessels, a more intuitive physical system. This analogy is illustrated in Fig.~\ref{model}B. Each synaptic variable is represented by the level of liquid in a beaker. The interactions between variables are mediated by tubes that connect the beakers. 
The first beaker (yellow) represents the synaptic weight. The synapse is potentiated by pouring liquid into it, whereas depression is implemented by removing liquid. As the liquid level deviates from equilibrium, the fluid flow through the tubes will tend to balance the level in all beakers. The balancing dynamics is fast when the beakers are small and the tubes large, but slow for large beakers and small tubes. A single synaptic modification is remembered as long as the liquid levels remain significantly different from equilibrium.

We now show how to construct the desired synaptic memory model by considering the analogous system of communicating vessels. An efficient memory system should have both long memory lifetimes (i.e.~long relaxation times) and a large initial memory strength, obtained with a relatively small number of variables (i.e.~number of beakers). It is possible to build a system in which the memory strength decays like a power law (approximately $1/\sqrt{t}$) and that only requires a number of variables that grows logarithmically with the memory lifetime. 

We will construct this system in three steps, progressively increasing the number of tuned parameters to improve memory performance. First, consider a series of identical beakers that are arranged in a linear chain and connected by a set of tubes with equal cross sections (see Fig.~\ref{model}C). When the first variable $u_1$ is perturbed, for example by adding liquid to the first beaker, liquid starts flowing to the other beakers, relaxing towards equilibrium. This relaxation dynamics is illustrated in Fig.~\ref{model}C for a system with 31 beakers. In the first plot, the liquid levels of all beakers are shown at three different times. The perturbation starts from the first beaker and then slowly spreads to all the other beakers. This process, which is analogous to heat diffusion (see Methods~\ref{diffusion}), is characterized by a decay of the perturbation that follows a power law ($1/\sqrt{t}$), at least for a time period that scales quadratically with the number of beakers, after which it becomes exponentially fast. This system has the desired decay properties, but it requires an unreasonably large number of beakers. A synapse based on this mechanism would require a number of biochemical processes (each process being equivalent to a beaker) that scales like the square root of the number of storable memories and can be as large as the square root of the total number of synapses.

Interestingly, it is possible to have a comparable memory performance with a significantly smaller number of variables. We can combine together multiple beakers, as shown in Fig.~\ref{model}D and construct a system with a number of beakers that scales only logarithmically with the memory lifetime. The first beaker remains the same as in the original linear chain. The next two are merged into a larger beaker with twice the cross-sectional area, which contains the same volume of liquid as the two original ones. Then, the next four beakers are combined together into a larger one, and we repeat this merging procedure until we reach the end of the chain.
At each step the number of original beakers that are combined doubles. This implies that the variables describing the system operate on different timescales that increase exponentially as one moves along the chain.

While this merging procedure dramatically reduces the number of beakers, the convergence to equilibrium is now significantly faster than before. In the original system, equilibrating two distant beakers takes a time that scales quadratically with the number of intermediate beakers. If these intermediate beakers are merged into one, the required time is drastically reduced, which leads to a much faster memory decay ($\sim 1/t$) than in the previous case, as illustrated in Fig.~\ref{model}D.

Fortunately it is possible to recover the slow decay, without increasing the number of beakers, by tuning the cross sections of the tubes, as shown in Fig.~\ref{model}E. When the identical tubes are replaced with progressively smaller ones (by powers of two), the decay slows down and follows $1/\sqrt{t}$ over a time period that grows exponentially with the number of beakers. This means that it is possible to construct an efficient synapse whose memory decays in the optimal way and that requires a number of biochemical processes that grows only logarithmically with the longest memory lifetime (see also Section~\ref{construction}). We now show that these features are preserved when we consider a population of synapses storing multiple memories, even if the synaptic dynamical variables can vary only in a limited range and their values can only be preserved with limited precision.

\begin{FPfigure}
	\begin{minipage}{\textwidth}
		\centering
		\includegraphics[width=0.9\textwidth]{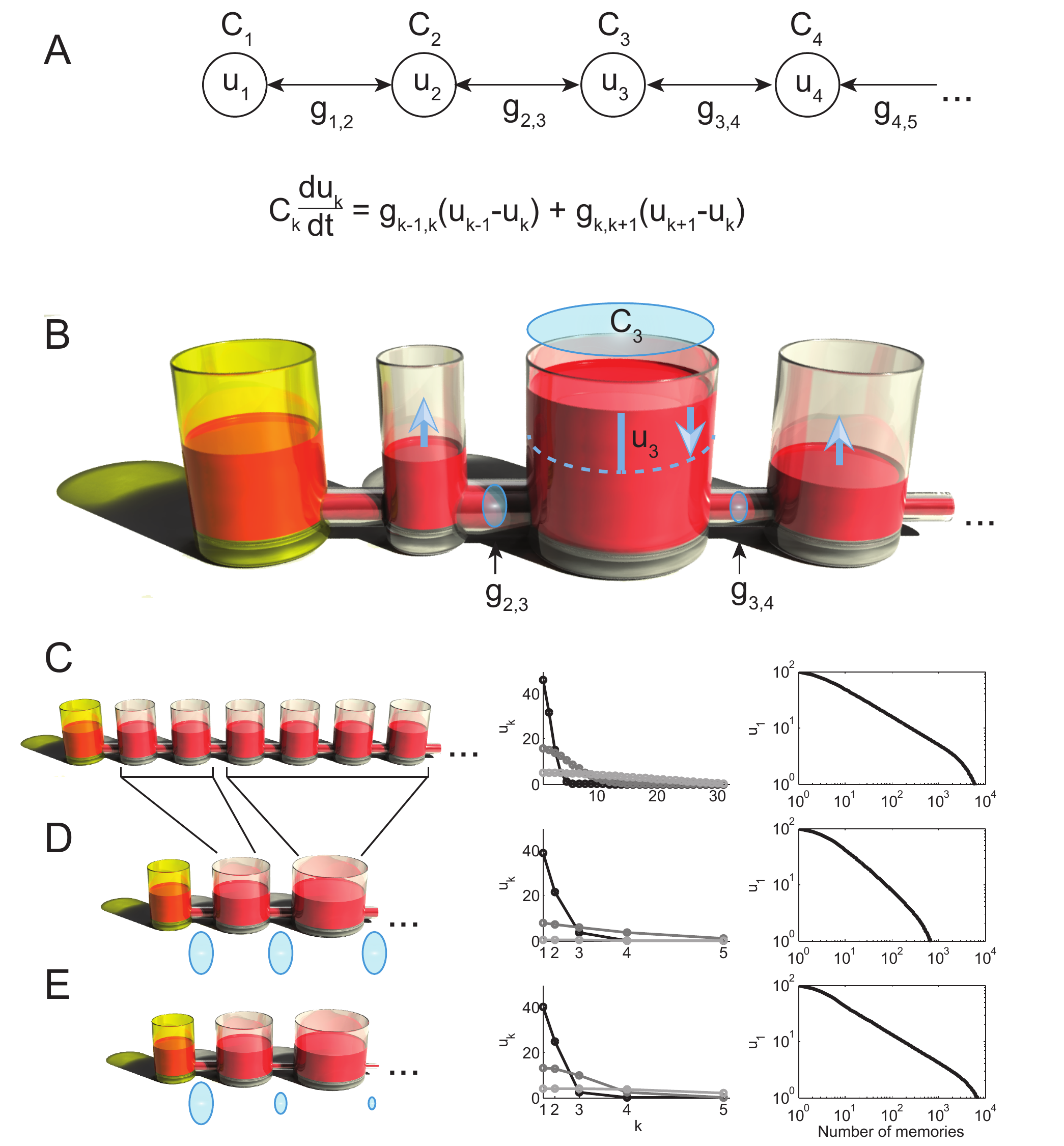}
	\end{minipage} 
	\caption[]{A. Schematic of a simple synaptic plasticity model. The dynamical variables $u_k$ represent different biochemical processes that are responsible for memory consolidation ($k=1,...,m$, where $m$ is the total number of processes). They are arranged in a linear chain and interact only with their two nearest neighbors (see differential equation), except for the first and the last variable. The first one interacts only with the second one (and is also coupled to the input), while the last one interacts only with the penultimate one. Moreover, the last variable $u_m$ has a leakage term that is proportional to its value (obtained by setting $u_{m+1}=0$). The parameters $g_{k,k+1}$ are the strengths of the bidirectional interactions (double arrows). Together with the parameters $C_k$ they determine the timescales on which each process operates. The first variable $u_1$ represents the strength of the synaptic weight. B. The schematic model of A behaves like a set of communicating vessels. The $u_k$ variables measure the deviation of the liquid level from equilibrium, shown in the third beaker as a blue dashed line. The $C_k$ represent the sizes (areas) of the beakers, and the coupling constants $g_{k,k+1}$ correspond to the cross-sections of the connecting tubes. Again, the liquid level in the first beaker (yellow) represents the synaptic strength. The last beaker is connected to a reservoir whose liquid level is always at equilibrium. This interaction represents the leak in the differential equation of $u_m$. C. Relaxation dynamics in a set of 31 identical beakers connected by tubes of equal size ($C_k=1, g_{k,k+1}=1/8$). A perturbation of the liquid level of the first beaker propagates to the others, slowly disappearing. The 31 $u_k$ variables are shown in the middle at three different times and the decay of $u_1$, which approximates a power law ($1/\sqrt{t}$), is plotted on the right on a log-log scale. D. A new set of beakers is obtained by merging those of panel C. The number of merged beakers progressively increases, leading to successively larger ones ($C_k=2^{k-1}$). The cross-sections of the tubes are still all identical (as indicated by the blue ovals). The number of variables is now significantly smaller, but the decay is too fast ($1/t$). E. Completely tuned set of communicating vessels: the sizes of the tubes connecting the beakers are progressively reduced to slow down the decay ($g_{k,k+1}=2^{-k-2}$), which now follows the desired $1/\sqrt{t}$ behavior as in C, but with a number of beakers that scales as the logarithm of the original number.
	}\label{model}
\end{FPfigure}

\clearpage


\section*{Discretization of the dynamical variables and scaling properties}

It is clearly unrealistic to assume that each dynamical variable $u_k$ can vary over an unlimited range and be manipulated with arbitrary precision when $u_k$ represents a physical quantity, such as the number of molecules in a particular state, which is typically relatively small given the size of a synapse (e.g.~there are only tens of CaMKII molecules at each synapse). Therefore, we discretize the $u_k$ and impose rigid limits on them. The dynamics of the model is now described by stochastic transitions between a discrete set of levels for each variable, arranged to approximate the continuous system constructed above.  At every time step, the $u_k$ are first updated as in the case of continuous variables described above, but then each variable is discretized by setting it stochastically to one of the two values in the discrete set that are closest to the updated value. The probabilities of ending up in each of the two values are chosen so that the average of the discretized $u_k$ matches the continuous $u_k$ (see Section~\ref{discretelinearchain} for details).

Assuming that memories are presented at a constant rate of one new uncorrelated memory per unit of time, we show in Fig.~\ref{scaling}A the SNR as a function of time for memory models in which the complexity of the synapse progressively increases -- the number $m$ of variables varies between 4 and 10. The curves are plotted on a log-log scale, so a straight (downwards) line corresponds to a power-law decay. In all cases, the SNR decays approximately as $1/\sqrt{t}$, as expected, over a time interval that increases exponentially with the complexity of the synapse. Then the decay accelerates and becomes exponential. Therefore, the corresponding memory lifetime increases exponentially with $m$ (see Fig.~\ref{scaling}B) up to a limit of order $N$, the total number of synapses. Conversely, increasing the number of synapses while keeping $m$ fixed, we find a memory lifetime that grows linearly with $N$ (see Fig.~\ref{scaling}D), until it saturates at the longest timescale of the memory system (which is exponential in $m$). This saturation can be avoided by adjusting the longest timescale appropriately, which leads to a memory lifetime scaling as $N/\log(N)$.

The memory lifetime in previous models of complex synapses with bounded weights scales at most as $\sqrt{N}$ \citep[see e.g.][]{FusiDrewAbbott2005}. A memory lifetime that scales (almost) linearly with the number of synapses constitutes a major improvement, especially in large neural systems. For the human brain, with $N \sim 10^{14}$, the memory capacity would potentially be extended by a factor of almost $10^7$. 
Importantly, this is achieved with a relatively small increase in the complexity of the synaptic machinery for memory consolidation, as $m$ grows only logarithmically with the memory lifetime. Moreover, the initial SNR, which is related to the amount of information stored per memory, has the same scaling with $N$ as in previous models (i.e. $\sqrt{N}$, see Fig.~\ref{scaling}E), and only decreases slowly with $m$ (as $1/\sqrt{m}$, see Fig.~\ref{scaling}C).



\begin{FPfigure}
	\begin{minipage}{\textwidth}
		\centering
		\includegraphics[width=6in]{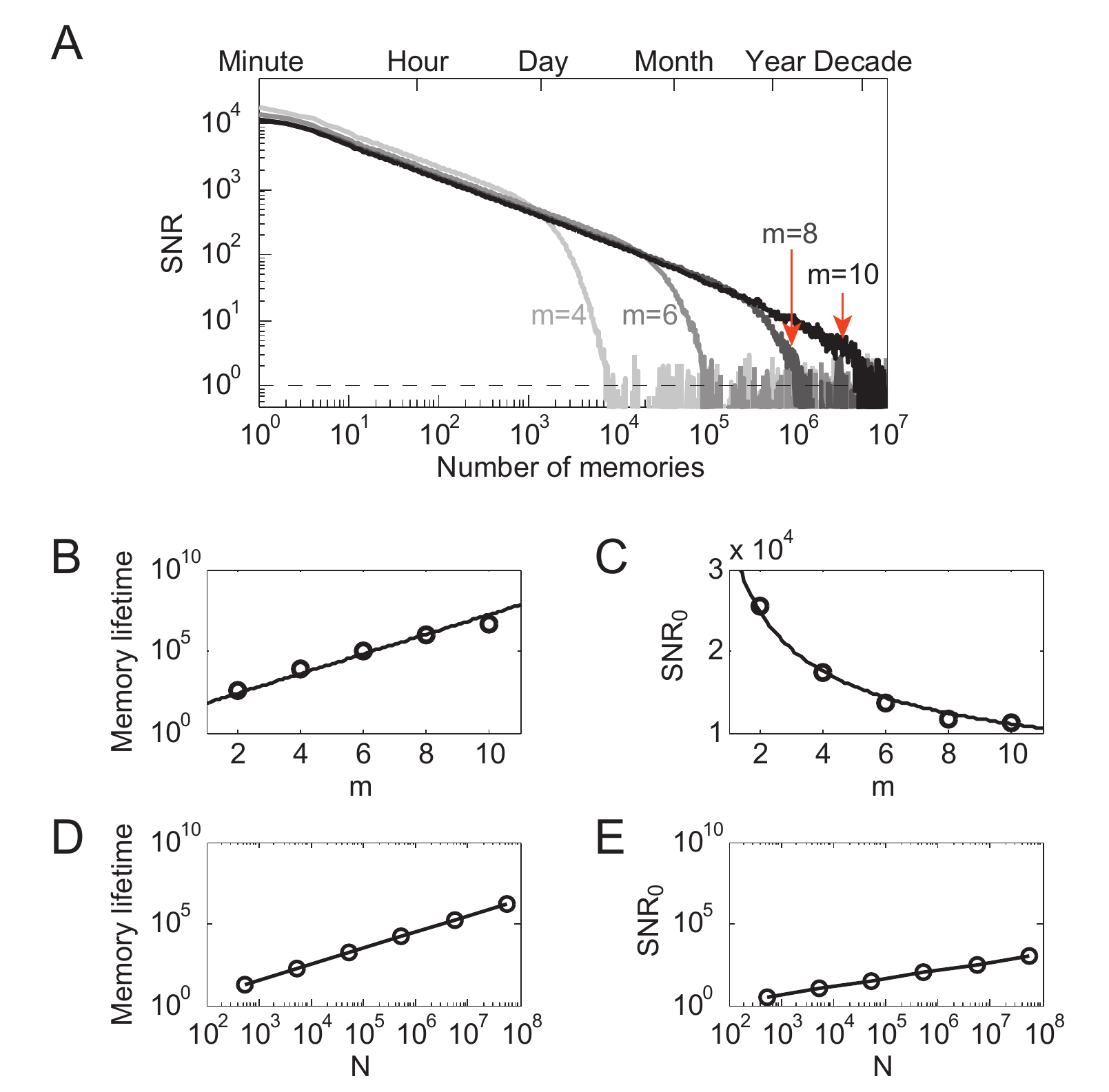}
	\end{minipage}
	\caption[]{Scaling properties of the synaptic model. A. Memory signal to noise ratio (SNR) as a function of the number of random uncorrelated memories that are stored after the tracked memory. The SNR is computed for a population of $N=5.4\times 10^9$ synapses. The scales of both axes are logarithmic. Different curves correspond to synaptic models with a different number of dynamical variables ($m=4,6,8,10$). $m$ is also the number of beakers in Figure \ref{model}. Each variable can vary on a discrete set of 40 equally spaced values. For all curves, the decay follows approximately a power law ($1/\sqrt{t}$) for a large number of memories and then becomes exponential where the curves visibly bend downwards. The range of the power-law decay increases exponentially with $m$, which is a measure of the complexity of the synapse. Memories are assumed to be stored at a constant rate of one new uncorrelated memory per unit time, which we choose to be one minute here, so that the SNR decay can also be expressed as a function of time (upper horizontal axis).  This choice of overall timescale is completely arbitrary and time is considered only to help the reader appreciate the wide span of memory lifetimes. The memory lifetime is defined as the time elapsed since storage (or number of subsequently stored memories) at which the SNR falls below some arbitrary threshold (dashed line). B. Memory lifetime vs $m$. The vertical axis is logarithmic, the horizonal one is linear, so the line that fits the simulation points represents an exponential growth.  C. Initial SNR, denoted by SNR$_0$, vs $m$. Both axes are linear. As $m$ increases, the initial SNR slowly degrades ($\sim 1/\sqrt{m}$). $N=5.4\times 10^9$ both in B and C. D. Memory lifetime vs $N$, the number of synapses on a log-log scale. The memory lifetime, which is proportional to the memory capacity (i.e.~the total number of memories that can be stored), increases linearly with $N$, as expected from the $1/\sqrt{t}$ decay of the SNR. This is a major improvement over previous synaptic models. E. Initial SNR vs $N$ on a log-log scale. SNR$_0$ grows like $\sqrt{N}$, as in the best previous synaptic models. $m=12$ for both D and E.
	}
	\label{scaling}
\end{FPfigure}

\section*{Robustness of the model}

\begin{figure}
		\centering
		\includegraphics[width=6in]{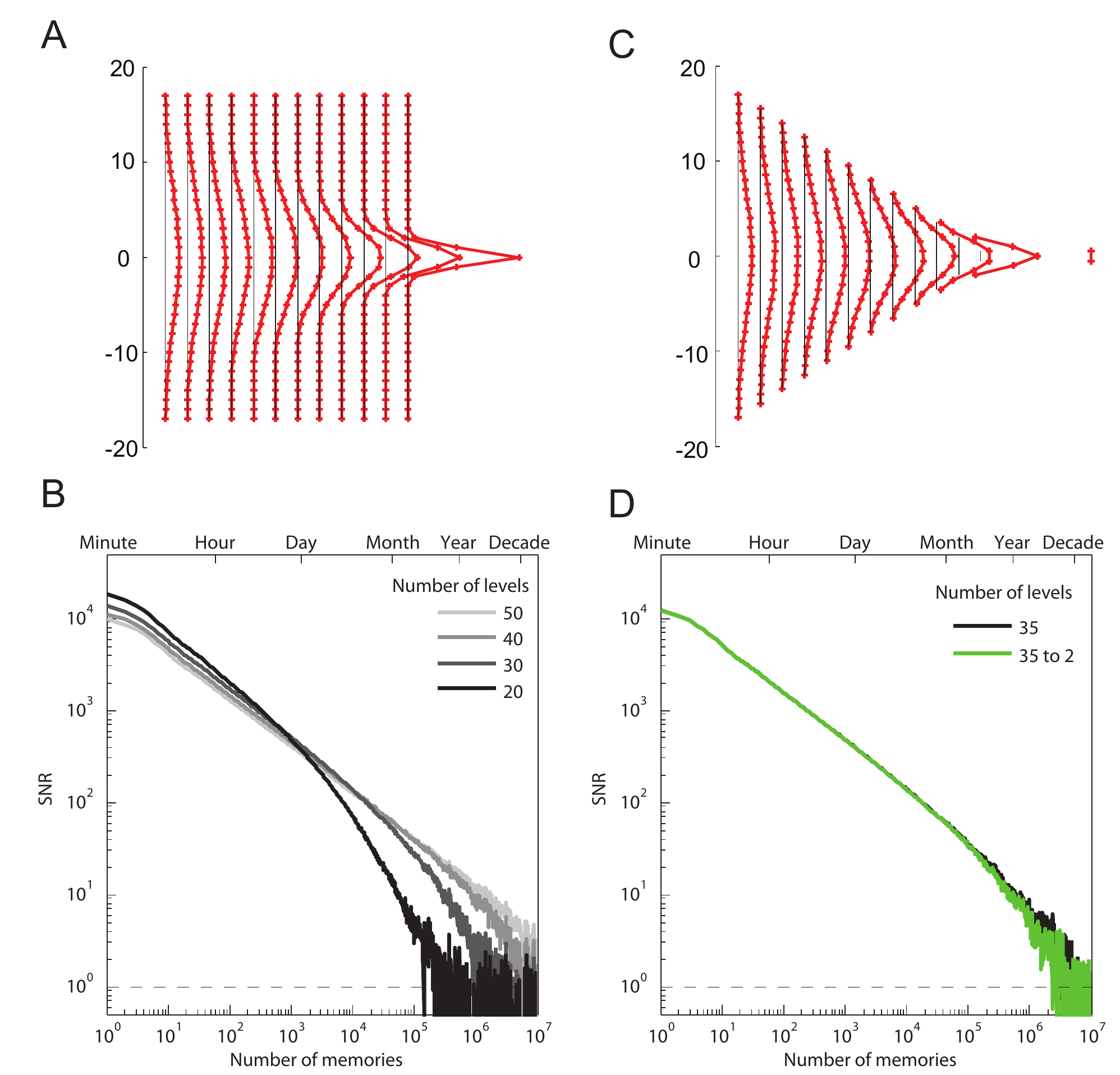}
	\caption[]{Robustness: Effects of different discretization schemes of the dynamical variables. A. Distributions of the $u_k$ variables for $k=1, 2,\ldots, m$ in a population of synapses at equilibrium. The synaptic model has $m=12$ dynamical variables with 35 discrete values each. The fastest variable is on the left. The distributions are approximately Gaussian and become progressively narrower for slower variables. B. SNR vs number of stored memories, as in Fig.~\ref{scaling}A, for discretizations with different numbers of levels (namely $20,30,40$ and $50$). C. Distributions of the $u_k$ variables when the number of discrete levels decreases progressively for the slower variables. The last variable has just two stable levels. The distributions are very similar to the case in which the boundaries and the number of levels are the same for all dynamical variables (panel A). D. SNR vs number of stored memories for constant (black) and decreasing number of discrete levels (green). The performance remains almost unaffected by the reduction in the number of levels.
	}
	\label{discrete}
\end{figure}

We now study systematically the effects of discretization on memory performance.
In Fig.~\ref{discrete}A we plot the distributions of the $u_k$ variables across the model synapse when each $u_k$  varies over a discrete set of 35 equally spaced values. The maximal and minimal values are rigid boundaries. All distributions are approximately Gaussian, with a width that is largest for the first variable $u_1$ and progressively decreases for the other $u_k$ as $k$ increases (see Suppl.~Info.~\ref{varianceandspatialcorr}). Since almost all values are well within the boundaries, the relaxation dynamics of the $u_k$ variables is very similar to the unbounded case and the SNR curve changes smoothly when we restrict the dynamical range even further 
(Fig.~\ref{discrete}B). Indeed, the width of the broadest distribution (that of $u_1$) scales only like 
$\sqrt{\log T}$, where $T$ is the longest timescale of the synapse (approximately $T \sim C_m/g_{m,m+1}=2^{2m+1}$).

Because the distributions are narrower for the slower dynamical variables, one may wonder whether the range and the number of levels could be progressively decreased as a function of $k$ without affecting the memory performance significantly. This is indeed the case, as demonstrated in Figs.~\ref{discrete}C,D. When the number of equally spaced levels decreases linearly with $k$, the distributions are very similar to the case in which the number of levels remains the same for all variables (Fig.~\ref{discrete}A). The memory performance is almost identical in the two cases (Fig.~\ref{discrete}D). This implies that the slower dynamical variables do not require as much precision as the fastest ones. Slower variables only need a number of levels that can be surprisingly small, just two for the slowest one in the example of Figs.~\ref{discrete}C,D. The slowest variables need to preserve their values over timescales of years, and this would likely be difficult to implement if a large number of values had to be distinguished. In contrast, bistable processes can easily be stable over very long time periods \citep{crick1984,miller05,Si2003}. For a small number of levels that is larger than two, one could combine multiple bistable processes or use slightly more complicated mechanisms \citep{Shouval2005}.



In Suppl.~Info.~\ref{robustness} we show that the model is robust not only to discretization, but also to
parameter variations, and can tolerate surprisingly large perturbations of the optimal beaker and tube sizes. The SNR of the perturbed model deviates from the SNR of the unperturbed model, but the deviation increases smoothly with the amplitude of the perturbations. Moroever, the SNR still decays approximately as $1/\sqrt{t}$. These results indicate that the model parameters do not need to be finely tuned. 

\section*{Generalizations of the model}

In the previous sections we considered synaptic models that can be represented by linear chains of dynamical variables. We focused on these models because their simplicity allowed us to illustrate the computational principles we used to design them. However, they appear too simple and orderly to accommodate the complexity and diversity of biological synapses. Here we show that it is possible to construct a broad class of synaptic models that are equivalent to linear chains in terms of memory performance. 
Such complex models can readily be constructed by starting from the undiscretized linear chain model depicted in Fig.~\ref{model} and then iteratively ramifying it. For example, the second beaker could be connected to two identical beakers instead of one, splitting the chain into two. Each of the two beakers would then be connected to a series of progressively larger ones. Pairs of corresponding beakers would have the same total capacity as the associated single beaker of the original chain. This ramification process can be iterated an arbitrary number of times, with any choice of relative importance weights assigned to different branches. Furthermore, such branches can merge again, leading to complex networks of interactions like the one shown in Fig.~\ref{gmodels}A, which are still equivalent to the original linear chain.

In Section~\ref{ramification} we show that if the cross-sections of the tubes are properly tuned these complex models have the same dynamics for the first beaker and therefore the same memory performance as the linear chain models. We also demonstrate that they are robust to relatively large perturbations, such as the complete loss of one interaction pathway, which can be partially compensated by parallel branches. Moreover, such complex synaptic models can readily reproduce various experimental observations, which include delayed LTP/LTD expression \citep[see e.g.][]{OConnor2005}, one form of metaplasticity \citep[e.g.][]{Abraham2008,OConnor2005} and spacing effects \citep[e.g.][]{Anderson2000}. 

Until now we have considered models in which the synaptic efficacy is instantaneously modified by adding or removing liquid from the first beaker. The long-term memory performance, however, remains basically unaltered when liquid is added or removed from other beakers instead, even though the expression of a synaptic modification is delayed by the time it takes the liquid to flow into the beaker representing the efficacy. This suggests that LTP and LTD induction protocols may affect distinct biochemical processes that correspond to different beakers in the model and do not need to operate directly on the same variable. Analogously, the synaptic efficacy does not need to be read out from the first beaker. It could be determined by another beaker or even be some function of the liquid levels of multiple beakers (see also Suppl.~Info.~\ref{readoutschemes}).
If the input beaker is not read out, very recent memories might not be immediately available for retrieval as the liquid has to propagate to the readout beakers first. 

Another natural consequence of the architecture of these synaptic models is metaplasticity, the dependence of plasticity on the history of previous synaptic modifications. 
Here, metaplasticity is an obvious consequence of the existence of hidden variables, represented by the
beakers that are not directly read out to determine the synaptic efficacy.
For example, synapses that undergo a long series of potentiating events become more resistant to depression \citep{OConnor2005}. A long series of LTP induction protocols can significantly increase the liquid levels in several beakers, making it more difficult to stabilize a subsequent synaptic depression. This is illustrated in Fig.~\ref{gmodels}B, in which we plotted the synaptic efficacy as a function of the time elapsed since an LTD induction protocol in two cases: in the first one LTD is preceded by a short series of LTP events, and the depression is relatively stable. In the second case LTD is preceded by a long series of LTPs and the synapse is only transiently modified even though there are now two LTD events. The different degrees of plasticity are determined by different initial conditions of the hidden variables (despite equal initial efficacies), which were set by the previous history of synaptic modifications.

In Fig.~\ref{gmodels}C we show that the model can also replicate the empirical phenomena known as spacing effects. The stability of memories that are stored repeatedly is known to depend on the spacing between the times of memorization. This phenomenon has been observed in several behavioral studies \citep[see e.g.][]{Anderson2000} and more recently in electrophysiology experiments on synaptic plasticity \citep[see e.g.][]{Carew1972,Zhou2003}. In these experiments, when the interval between repetitions is too short or too long, the memories are less stable than in the case in which the repetitions are properly spaced. Our explanation for these observations is surprisingly simple. Consider the analogy with communicating vessels when a synapse is repeatedly potentiated. In the case of long lags, the liquid added during potentiation has time to almost settle to equilibrium between repetitions, leading to little accumulation of synaptic modifications. In the case of massed repetitions on the other hand, one of the dynamical variables may hit its upper bound, which would correspond to liquid spillover in our analogy. The overall effect of potentiation would also be reduced by this loss of liquid.


\begin{FPfigure}
	\begin{minipage}{\textwidth}
		\centering
		\includegraphics[width=6in]{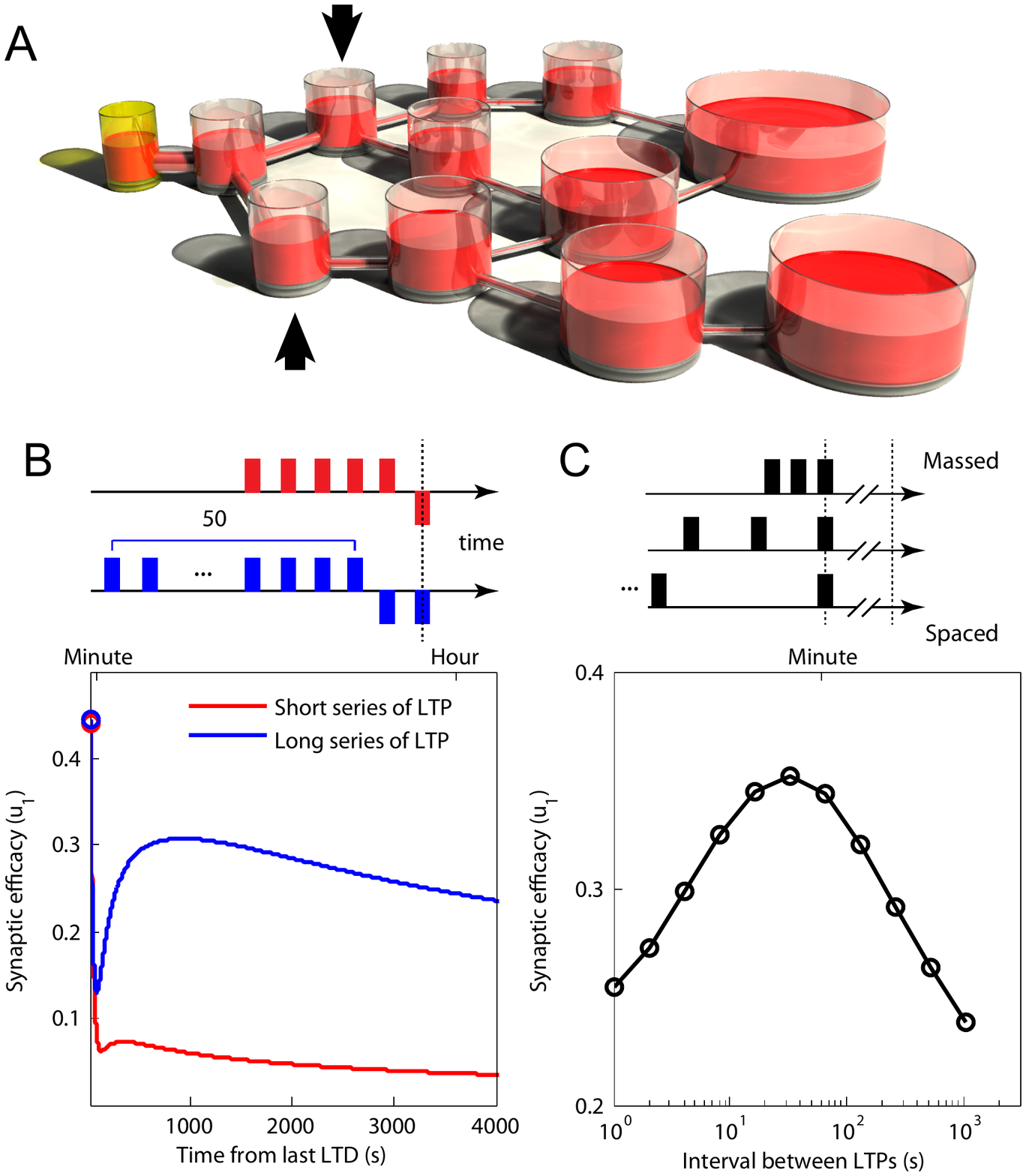}
	\end{minipage}
	\caption[]{A broad class of complex synaptic models with equivalent memory performance. A. A generalization of the model shown in Fig.~\ref{model}, in which each dynamical variable is coupled to two or more other variables. For simplicity we consider continuous dynamical variables (i.e.~not discretized). The model is constructed iteratively starting from the linear chain of beakers of Fig.~\ref{model}. For example, the second beaker is now connected to two beakers on the right. Analogous splittings and mergings lead to the set of beakers of the figures. When the cross-sections of the tubes are properly tuned, the memory performance of the model is the same as for the original linear chain of beakers. B. Metaplasticity: the dynamics of the synaptic efficacy depends on the history of synaptic modifications.  Red: the synapse is depressed at time 0 and the depression is preceded by a short series of 5 LTP events. In this case the last LTD event is still effective and long-lasting. The unit of time is one second. Blue: depression at time 0 is preceded by a long series of 50 LTP events and another LTD event. The depression of the synapse is only transient revealing that the synapse is more resilient to long-term changes than in the case in which depression is preceded by a short series of LTP events. The number of LTP/LTD events has been chosen so that the initial efficacy is approximately the same in the two cases. C. Spacing effect: synaptic efficacy 100 time steps after the end of a sequence of three LTP events which have been spaced differently for different points. Massed repetitions of LTP events (short intervals) and distributed repetitions (long intervals) are less effective than properly spaced ones. The optimal interval is around 40 time steps. The unit of time is again one second, to match the timescales of the experiments described in \citet{Zhou2003}.  }
	\label{gmodels}
\end{FPfigure}

\section*{Testable predictions}

One of the testable quantitative predictions of the theory concerns the rate of decay of memory traces. A power-law decay of the memory SNR approximating $1/\sqrt{t}$ is a signature feature of the models that we discussed. Although it is known that memory decay can be described by power laws in psychology experiments \citep[see e.g.][]{WixtedEbbesen1991,WixtedEbbesen1997}, the power varies significantly from experiment to experiment, and it is difficult to draw conclusions. This variability is probably due to the fact that in most of the experiments the memories are not random and uncorrelated, as subjects often experience the same or similar episodes multiple times and can even internally rehearse previously stored memories.
Consequently, the memory decay depends on the specific statistics of the memories, their relative importance, and the rate at which they are rehearsed or re-experienced. Complex system level processes that deal with such structured memories were not incorporated in our model and in any case could be difficult to control in experiments.

\begin{figure}
	\begin{minipage}{\textwidth}
		\centering
		\includegraphics[width=6in]{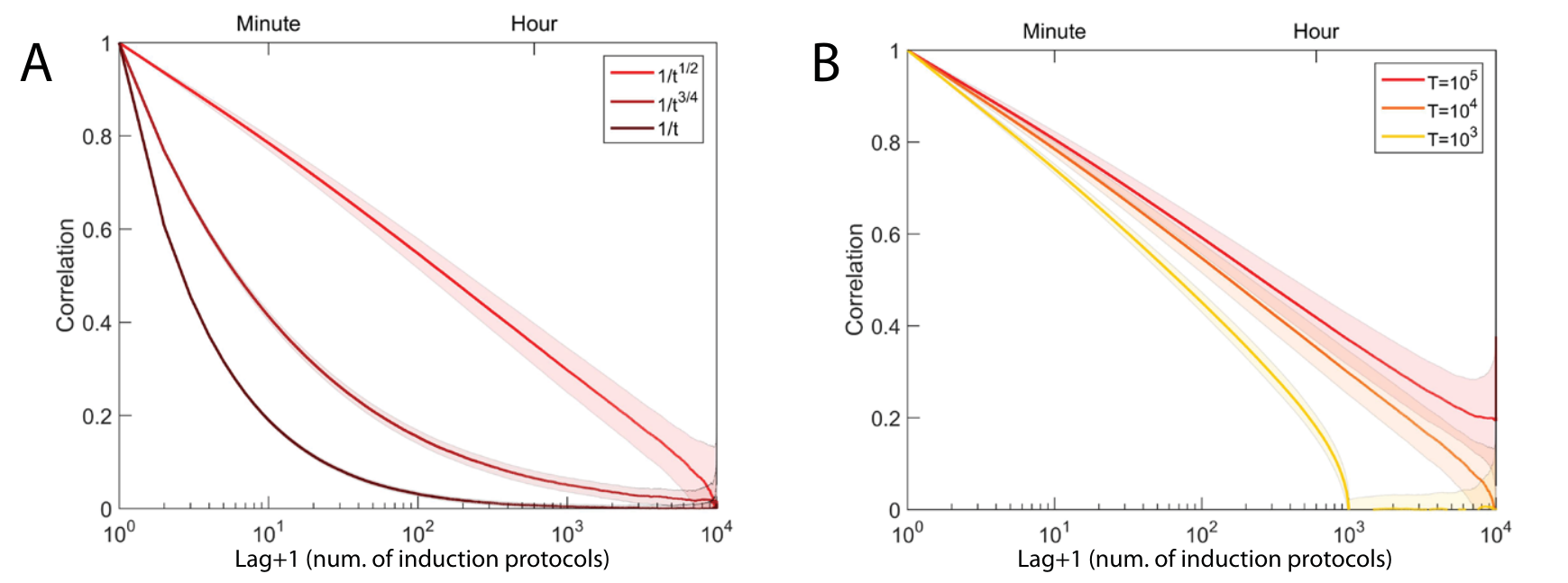}
	\end{minipage}
	\caption[]{Testing the model in experiments. A. Autocorrelation of synaptic efficacy in a simulated experiment in which a synapse undergoes a long random series of 10000 LTP and LTD protocols. Here we assumed a rate of 10 protocols per minute. On the lower horizontal axis we represent the lag expressed in terms of the number of protocols and on the upper axis we represent the time lag. Notice that in both cases the scale is logarithmic. The three curves represent the autocorrelation functions for three different models. Our proposed model (light red) has a distinctive decay, which appears almost as a straight line on a log-linear plot. Other models, with faster decays of the signal to noise ratio, exhibit autocorrelations with a significantly steeper falloff. The shaded areas represent the standard error for 20 repetitions of the experiment. B. Dependence of the autocorrelation function on the longest timescale of the memory system under consideration. The different lines, again plotted on a log-linear scale, correspond to progressively increasing longest timescales. In the limit of very large timescales, this line would become horizontal.
	}
	\label{autocorr}
\end{figure}

A feasible experiment to test our theory would be to repeatedly modify a single synapse (or a population of synapses) and observe how the autocorrelation of the synaptic efficacy decays with time. A balanced random series of LTP and LTD protocols can induce multiple changes in the synaptic efficacy. We expect that the observed autocorrelation would be very broad and its decay only logarithmic on long timescales (shorter than the memory lifetime; see Methods~\ref{autocorrsection} and Suppl.~Info.~\ref{autocorrest} for details). Such a logarithmic decay is a distinctive feature of models with a signal to noise ratio that approximates $1/\sqrt{t}$. As shown in Fig.~\ref{autocorr}A, the autocorrelation is approximately a straight line when plotted against the logarithm of the time lag. Models with faster memory decay ($1/t^{3/4}$ and $1/t$ are shown in the figure) are characterized by autocorrelation functions that decay significantly faster, with a prominent positive curvature. Interestingly, the slope of the line depends on the longest timescale of the memory system under consideration. As this timescale increases, the slope decreases, and the line becomes progressively more horizontal (see Fig.~\ref{autocorr}B).

While there are several technical issues complicating such an experiment, we believe that none of them are insurmountable: First, the duration of the experiment should be long enough to cover at least three orders of magnitude (e.g.~with $1000$ brief induction protocols). Second, LTP and LTD should be suitably balanced. Since one of the two protocols might be more effective than the other, some calibration would be required to avoid imbalance. Unfortunately, the calibration procedure may require a time that is as long as the duration of the experiment throughout which the autocorrelation is measured, as balance should be achieved on all timescales that are considered.



\section*{Discussion}

We have presented a broad class of synaptic models that exhibit a huge memory capacity. These models show that complexity, which is widely observed in all types of biological synapses, is important to achieve long memory lifetimes and strong initial memory traces. Complexity was already shown to be beneficial in previous models, both for synaptic  \citep{FusiDrewAbbott2005} and for systems level memory consolidation \citep{Roxin2013}. In both cases the memory traces were initially stored in fast variables and then progressively transferred to slower variables. Multiple timescales and memory transfer were the two key ingredients needed to achieve simultaneously slow decays of memory traces and strong initial signals.
A $1/t$ decay, with $t$ the age of the memory, led to initial memory traces and memory lifetimes whose magnitudes scale as $\sqrt{N}$, where $N$ is the total number of synapses. We showed here that it is possible to combine the same key ingredients to drastically extend the memory lifetime without sacrificing the initial strength of the memory traces and without substantially increasing the complexity of the synapse (e.g.~the number of dynamical variables). Indeed, the model presented here exhibits a significantly slower decay, approximately $1/\sqrt{t}$, which permits memory lifetimes that scale almost like $N$ instead of $\sqrt{N}$, and initial SNRs that are basically the same as in the old models (see Suppl.~Info.~\ref{comparison} for a direct comparison between models). When considering large systems like the human brain, this is potentially a huge improvement, that has been obtained by introducing bidirectional interactions between fast and slow variables.

Note that in our model the interactions between fast and slow variables are significantly more important than in previous models. In Suppl.~Info.~\ref{independent} we show that it is possible to build a system with non-interacting variables that exhibits a $1/\sqrt{t}$ decay. However, this requires disproportionately large populations of slow variables, which greatly reduce the strength of the initial SNR. In the best case, the memory lifetime scales only like $\sqrt{N}$, and the initial SNR like $N^{1/4}$. Both quantities are substantially worse than in our model with interacting variables. This is not the case for previous models, for which the advantage of interactions was significant, but the scaling properties of both the memory lifetime and the initial SNR were approximately the same in the interacting and non-interacting case.

The proposed synapses are complex, as they require processes that operate on multiple timescales, but the number of these processes is relatively small and scales only logarithmically with the memory lifetime.
This is achieved by properly spacing the characteristic timescales.
As the dynamical variables can all be varied independently, the space of all possible states of each synapse can be huge and grows exponentially with the number of variables (see Suppl.~Info.~\ref{markov}). This is known to allow for slow memory decay~\citep{Lahiri2013}.

The improved performance obtained with slow decays requires some degree of tuning. As we showed, the model is robust to certain types of perturbations of the parameters, but significantly less so to others. Specifically, when described as a set of communicating vessels, the model can tolerate surprisingly large variations of the beaker and tube sizes. However, the memory performance decreases drastically when potentiations are not balanced with depressions (see Suppl. Info.~\ref{robustness}), or when the set of communicating vessels has leaks that are larger than the one of the last beaker.
The necessary forms of tuning increase the memory performance by several orders of magnitude, and therefore are probably encoded genetically and maintained actively by homeostatic mechanisms. A failure of these mechanisms can lead to dramatic consequences. The model sensitivity to the potentiation/depression imbalance could be related to the severe degradation of memory performance observed in the early stages of Alzheimer's disease, when depression becomes significantly more effective than potentiation \citep[see e.g.][]{Shankar2008}.

\subsection*{Optimality of the model}

As previously noted, the approximate $1/\sqrt{t}$ decay of the memory trace is the slowest allowed among power-law decays. Slower decays lead to synaptic efficacies that accumulate changes too rapidly and grow without bound. Interestingly, one can prove (see Suppl.~Info.~\ref{optimization}) that the $1/\sqrt{t}$ decay maximizes the area between the log-log plot of the signal to noise ratio and the threshold (i.e.~the area between the SNR curve in Fig.~\ref{scaling}A and the threshold).
This statement is true not only when one restricts the analysis to power laws, but also when all possible decay functions are considered. One might wonder what would be the rationale behind maximizing the area under the log-log plot of the SNR. The intuitive reason can be summarized as follows: while we want to have a sizable SNR to be able to retrieve a memory from a small cue \citep[see e.g.][]{Krauth1988}, we do not want to spend all our resources making an already large SNR even larger. Thus we discount very large values by taking a logarithm. Similarly, while we want to achieve long memory lifetimes, we do not focus exclusively on this at the expense of severely diminishing the SNR, and therefore we also discount very long memory lifetimes by taking a logarithm. While putting less emphasis on extremely large signal to noise ratios and extremely long memory lifetimes is very plausible, the use of the logarithm as a discounting function is of course arbitrary.

It is interesting to consider also the case in which the SNR is not discounted logarithmically, i.e.~when one wants to maximize the area under the log-linear plot of the SNR. In this situation, the optimal decay is faster, namely $1/t$, as in some of the synaptic models previously considered \citep{Roxin2013,FusiDrewAbbott2005}.

\subsection*{Biological interpretations}

To understand how the proposed model can be implemented by biological processes, it is important to discuss the possible interpretations of its dynamical variables. One possibility is that the variables $u_k$ represent the deviations from equilibrium of chemical concentrations (see Suppl.~Info.~\ref{chemical} for details). The timescales on which these variables change would then be determined by the equilibrium rates (and concentrations) of reversible chemical reactions. However, for the slowest variables, which vary on timescales of the order of years, it is probably necessary to consider biological implementations in which each $u_k$ corresponds to multiple interacting processes. For example, we showed that the slowest variable can be discretized with only two levels, and hence it could be implemented by a bistable process, which would allow for very long timescales \citep{crick1984,miller05,Si2003}. These biochemical processes could be localized in individual synapses, and recent phenomenological models indicate that at least three such variables are needed to describe experimental findings \citep{Ziegler2015}. However, these processes could also be distributed across different compartments of a neuron, across different neurons in the same local circuit or even across multiple brain areas. If two coupled $u_k$ variables reside in different neurons, their interactions must be mediated by processes that likely involve neuronal activity, such as the widely observed replay activity, as proposed in \citet{Roxin2013}. In the case of different brain areas, the synapses containing the fastest variables might be in the medial temporal lobe, e.g.~in the hippocampus, and the synapses with the slowest variables could reside in the long-range connections in the cortex. In all these cases the parameters $N$ and $m$ of the model would have a different interpretation that depends on the specific architecture of the modeled neural circuits, but the scaling properties of the system would be as optimal as in the case that we discussed.

\subsection*{Memory retrieval in simple neural circuits}

The ideal observer approach allowed us to analyze the scaling properties of memory systems with hardly any assumptions about the architecture of the neural circuit, the specific learning rule and the neural representations. However, it is important to test whether these scaling properties are preserved in specific simulated neural circuits. In Suppl.~Info.~\ref{memretrievalsupp} we report the analysis of two simple cases of memory retrieval, which have been used as memory benchmarks in the past. The first one is a simple feedforward perceptron storing random patterns. The second one is a fully connected recurrent neural network similar to the one proposed by Hopfield \citep{Hopfield1982}, whose memory capacity is estimated both in full simulations of the dynamical network and theoretically, as in \cite{AmitFusi1994}, where the signal-noise ratio analysis is basically equivalent to the ideal observer approach used in our article. In both cases, the ideal observer approach predicts that the number of storable memories scales linearly with the number of neurons $N_n$. Note that in the recurrent network the total number of synapses is of $\mathcal{O}(N_n^2)$. However, not all of these synapses receive independent inputs, as different neurons read out presynaptic activity patterns that are highly correlated, since they differ only by one neuron. As the ideal observer approach assumes that the synaptic modifications are independent, one should consider only the $N_n$ independent synapses.

The scaling is verified in simulations in Suppl.~Fig.~\ref{memretrfracvsage} in the case of the perceptron. Interestingly, in these neural circuits we can also study the ability of the readout neuron to generalize. In the case of the perceptron we trained the readout neuron to classify random input patterns and then retrieved memories by imposing on the input neurons degraded versions of the stored patterns. The ability to generalize can be expressed in terms of the minimum overlap between the input and the memory to be retrieved that can be tolerated (i.e.~that produces the same response as the stored memory). This overlap is directly related to the SNR, as previously known (see e.g. \cite{Krauth1987}), and in Suppl.~Info.~\ref{memretrievalsupp} we show analytically and in simulations that in our case it scales like $1/\mathrm{SNR}$. This demonstrates the importance of large SNRs that are significantly above the minimum value which is required to retrieve memories when the cues are undegraded. Large SNRs allow for a significantly larger ability to generalize.

\subsection*{Sparseness and synaptic complexity}

Sparseness is known to be important for increasing the memory capacity of neural circuits such as recurrent neural networks, both for synapses that can vary in a unlimited range \citep{TsodyksFeigelman1988} and for bounded, bistable synapses \citep{AmitFusi1994}. In both cases the number of memories that can be stored scales almost quadratically with the number $N_{n}$ of neurons when the representations are extremely sparse (i.e.~when $f$, the average fraction of active neurons, scales approximately as $1/N_{n}$). This is a significant improvement over the linear scaling obtained for dense representations. However, this capacity increase entails a reduction in the amount of information that is stored per memory.

The synaptic model we propose can also benefit from sparsification, as discussed in Suppl.~Info.~\ref{sparsereps}, where we show that the SNR increases by a factor of up to $\mathcal{O}(1/{\sqrt{f}})$ when a network similar to the one discussed in \citep{AmitFusi1994} is considered. The beneficial effects of sparseness that led to this improvement in memory performance in \citep{AmitFusi1994} were at least threefold: the first one was a reduction in the noise, which occurs under the assumption that during retrieval the pattern of activity imposed on the network reads out only the $f\,N_n$ synapses (selected by the $f\,N_n$ active neurons) that were potentially modified during the storage of the memory to be retrieved. The second one was the sparsification of the synaptic modifications, as for some learning rules it is possible to greatly reduce the number of synapses that are modified by the storage of each memory (the average fraction of modified synapses could be as low as $f^2$). This sparsification was almost equivalent to changing the learning rate, or to rescaling forgetting times by a factor of $1/f^2$. The third one was a reduction in the correlations between different synapses, which is also discussed in the next subsection. 

Our model can also benefit from these effects and from others that are discussed in Suppl.~Info.~\ref{sparsereps}. To quantify the memory improvement we need to specify how the synapses are modified and then read out. In Suppl.~Info.~\ref{sparsereps} we present the analysis of two different learning rules, and show that when $f \sim 1/N_n$, the memory capacity scales approximately quadratically with $N_n$, as in \citep{AmitFusi1994}, but with an initial SNR that is $\mathcal{O}(N_n)$ times larger for our proposed model.

It is important to note, however, that the sparseness $f$ has to scale with the number of neurons of the circuit in order to achieve a superlinear scaling of the capacity. While $f \sim 1/N_n$ may be a reasonable assumption which is compatible with electrophysiological data when $N_n$ is the number of neurons of the local circuits, this is no longer true when we consider neural circuits of a significantly larger size $N_n$. Moreover, sparseness can also be beneficial in terms of generalization, but only if $f$ is not too small \citep[see e.g.][]{BarakRigottiFusi2013}.
For these reasons, sparse representations are unlikely to be the sole solution to the memory problem. Nevertheless, plausible levels of sparsity can certainly increase the number of memories that can be stored, and this advantage can be combined with those of synaptic complexity.

\subsection*{Limitations of our theoretical framework}

Our estimates of memory capacity are based on the ideal observer approach, i.e.~the point of view of an observer who can measure the efficacies of all synapses to retrieve memories. This is clearly unrealistic, as an individual neuron can read out directly only the synapses on its dendritic tree. In the brain the readout is probably implemented by complex neural circuitry, and estimates of the strength of the memory trace based on the ideal observer approach  provide us with an upper bound on the memory signal. We validated our results in two realistic local circuits, but it remains unclear how to perform the validation in large neural systems respecting the observed sparse connectivity and modular organization of the brain. The scalability of such large systems has been studied only in very specific cases \citep[see e.g.][]{Okane1992,Roudi2007} and is an important future direction for our work.

A second limitation, related to the first, arises from the assumption of random and uncorrelated synaptic modifications. Although it is reasonable to assume that the brain processes information to be stored such as to memorize only what is not already in memory, it is known that synaptic modifications are correlated, even when memories are not \citep[see e.g.][]{AmitFusi1994,Savin2014}. For example, synapses on the same dendritic tree share a postsynaptic neuron, and for this reason their efficacies are correlated by many learning rules.
Fortunately, the disruptive effects of these correlations seem to disappear when neural representations are sparse \citep{AmitFusi1994}, as we have seen for a specific neural circuit in Suppl.~Info.~\ref{sparsereps}. Sparsification of the neural representations is not the only way to decorrelate synaptic modifications. The initiation of long term synaptic modifications typically requires the coincidence of relatively rare events. It is not unreasonable to think that these mechanisms can also greatly contribute to the decorrelation of synaptic modifications. If this is the case, the theoretical framework that we developed will be applicable to a large number of memory systems.

\subsection*{Acknowledgements}

We are grateful to L.F.~Abbott and U.S.~Bhalla for many useful comments on the manuscript and for interesting discussions. This work was supported by the Gatsby Charitable Foundation, the Simons Foundation, the Swartz Foundation, the Kavli Foundation and the Grossman Foundation. The illustrations of the beakers were generated using the free ray tracing software POV-Ray.




\clearpage
\setcounter{figure}{0}
\renewcommand{\thefigure}{M\arabic{figure}}
\setcounter{page}{1}
\renewcommand{\thepage}{M-\arabic{page}}
\renewcommand\thesection{M}
\setcounter{subsection}{0}
\section*{Experimental Procedures and Methods}

\subsection{Formal definition of memory signal and noise}
\label{snrdef}

We assume that memories are stored through synaptic modification, with each new memory being encoded in a change in the efficacies of (a subset of) the synapses of a neural network. To formalize this problem we will represent each memory as a random binary pattern $\Delta w_\frak{ij}(t) = \pm 1$ of desired  modifications (with +1 representing potentiation and -1 depression) of the synaptic weights between neurons labeled $\frak{j}$ and $\frak{i}$.
We will consider the components of $\Delta w_\frak{ij}(t)$ to be uncorrelated (both across different memories and different synapses in a certain set), as would be the case if a suitable preprocessing step had decorrelated a stream of incoming patterns for optimal compression. 

Note that we are not considering any particular network architecture and learning rule, but instead we are working with synaptic modifications directly, thus sidestepping the learning rule that would determine them from the activities of pre- and postsynaptic neurons.
This makes sense in the context of the ideal observer approach, where the underlying assumption is that all the information stored in the synaptic weights can be recovered, but of course it must be stressed that it is not obvious a priori whether there exists a network architecture that can in fact read out this information (see also the Discussion section). 

Nevertheless, classical memory models support the notion that the ideal observer approach correctly captures the scaling behavior of the achievable memory performance. For example, in the standard Hopfield model \citep{Hopfield1982} 
the desirable modifications for a set of synapses that share a postsynaptic neuron would be uncorrelated (as assumed above) and a simple signal to noise analysis using the ideal observer approach correctly predicts a memory lifetime that scales linearly with the number of neurons.

If we index the set of $N$ synapses under consideration by $a$ (instead of $\frak{i}$ and $\frak{j}$), the signal relevant for the retrieval of a particular memory that was stored at time $t'$ is given by the overlap between the pattern of the associated (desirable) synaptic modifications $\Delta w_a$ and the current state of the synaptic weights $w_a$ at time $t$:
\[ \label{DefSignal}
\mc{S}_{t'}(t) \equiv {1 \over N} \Big\langle  \sum_{a=1}^N w_{a}(t)\,  \Delta w_{a}(t') \Big\rangle \ .
\]
Here angle brackets indicate an average over the ensemble of random uncorrelated patterns that form the sequence of memories impinging on this set of synapses, and we have assumed for simplicity that the expectation value of $\Delta w_{a}$ vanishes (i.e.~the inputs are balanced), otherwise a term proportional to this expectation value would have to be subtracted from the above. 

Similarly, the corresponding (squared) noise term, again for the pattern stored at time $t'$, is given by the variance of this overlap
\[
\mc{N}^2_{t'}(t) \equiv  \left\langle {1 \over N^2} \Big{(} \sum_{a=1}^N w_{a}(t)\,  \Delta w_{a}(t') \Big{)}^2 \right\rangle - \mc{S}^2_{t'}(t)  \ . \nn
\]
The quotient of the signal and its standard deviation, the signal to noise ratio, is the key quantity to consider when assessing the possibility and fidelity of recall of a previously stored memory. While we have considered a particular pattern stored at time $t'$, we will assume that all memories are initially encoded with the same strength (though it is easy to generalize to a distribution of initial strengths), so that there is nothing special about any one memory. In this context, if the distribution of the synaptic weights reaches a steady state (as it does in the cases we are interested in), the signal to noise ratio really only depends on the time $t-t'$ elapsed since storing the memory in question (i.e.~the age of the memory). Accordingly, we will write it simply as a function of this time difference, which for a wide range of models will be monotonically decreasing.

A good memory system is one that has a large initial signal to noise ratio, such that recent memories can easily be retrieved (using only a small, i.e.~potentially highly corrupted, cue), and a long memory lifetime. The latter is defined as the time elapsed until the signal to noise ratio drops below a certain retrieval threshold, the minimum value of $\mc{S/N}$ at which recall is still possible. The precise value of this threshold will depend on the details of the network architecture and the retrieval dynamics, but as long as it is of order unity this will not affect the scaling results derived below, and thus in what follows we will simply set it to one unless otherwise noted. If the rate of memory storage is constant, the memory lifetime is proportional to the capacity of the system, i.e.~the total number of memories that can be recalled at a given time.
The tradeoff between the two goals of large initial signal and long memory lifetime will be discussed in detail below, and will eventually lead us to optimizing an appropriately defined area under the signal to noise curve that captures the joint target of having as large a signal to noise ratio as possible for as long as possible.

\paragraph{Desiderata for a useful synaptic memory model}

Our aim here is to build a model of long-term memory that exhibits a number of properties we consider essential. We would like our model synapse to be able to learn online (one pattern at a time), and forget gradually and smoothly \citep[without phase transition such as the catastrophic forgetting in standard Hopfield-type models, see e.g.][]{Amit1989}. In addition to exhibiting a large initial signal to noise ratio and long memory lifetime, the synaptic weights should reach a steady state distribution (given constant input statistics) that has support in only a small range of values (i.e.~that does not allow for arbitrarily large weights, or equivalently, weights in a finite range that must be read out with arbitrarily high precision). Note that one can easily obtain a model with bounded synaptic weights by restricting (hard-limiting) the range of a standard unbounded synapse (with plasticity events of unit magnitude) to values of order $\sqrt{N}$ \citep{Parisi1986,FusiAbbott2007}, which is still an unrealistically large number. We will consider much more tightly bounded synaptic weights.
Finally, all this has to be achieved while keeping the complexity of the model rather small, i.e.~avoiding overly baroque internal mechanisms involving too many variables.

\subsection{Abstract models with linear superpositions of memories}
\label{optimaldecays}

\paragraph{Basic assumptions}


In order to build an efficient synapse with bounded weights we will start by considering a continuous synaptic variable with an additive plasticity rule and a time-dependent kernel $r(t-t')$, which we take to be the same for all synapses and plasticity events (i.e.~across all stored patterns):
\[ \label{KernelAnsatz}
w_{\frak{ij}}(t) = \sum_{t' < t} \Delta w_{\frak{ij}}(t')\, r(t-t') \ .
\]
By additive plasticity rule we simply mean that the efficacy $w_{\frak{ij}}$ is a weighted sum over past plasticity events, which we take to be of fixed magnitude $\Delta w_{\frak{ij}}(t) = \pm 1$ (with a plus sign for potentiation and minus for depression). The $\Delta w_{\frak{ij}}(t)$ may be computed from the neural activations $\xi_\frak{i}$ corresponding to the patterns we want to store. For example, they could be determined according to a covariance rule $\Delta w_{\frak{ij}}(t) \propto (\xi_\frak{i} - \langle\xi_\frak{i}\rangle)(\xi_\frak{j} -\langle\xi_\frak{j}\rangle)$, where the $\xi_\frak{i} = \pm 1$ are binary with equal probability for both values (such that $ \langle\xi_\frak{i}\rangle = 0$). Recall could be achieved by the network dynamics \citep[of an auto-associative, Hopfield-type network, see][]{Hopfield1982} that completes the stored pattern of neural activations from a partial (or corrupted) cue $\tilde{\xi}_\frak{i}$.

However, we deliberately divorce our analysis from the choice of learning rule and the network dynamics, by focusing on a subset of synapses that receive statistically independent inputs and taking an ideal observer approach. Successful retrieval of a previously stored memory then requires the signal to noise ratio of this set of synapses to be larger than a certain threshold (which we will set to one).

We are assuming that potentiation and depression events are equally likely\footnote{If this was not the case a homeostatic mechanism would be needed to adjust the relative magnitude of these types of plasticity events in order to achieve a steady state without introducing any explicit bounds on the synaptic variables $w_{\frak{ij}}$. More generally one could imagine a distribution over magnitudes of plasticity events, and again the existence of an equilibrium without explicit bounds on the weights would require a balance condition, namely that the expectation value of the initial size of plasticity events vanishes. Another conceivable generalization would be to introduce different kernels for potentiation and depression events.}, and uncorrelated between different synapses and memories.  In other words, we consider storing random patterns of synaptic modifications in which each bit of each memory can be thought of as determined independently by the flip of an unbiased coin.

\paragraph{Signal to noise ratio} We have introduced a time-dependent kernel $r(t-t')$ above since otherwise the synaptic weight would grow without bound as more and more patterns are stored. This can avoided, however, if $r(t-t')$ decays sufficiently fast as a function of the age of the corresponding memory (i.e.~the time elapsed since storage).

Following the definition of eqn.~(\ref{DefSignal}), the signal (at time $t$) associated with a particular memory is given by the overlap of the corresponding pattern of synaptic modifications (stored at time $t'$) with the current synaptic weights, which using the ansatz~(\ref{KernelAnsatz}) leads to
\[
\mc{S}(t-t') = {1 \over N} \Big\langle  \sum_{a=1}^N w_{a}(t)\,  \Delta w_{a}(t') \Big\rangle = r(t-t') \ , \nn
\]
where the neuronal indices $\frak{i}$ and $\frak{j}$ have now been replaced by a single synaptic index $a$, ranging over the set of synapses under consideration. Combining this with the corresponding noise term, we obtain the signal to noise ratio
\[ \label{SNRKernel}
\mc{S/N}(t-t') = \sqrt{N r^2(t-t')\over \sum_{t'' < t, t'' \neq t'} r^2(t-t'')}  \ .
\]
It will be convenient in what follows to approximate the sum in the denominator by an integral over the full range of past $t''$ values (see also Section \ref{autocorrsection} for details), neglecting the small correction that arises from the fact that there is a term corresponding to $t'' = t'$ missing in the sum (since this term is the signal, rather than part of the noise). The noise will then be represented by an integral of the form $\int_1^\infty r^2(t)\, {\rm d}t$, and thus if the decay kernel is a power law $r(t) = t^{-\gamma}$ is it clear that we must have $\gamma > 1/2$ or else this integral will not converge. Crucially, the divergence of this noise integral also implies that the variance of the synaptic weight would blow up, so that even if we regularized the integral appropriately for $\gamma < 1/2$, the resulting range of synaptic efficacies would be large. Therefore, the slowest power-law decay we can afford is $r(t) = t^{-1/2}$, which is the critical case in which the synaptic variance just starts to diverge (see also Suppl.~Info.~\ref{optimization}).

\subsection{Constructing models by coarse-graining random walks}
\label{construction}

Here we describe the procedure for building a model of a complex synapse that implements the required forgetting curve ($1/\sqrt{t}$) in a natural and parsimonious fashion. We will begin with general considerations of random walks and diffusion processes, and then refine as well as generalize the model step by step, throughout Sections \ref{discretelinearchain} and \ref{ramification}. 

The present section serves primarily to provide a more systematic background for the model construction steps leading from Fig.~\ref{model}C to Fig.~\ref{model}E, and furnish some mathematical details. Reducing the analogy of fluid flowing through a system of communication vessel to its most basic ingredients, we will consider a random walk of particles (which can be thought of as the molecules in the liquid) along a chain of discrete sites (which correspond to the beakers). Even though more abstract and general, this construction is equivalent to that of the main text  in the particular case discussed there.

See also Section \ref{diffusion} for an alternative point of view using the (approximately equivalent) language of diffusion processes, which leads to a particularly simple description of the proposed synaptic dynamics that allows for analytical derivations of a number of important properties of the model.

\paragraph{Linear chain models}


Consider a random walk of particles on a semi-infinite chain in discrete time steps. We denote the number of particles at location $j$ at time $t$ by $v_j(t)$ for $j=1\ldots\infty$. (Note that this number can be negative; we can think of the particle number as being measured relative to a constant background density.) At every time step each particle has a finite probability of moving one step to the left or to the right. This probability is the same for both directions and for all locations except $j=1$, which has no left neighbor. It is easy to see that for such a stochastic process the time derivative of the particle numbers is equal to a discrete Laplacian: $\mr{d}v_j/\mr{d}t \propto v_{j-1} - 2\, v_j + v_{j+1}$ for $j>1$. In other words, we have a spatially discretized diffusion process with constant diffusivity (see top panel of Fig.~\ref{SimpleGraph}).

In order to make contact with systems of exponentially varying diffusivities that we are interested in, we will now consider discretizing the above random walk even further, on a coarser scale. 
We introduce a new set of coarse-grained variables $u_i$ which are located at positions $j=2^{i-1}$ on top of $v_{2^{i-1}}$, i.e.~they are exponentially spaced. Our goal is to find an effective, approximate description of the system in terms of the $u$ variables alone, where we think of each $u_i$ as reflecting the average behavior of the system in the interval between its own location and that of its right neighbor $u_{i+1}$.

We can achieve this by assuming that the particle density profile is piecewise linear, with kinks located only at positions $j=2^{i-1}$, such that all the curvature (which drives diffusion) is concentrated there. We can then use simple linear interpolation to eliminate all the $v_j$ from the equations of motion except those that coincide with the $u_i$.  This would lead to the following expressions:
$\mr{d}v_{2^{i-1}}/\mr{d}t \propto 2^{-i+2}(v_{2^{i-2}} - v_{2^{i-1}}) - {2^{-i+1}}(v_{2^{i-1}} - v_{2^i})$ for $i=2,3,4\ldots$, while the time derivatives of the other $v_j$ (for which $j$ is not a power of two) would vanish, since they are situated in regions of linear particle density.

However, for the piecewise linear approximation to be self-consistent (i.e.~still applicable at the next time step) changing the particle number at the end of a line segment must be accompanied by an appropriate change everywhere along the segment to maintain linearity.
In other words, the time derivative of the endpoint $v_{2^{i-1}}$ must be distributed among all variables along the line segment.
Thus if our effective variable $u_i$ is proportional to $v_{2^{i-1}}$, its time derivative has to be proportional to the average derivative along the line segment to its right\footnote{The details of this coarse-graining procedure are a matter choice, of course. The mathematically inclined reader will notice that it would be more appealing to have a symmetric prescription in which we average over the left and right line segments, or alternatively to think of the $u_i$ variables as living in the middle of a line segment. This would merely change the overall timescale of variation that is unimportant here, so for ease of illustration we stick with a one-sided prescription.}.

There are $2^{i-1}$ variables on this line segment and denoting the constant of proportionality by $\al/2$ this leads to $\mr{d}u_{i}/\mr{d}t =  2^{-2i+2} \al\,  (u_{i-1} - u_i) - 2^{-2i+1} \al\, (u_i - u_{i+1})$, which describes a discretized diffusion process on a logarithmic scale (i.e.~as if viewed on a plot in which the spatial axis is logarithmic).

In such a random walk model a plasticity event would correspond to adding or removing a particle from the leftmost location, which modifies the equation for $i=1$. If we denote this time-dependent input of unit magnitude (and sign discriminating potentiation/depression) by $\mc{I}$ we find  $\mr{d}u_{1}/\mr{d}t =  \mc{I} - 2^{-1} \al\, (u_1 - u_2)$. Similarly, if the chain is not semi-infinite the equation for the last ($m$th) variable will only contain a coupling to its (sole) left-hand neighbor, but we can add a leak (exponential decay) term to it to render the variances of all particle numbers finite. This is easily achieved by simply setting the value of the (non-existent) right hand neighbor to zero, such that
$\mr{d}u_{m}/\mr{d}t =  2^{-2m+2} \al\,  (u_{m-1} - u_m) - 2^{-2m+1} \al\, u_m$.

\paragraph{Model with different ratios of timescales}

While above and in the main text we have chosen parameters that vary as powers of two for ease of illustration, this can easily be generalized to arbitrary exponents
\[ \label{introduce_n}
{\mr{d}u_{i} \over \mr{d}t } =  n^{-2i+2} \al\, (u_{i-1} - u_i) - n^{-2i+1} \al\, (u_i - u_{i+1}) \ ,
\]
which still approximates the desired $t^{-1/2}$ behavior of the Green's function for arbitrary real-valued $n>1$, with ratios of successive timescales of $\mathcal{O}(n^2)$. The tradeoff in the choice of $n$ is that for large $n$ this approximation is not very good (since a superposition of a small number of exponentials leads to a rather bumpy surrogate for a power law), while for $n$ only slightly bigger than unity a large number $m$ of variables are needed to cover a given range of timescales, say between one and $T$, namely $m \sim \log{T}/(2 \log{n})$. 

Note that even within the space of linear (and first order in time) equations with nearest neighbor interactions on a chain we could generalize eqn.~(\ref{introduce_n}) even further by introducing different ratios of successive timescales instead of just one global parameter $n$, while still approximating the inverse square root Green's function.

\subsection{Continuum space limit and diffusion equation}
\label{diffusion}

In the preceding section we discussed a set of first order differential equations describing a random walk of a large number of particles (or equivalently the flow of water between connected beakers).
In this construction space was discrete from the beginning (represented by a number of sites or beakers), but we could have chosen instead to step back even further and start from a model in which space is continuous. This even simpler model, which is highly instructive and allows for an intuitive explanation of important properties such as the $1/\sqrt{t}$ decay, connects the proposed synaptic dynamics to heat diffusion on a line (e.g.~along a thermally insulated wire).

Consider the one-dimensional diffusion equation (with $u(x',t)$ interpreted as the temperature profile along a homogeneous rod)
\[ \label{SimpDiffusion}
{\partial u \over  \partial t} = D\, {\partial^2 u \over \partial x'^2} \ .
\]
Its Green's function for a $\delta$-function input (of one unit of heat energy) at time $t=0$
\[ \label{Greenfunc}
G_u(x',t) = {1 \over \sqrt{4 \pi D t}}\,  e^{-{x'^2 \over 4 D t}} \ ,
\]
decays as $1/\sqrt{t}$ at the origin (i.e.~at $x' = 0$, where the $\delta$-function is localized). Thus if we represent the input to the system by such an instantaneous pulse, the correct decay of the signal is already built in, as long as we read out the synaptic weight at $x'=0$. Since the equation is linear, we can simply superimpose Green's functions for a sequence of such inputs (positive for potentiation and negative for depression), and they will behave as required by eqn.~(\ref{KernelAnsatz}). 

Even though the Green's function we wrote here is for an infinite line, it is symmetric around the origin, and thus we can simply fold the system in half (leading to a Neumann boundary condition) and use the same Green's function (up to a factor of two) on the semi-infinite line. A $\delta$-function input at the origin will then evolve into a half Gaussian bump that will gradually spread out (the peak remaining at the origin) with a standard deviation that grows in proportion to $\sqrt{t}$.

To revert back to the system of communicating beakers described above, we simply have to spatially discretize this diffusion process by chopping up the rod into finite chunks and considering the resulting interactions of the average temperatures of those chunks. The piece closest to the origin corresponds to the synaptic weight, while the other ones give rise to the hidden variables. If all those chunks have the same (say unit) size, this will lead to the system shown in Fig.~1C. While it has the correct decay behavior, the system cannot be of infinite extent. There will be some finite number $m$ of separate chunks, and when the width of the Gaussian bump becomes comparable to the total size of the system, the $1/\sqrt{t}$ decay of the Green's function (\ref{Greenfunc}), which assumes an infinite system, will break down. In other words, if there is a second boundary, we have to choose a boundary condition there, which will modify the power law decay on a timescale $T\sim m^2$. Thus if want to achieve an extensive memory lifetime $T\sim N$, the number of variables that would be required is $m \sim \sqrt{N}$, which is unrealistically large. 

Note that we have assumed that the system is purely diffusive and free of any drift term. If that was not the case, the situation would be even worse, since the peak of the Green's function would move at a finite velocity and hit the second boundary at a time $T\sim m$, so that we would need even more variables ($m \sim N$) to obtain an extensive memory lifetime.

Fortunately, drastically reducing the number of required variables while maintaining a close approximation to power-law decay is not difficult. 
Recall that the (thermal) diffusivity $D$ in eqn.~(\ref{SimpDiffusion}) in general is a ratio of a thermal conductivity $g(x)$ and a heat capacity $C(x)$, and that those can be spatially varying, which leads to the more general diffusion equation  
\[
{\partial u \over  \partial t} = {1 \over C(x)} \,{\partial \over \partial x} \left(g(x)\,{\partial u \over \partial x}\right) \ . \nn
\]
If we break the homogeneity of the system by introducing exponentially varying parameters $C(x) \sim e^{\beta x}$ and  $g(x) \sim e^{-\beta x}$, we obtain the differential equation
\[ \label{BetaDiffusion}
{\partial u \over  \partial t} = {D \over \beta^2 e^{\beta x}} \,{\partial \over \partial x} \left(e^{-\beta x}\,{\partial u \over \partial x}\right) \ ,
\]
parameterized by positive constants $D$ and $\beta$, which has a Green's function given by
\[
G_u(x,t) = {1 \over \sqrt{4 \pi D t}}\,  e^{-{(e^{\beta x}-1)^2 \over 4 D t}} \ . \nn
\]
It describes a signal (in the form of a temperature difference) that propagates only very slowly towards larger $x$. This is because the thermal conductivity 
decreases exponentially while the heat capacity increases with $x$, and thus an input given by a certain amount of heat energy at $x=0$ will lead to a noticeable temperature difference at finite $x$ only at exponentially large times, when $t \sim (e^{\beta x}-1)^2/(4 D)$. Therefore, to reach an extensive memory lifetime the largest value of $x$ we need to consider, which is proportional to $m$, will now only scale as $\log N$. 

Throughout this diffusion process, the heat energy $Q = \int \mathrm{d}x\, C(x)\, u(x,t)$ is a conserved quantity, modulo a leakage term potentially introduced by the second boundary condition at $x \sim m$. Spatially discretizing this system as above leads to the model of communicating vessels shown in Fig.~1E, which achieves the correct power-law decay and extensive memory lifetime with only a logarithmic number of variables.

Note that the two continuum models (\ref{SimpDiffusion}) and (\ref{BetaDiffusion}) we have discussed in this section are in fact equivalent under the change of variables $x' = e^{\beta x} -1$. This implies that there is another way of arriving at the simple linear chain model we want. We can start from a homogeneous diffusion process (constant diffusivity), but instead of discretizing space on a linear scale (into chunks of equal length) we can discretize on a logarithmic scale (i.e.~divide the system into chunks of exponentially increasing size). This is precisely in the spirit in which we have described the transition from the homogeneous random walk (communicating vessels of constant size) to the desired linear chain model in Fig.~1 and Section \ref{construction}. Both are approximations to a simple one-dimensional diffusion process, but spatially discretized in different ways, with the latter being much more efficient in terms of the number of variables needed.

\subsection{Detailed description of the linear chain model in discrete time and with quantized variables}
\label{discretelinearchain}

While above we have written equations for a continuous time system, it is a simple matter to discretize time, as is appropriate for a incoming stream of temporally discrete patterns representing different experiences to be stored. We will choose one time step to correspond to one such memory and write
\[\label{evolgen}
u_{i}(t+1) &=& u_{i}(t) +  n^{-2i+2} \al\, (u_{i-1}(t) - u_i(t)) - n^{-2i+1} \al\, (u_i(t) - u_{i+1}(t)) \ .
\]
Again, the last equation (for $i=m$) is obtained from this by simply setting $u_{m+1}=0$ for all times (thus introducing an exponential decay term on very long timescales), while the first equation (for $i=1$) is modified by introducing the binary input of unit magnitude $\mc{I}(t)$:
\[\label{evol1}
u_{1}(t+1) = u_{1}(t) + \mc{I}(t) - n^{-1} \al\, (u_1(t) - u_2(t)) \ .
\]
The value of $\al$ is a free parameter in these equations that determines the overall timescale of the dynamics (but should be chosen small enough such that the transition matrix on the left hand side of these equations has no negative eigenvalues, which could lead to oscillations). We will take $\al = 1/4$ below and in all numerical experiments. 

Having discretized time we are now left with a model of a complex synapse consisting of a small number $m$ of coupled variables operating in discontinuous time steps (one step per incoming memory) according to the deterministic (given $\mc{I}(t)$) dynamical eqns.~(\ref{evolgen}) and (\ref{evol1}). However, the values of these variables are still continuous, and in the next step we will discretize those as well, thus turning the model into a Markov chain (with inputs given by the random patterns to be stored and stochastic transition dynamics for the $u_i$).   

In order to achieve this discretization, we will simply declare that every variable can only take one of a finite number of values (which we will refer to as levels) at every time step. We assume that these levels are integer-spaced and distributed symmetrically around zero (such that for an odd number of levels the allowed values are integers, while for an even number they are odd multiples of one half), though the algorithm described below can easily be generalized to arbitrary choices of discrete levels. 

For every time step, we first compute the right hand sides of eqns.~(\ref{evolgen}) and (\ref{evol1}), with the $u_i(t)$ from the previous time step  entering as (half) integers (and similarly the input $\mc{I}(t)$ as $\pm 1$). If the resulting $u_i(t+1)$ happens to coincide with one of the quantization levels there is nothing further to be done, but in general the result will fall in between two levels, and in that case we have to decide which of the two neighboring levels will be the new state of that variable. This can be done by independently flipping a biased coin for each such decision, with the odds ratio of the coin (corresponding to one or the other level being chosen) equal to the inverse ratio of the distances from the desired (non-quantized) value to the respective levels, such that the closer one of the neighboring levels will be more likely.  

The number of levels for each variable is finite, and if the right hand sides of eqns.~(\ref{evolgen}) and (\ref{evol1}) lead to a desired update for any variable $u_i$ that would cause it to become larger than the value of its highest level we set it to this level with probability one (and similarly for the lower end of its dynamical range).

This is the fully discretized, stochastic model that we use for simulations, in particular those shown in Figs.~\ref{scaling}, \ref{discrete}, \ref{robust} and \ref{SNRSign}.
It should be stressed, however, that the quantization of the variables is neither necessary for the model to work, nor required for a plausible biophysical implementation. In fact the signal to noise ratio will be somewhat higher without the additional noise introduced by the stochasticity of the random choices between nearby levels (though the scaling behavior appears to be the same).  
However, even though we don't need stable, discrete levels, we do perform this quantization in order to emphasize the fact that the variables never need to be kept track of with high precision (as long as there is no systematic drift) and that there is no biologically implausible information hidden in exactly read out continuous variables.



\subsection{Models on more general graphs}

\label{ramification}

Using the procedure of coarse-graining a random walk with exponentially spaced variables described in Section \ref{construction}, we can construct a broad family of models that are equivalent to the above nearest neighbor chain models, but whose topology is that of an arbitrarily complex graph instead of a one-dimensional chain.

\begin{figure}
	\begin{minipage}{\textwidth}
		\centering
		\includegraphics[width=6in]{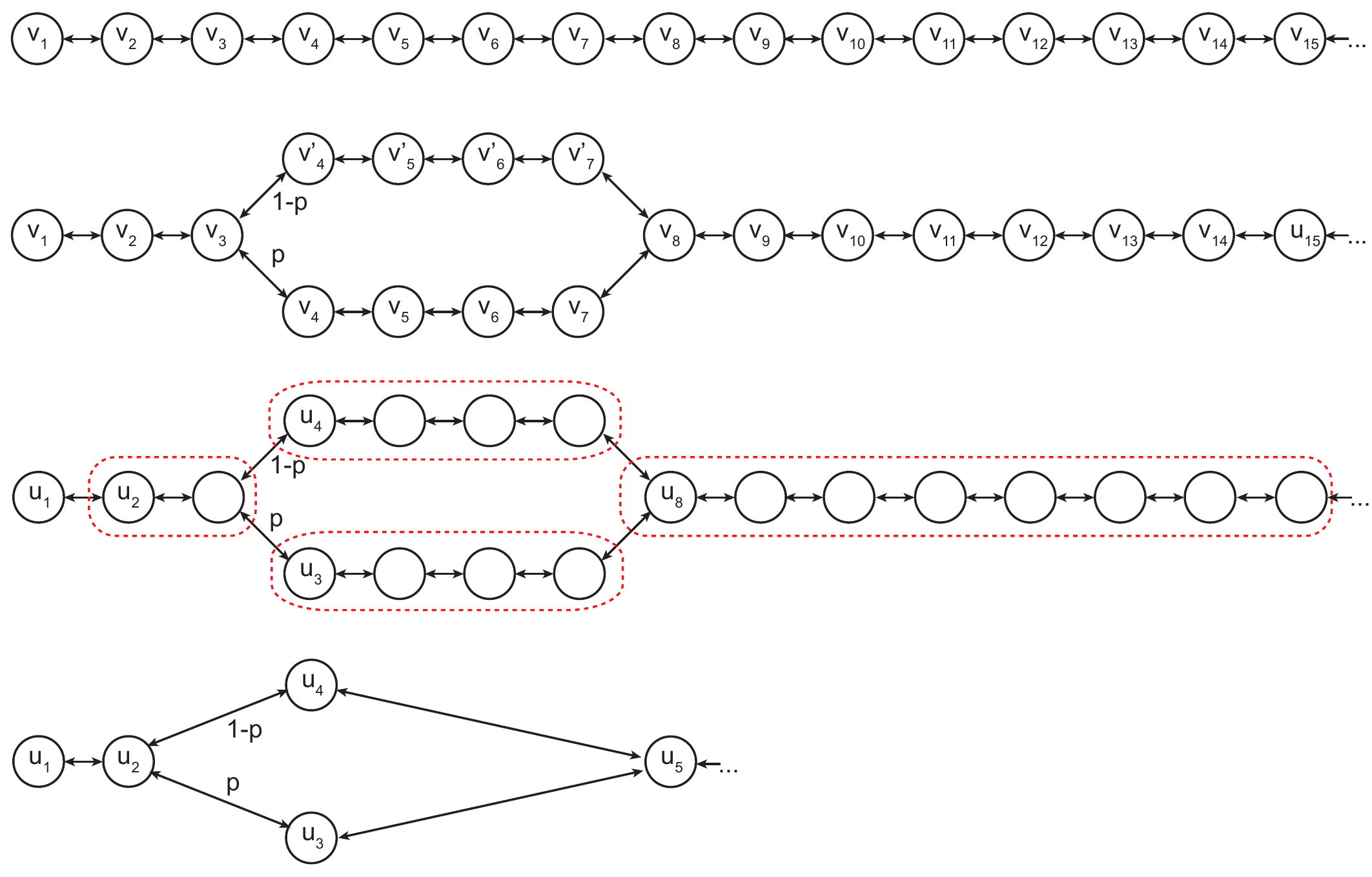}
	\end{minipage}
	\caption[]{Constructing a model with the topology of a simple graph with a single loop. From top to bottom: 
		A simple, unbiased random walk on a homogeneous, semi-infinite chain, where each circle denotes a $v_j$ variable representing the local particle density. A part of the chain is divided into two parallel branches, with assigned importance weights that add to one (determined by the branching probability $p$).
		Coarse-graining using intervals of exponentially increasing length as in Fig.~\ref{model} and Section~\ref{construction} (i.e.~merging variables within regions outlined in red and rescaling coupling constants between them) leads a simple model in terms of the $u_i$ variables, in which $u_1$ has the same dynamics as in the single chain model.
	}
	\label{SimpleGraph}
\end{figure}

Let us illustrate this with a simple example, a graph containing a single loop (see Fig.~\ref{SimpleGraph}). 
We start again from a homogeneous chain of $v_j$ variables, which will be coarse-grained over intervals of exponentially increasing length $\ell$ (growing as powers of two) to obtain the effective $u_i$ variables that appear in our model. 
However, before coarse-graining, we split the chain in two and assign arbitrary importance weights ($p$ and $1-p$, with $0 \le p \le 1$) to the two branches.
Since the dynamical equations are linear, this can be done in such a way that in their influence on the input/output variable $u_1$ (whose behavior directly determines the synaptic efficacy) the combined dynamics of these two branches remains indistinguishable from that of a single branch of unit weight, regardless of the value of $p$.


In the random walk picture of the resulting model, when a particle takes a step to the right from $u_2$, it chooses the lower branch with probability $p$ and the upper one with probability $1-p$. This means that compared to the original single chain the couplings of $u_2$ to its right-hand neighbors $u_3$ and $u_4$ are effectively multiplied by these probability factors (the connecting tubes are narrower, but their sizes add up to that of the original tube connecting $u_2$ and $u_3$ in the single chain). On the other hand, the capacities for $u_3$ and $u_4$ to absorb particles are also multiplied by these weight factors (the sizes of the corresponding beakers are reduced to maintain the same overall capacity), such that all factors of $p$ or $1-p$ cancel in their equations of motion. When the two branches recombine at $u_5$, their importance weights add to unity, and while the couplings to the left still carry the branching factors, the capacity of $u_5$ does not (since this beaker is the same is in the original chain). This leads to the set of equations
\[
{\mr{d}u_{2} \over \mr{d}t } &=& {\al \over 4} (u_1 - u_2) + {\al\, p \over 8} (u_3 - u_2) + {\al\, (1-p)\over 8} (u_4 - u_2) \ ,\nn \\
{\mr{d}u_{3} \over \mr{d}t } &=& {\al \over 16} (u_2 - u_3) + {\al \over 32} (u_5 - u_3) \ , \nn \\
{\mr{d}u_{4} \over \mr{d}t } &=& {\al \over 16} (u_2 - u_4) + {\al \over 32} (u_5 - u_4) \ , \nn \\
{\mr{d}u_{5} \over \mr{d}t } &=& {\al\, p \over 64} (u_3 - u_5) + {\al\, (1-p) \over 64} (u_4 - u_5) + {\al \over 128} (u_6 - u_5) \ .\nn
\]
while the equation for $u_1$ will be the same as in eqn.~(\ref{evol1}) for $n=2$. Note that e.g.~the equation for $u_3$ is the same as for the single chain according to eqn.~(\ref{evolgen}), and if the two chains didn't merge again at $u_5$ similarly all variables to the right of $u_2$ would obey the same equations as the corresponding ones in the original model.
It is easy to see that $p\, u_3 + (1-p)\, u_4$ is the only combination of the branched variables that affects their neighbors, and it is this combination that also appears in the conserved quantity of the system.

We can easily generalize this procedure to construct more complex graphs with multiple (perhaps partially or fully recombining) branches that are still completely equivalent to a linear chain for any choice of branching probabilities as long as corresponding variables on parallel branches (vertically aligned in the diagrams) are coarse-grained using intervals of the same length.

Even more generally, we can construct a large class of models that are only approximately equivalent to a linear chain, by choosing different discretization intervals on parallel branches, and choosing different patterns of separating and recombining branches that are not commensurate with exponentially increasing spacing. This will still give an efficient approximation to the desired behavior for $u_1$ as long as the mean interval size (e.g.~taking a weighted average over parallel branches) increases roughly exponentially going from left to right.

\begin{figure}
	\begin{minipage}{\textwidth}
		\centering
		\includegraphics[width=4.5in]{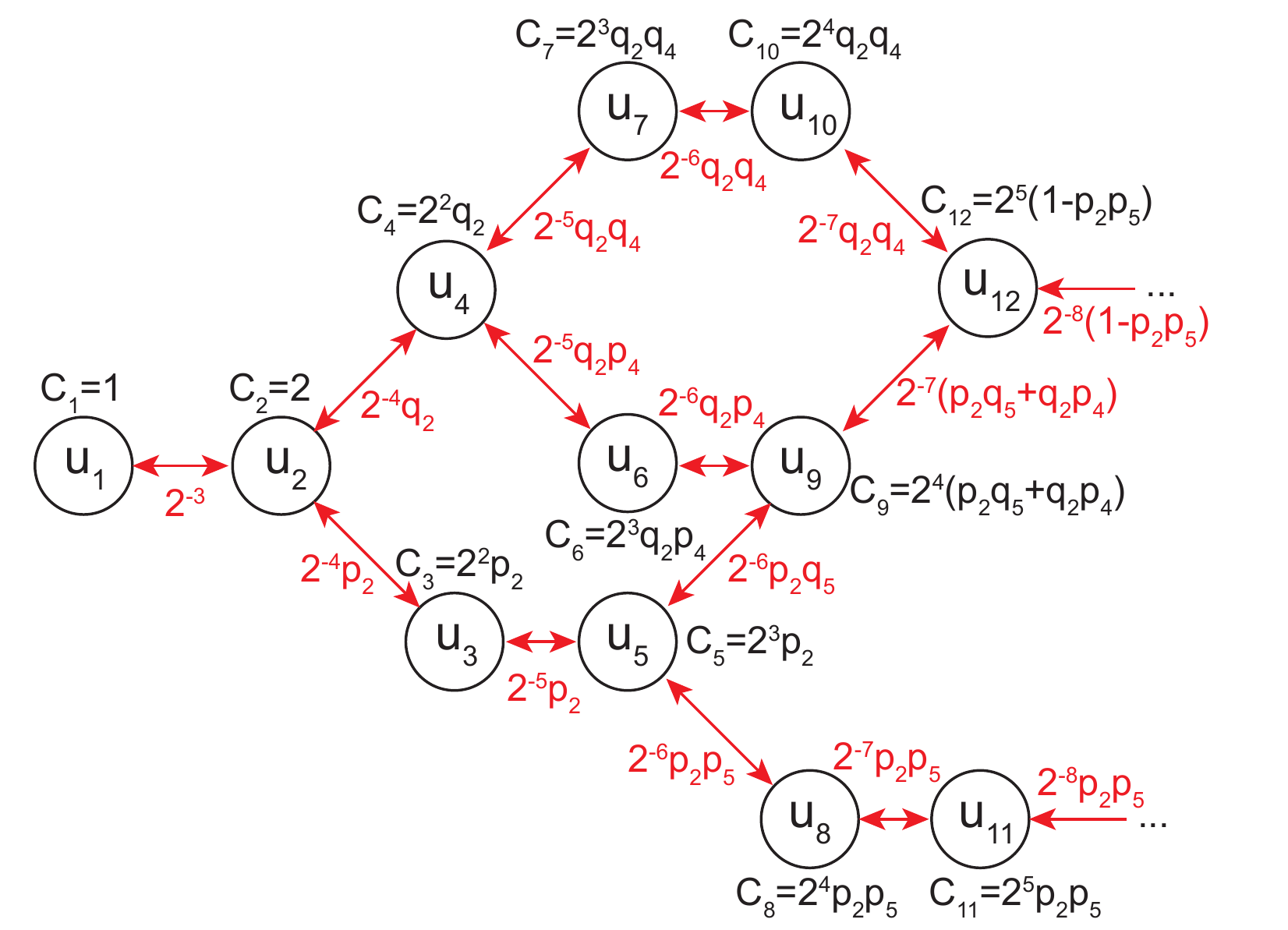}
	\end{minipage}
	\caption[]{Diagram of a more complicated graph model with multiple partially recombining branches, corresponding to the network of beakers shown in Fig.~\ref{gmodels}A. We label the $u_i$ variables from left to right and bottom to top, with branching probabilities $p_i$ carrying the index of the associated $u_i$, and $q_i$ defined as $1-p_i$. The capacities $C_i$ are shown above the corresponding variables, and the coupling constants $g_{ik}$ are given in red next to the arrows connecting pairs of interacting variables (or variables to the reservoir). Using these parameters in eqn.~(\ref{GraphLap}), the dynamics of $u_1$ is again equivalent to that in a single chain model.}
	\label{ComplexGraph}
\end{figure}

Let us now describe an algorithm for choosing appropriate parameters for such a model (which subsumes the simple example discussed above). We are given an undirected graph with junctions at the locations of (some of) the $u_i$ and a set of branching probabilities $p_i$ for each dividing junction (in general there may be more than two branches emerging at one junction, in which case one could use probabilities $p_{ik}$ for the branch connecting $u_i$ to $u_k$). As above, we can formalize the problem in terms of coupling constants $g_{ik}$ between variables and their neighbors, and capacities $C_i$, which need to be chosen such as to endow the system with the correct ($\sim 1/\sqrt{t}$) Green's function for the first variable (see Fig.~\ref{ComplexGraph}). The right hand side of the equations of motion will take the form of a weighted Laplacian on the graph:
\[ \label{GraphLap}
{\mr{d}u_{i} \over \mr{d}t } &=& {1 \over C_i} \sum_{k\, \mr{adjacent\, to}\, i} g_{ik}(u_k - u_i) \ ,
\]
where $g_{ik} = g_{ki} = \al\, \omega_{ik}/(2 \ell_{ik})$ and $C_i = \sum_{\mr{right\, neighbors}\, k} \omega_{ik} \ell_{ik}$ (summing over edges to adjacent variables $u_k$ to the right of $u_i$, unlike in eqn.~(\ref{GraphLap}) where the sum is over all adjacent variables). Here $\ell_{ik}$ is the length of the interval (the number of steps in the random walk construction of Fig.~\ref{SimpleGraph}) between neighboring variables $u_i$ and $u_k$, while $\omega_{ik}$ is the importance weight of the branch connecting them. The branch weights $\omega_{ik}$ are computed by traversing the graph from the input to the right, multiplying by the appropriate branching probability $p$ whenever a dividing junction is encountered, and adding the corresponding weights when two branches recombine (such that the $\omega_{ik}$ always add to one along any vertical cut through the graph). To illustrate this simple prescription, we have spelled out in the schematic of Fig.~\ref{ComplexGraph} the capacities and coupling constants for the case of the complex model of Fig.~\ref{gmodels}A.

As above, the equation for the first variable $u_1$ will be modified by the addition of the input term, and the equations for the rightmost variables by the addition of exponential decay terms (couplings to the reservoir). Note that if there are multiple branches extending all the way to the right, appropriate leakage terms should be added to all of them.

One should bear in mind that we have set $n=2$ in this section only for illustrative purposes, to provide an intuitive connection to integer-spaced random walks, but the construction of this family of models on graphs can be extended to arbitrary real-valued $n$ without difficulty. 
The distances $\ell_{ik}$ that determine the parameters $C_i$ and $g_{ik}$ don't have to be integers. They would be proportional to $n^i$ in the linear chain model (for which the right neighbors always have $k=i+1$ and the importance weights $\omega_{ik}$ along the sole branch are all equal to one), such that eqn.~(\ref{GraphLap}) agrees with eqn.~(\ref{introduce_n}) up to a trivial rescaling of the overall timescale $1/\al$. 
For a more general graph laid out as in Fig.~\ref{SimpleGraph}, we can take $\ell_{ik}$ to be the horizontal distance between adjacent variables, assuring that the total distance between two given nodes is the same along any branch connecting them.


------------------------------------------------------------
\clearpage
\setcounter{figure}{0}
\renewcommand{\thefigure}{S\arabic{figure}}
\setcounter{page}{1}
\renewcommand{\thepage}{S-\arabic{page}}
\renewcommand\thesection{S}
\setcounter{subsection}{0}

\section*{Supplemental Information}

\subsection{Linear superpositions of memories: optimal decays}
\label{optimization}

Using the abstract model of Section~\ref{optimaldecays} we can consider optimizing (certain objective functionals depending on) the signal to noise ratio of eqn.~(\ref{SNRKernel}) over all possible decay functions $r(t)$. This can be achieved using an integral approximation and variational arguments as follows.

\paragraph{Maximizing the signal to noise ratio}

We want to maximize the doubly logarithmic area $\mc{A}(T)$ 
under the signal to noise curve (normalized here for a single synapse)
\[
\mc{A}(T) = \int_0^{\log T} \mr{d}(\log t)\ \log\left[ {r(t) \over \sqrt{\int_1^T \rm{d} t' \, r(t')^2}} \right]  = \int_1^T {\mr{d} t \over t} \log r(t) \ - {\log T \over 2}  \log\left[ \int_1^T \mr{d} t \, r(t)^2 \right] \ ,  \nn
\]
on a fixed time interval $1 \le t \le T$, under the assumption that at time $T$ the signal to noise ratio will still be above (or at) the retrieval threshold (since we really only care about the area above the threshold, and without this assumption we would have to take the positive part of the logarithm of $\mc{S/N}$ over the threshold before integrating). A simple variational argument proceeds by perturbing $r(t)$ by an arbitrary function of bounded range $\eta(t)$ with a small coefficient $\ep$, i.e.~$r(t) \to r(t) + \ep\, \eta(t)$, and equating to zero the variation
\[
{\delta \mc{A}(T) \over \delta \ep} = \int_1^T {\mr{d} t \over t} {\eta(t) \over r(t)} -  {\log T \over 2}  {\int_1^T \mr{d} t \,2\, r(t) \eta(t) \over \int_1^T \mr{d} t' \, r(t')^2} = \int_1^T \mr{d} t \, \eta(t) \left[ {1 \over t \, r(t)} - {\log{T} \, r(t) \over \int_1^T \mr{d} t' \, r(t')^2} \right]=  0 \ , \nn
\]
Since this has to hold for all $\eta(t)$ it implies $r(t) = \sqrt{ \int_1^T \mr{d} t' \, r(t')^2 }\Big/\sqrt{t\, \log T}$ and thus (noting that the integral on the right hand side is simply a number, which has to be finite for the solution to be valid) we have $r(t) \sim t^{-1/2}$.

\paragraph{Alternative optimization using Lagrange multipliers}

A slightly more standard approach to optimizing our objective, which ultimately leads to the same result, but avoids nested integrals, is to use a Lagrange multiplier $\la$ to keep the noise integral fixed, while maximizing the (logarithm of the)  signal. The modified utility functional reads
\[ \label{AltOptLagMul}
\hat{\mc{A}}(T)  = \int_1^T {\mr{d} t \over t} \log r(t) \ - \la \left[ \int_1^T \mr{d} t \, r(t)^2 - \hat{\mc{N}}^2 \right] \ , 
\]
Using the same variational expansion $r(t) \to r(t) + \ep\, \eta(t)$ we find that
\[
{\delta \hat{\mc{A}}(T) \over \delta \ep} = \int_1^T {\mr{d} t \over t} {\eta(t) \over r(t)} - 2\, \la \int_1^T \mr{d} t \, r(t) \eta(t)  = \int_1^T \mr{d} t \, \eta(t) \left[ {1 \over t \, r(t)} - 2\,\la\, r(t) \right] = 0\quad \Rightarrow \ r(t)= {1 \over \sqrt{2\, \la\, t}} \ .\nn
\]
The value of the Lagrange multiplier is fixed by $\hat{\mc{N}}^2 = \int_1^T \mr{d} t \, r(t)^2 = \log T /(2 \la)$.

\paragraph{Scaling behavior of the optimal solution}

For an inverse square root time dependence of the memory trace, the resulting signal to noise ratio for $N$ synapses behaves as $\mc{S/N}(t) =  \sqrt{N /(t \log T) }$\,, which means that the memory lifetime $t^*$ at which it drops to one is $t^* =  N / \log T $. In particular, if we choose $T = N$, we would obtain a memory lifetime $t^* = N / \log N$, which is however not quite consistent with our assumptions, since $t^* < T$, i.e.~the signal to noise ratio drops below the threshold before the cutoff time. On the other hand, choosing $T = N/\log N$ leads to $t^* = N / (\log N - \log\log N) = N / \log N +N (\log\log N)/(\log N)^2 + \ldots$,
which is consistent as we now have $t^*>T$.
The optimal choice for $T$, where it is equal to $t^*$, is determined by the equation $T = N / \log T$, which does not have a solution in terms of elementary functions. However, we can expand the resulting $t^*$ for large $N$ and find that the first two terms are the same as above, and thus we conclude that  we can achieve a memory lifetime $t^* = \mc{O}(N/\log N)$. The coefficient in front of $N/\log N$ depends on the value of the retrieval threshold of course (and is one for our arbitrary choice of unit threshold).
The corresponding initial signal to noise ratio scales as $\mc{S/N}(t=1) = \mc{O}(\sqrt{N/\log N})$.

\paragraph{Alternative regularizers}

We have optimized the area under the signal to noise curve in a finite interval $1 \le t \le T$, which may not appear to be exactly what we want, since it involves choosing a cutoff $T$ (and for a specific application it may not be clear to begin with what $T$ should be). However, since the optimal solution $r(t) \sim t^{-1/2}$ is independent of $T$, we are free to just set it a posteriori to any value smaller than or ideally equal to the memory lifetime.

Nevertheless, recalling that we want to construct a time-invariant system, we should really be considering the area above the threshold on an semi-infinite interval $t \ge 1$, in which case there is an issue with naively extending the solution $r(t) \sim t^{-1/2}$ to infinitly large $t$. The cumulative noise term $\hat{\mc{N}}^{2} = \int \mr{d} t \, r(t)^2$ now receives contributions from all previously stored memories (even arbitrarily old ones whose signal to noise ratio has dropped below threshold) and thus diverges logarithmically as the range of integration is extended to infinity, which means that the solution needs to be regularized to be viable. Of course everything would be perfectly consistent if we simply declared that $r(t)$ decayed as the inverse square root of $t$ until time $T$, then instantaneously dropped to zero and remained there for all times larger than $T$ (i.e. $r(t)=0$ for $t>T$). This is what we have effectively assumed in our derivation above when we set the integration range of the noise term to extend up to the cutoff time $T$ only. However, such discontinuous behavior does not seem very natural, which is why in what follows we will be considering other regulators that allow us to smoothly extend $r(t)$ as $t \to \infty$.

One simple regulator involves modifying our solution by an exponential with a long timescale $\tau$ (comparable to $T$ above), such that $r(t) = t^{-1/2}\, \exp(-{t-1 \over 2 \tau})$ decays rapidly for $t \gg \tau $. The resulting signal to noise ratio, after performing the sum in eqn.~(\ref{SNRKernel}), is given by
\[
\mc{S/N}(t) =  \sqrt{N \over  - t\,e^{t/\tau}\log(1-e^{-1/\tau}) - 1}\ \xrightarrow{\tau \to \infty}\  \sqrt{N \over  t \log \tau} \ .\nn
\]

Alternatively, we could modify the power law to decay just slightly faster than the inverse square root of time, such that $r(t) = t^{-(1+\ve)/2}$, and therefore
\[
\mc{S/N}(t) =  \sqrt{N \over  t^{1+\ve}\, \zeta(1+\ve) - 1}\ \xrightarrow{\ve \to 0}\  \sqrt{N \ve \over  t} + \mc{O}\left(\sqrt{N \ve^3 \over t}\,  \log  t\right) \ ,\nn
\]
where $\zeta$ is the Riemann zeta function. In both cases, upon setting $\tau = N$ and $\ve = 1/\log N$, respectively, we find that the leading order scaling behavior remains $\mc{O}(\sqrt{N/\log N})$ for the initial signal to noise ratio, and $\mc{O}({N/\log N})$ for the memory lifetime.

\paragraph{Different choices of objective functionals} 

One might wonder why we choose to optimize the area under the log-log plot of the signal to noise ratio, rather than some other functional of the signal to noise curve. The intuition behind this is clear: while we want to have a big $\mc{S/N}$ to be able to retrieve a memory from only a small (highly corrupted) cue, we do not want to spend all our resources making an already large signal to noise ratio even larger, and thus we discount very large values by taking a logarithm. Similarly, while we want to achieve long memory lifetimes, we do not focus exclusively on this at the expense of severely diminishing $\mc{S/N}$, and therefore we also discount very long memory lifetimes by a taking a logarithm\footnote{In particular, there would be no point in having memory lifetimes longer than the lifetime of the animal in question, which can also be encoded by an appropriate choice of cutoff $T$ above.}. While putting less emphasis on extremely large signal to noise ratios and extremely long memory lifetimes is very plausible, the use of the logarithm as a discounting function is of course simply a matter of mathematical convenience, and one could certainly imagine choosing a different functional form of the objective to be optimized.

Let us briefly discuss what would happen if one did not employ such a discounting procedure. First, one might be tempted to simply maximize the memory lifetime, i.e.~not discount very long lifetimes. This would be appropriate if we did not care about memory retrieval from small cues, which require large signal to noise ratios, but we were in a (somewhat artificial) situation where we only ever needed to recover memories from a cue of a fixed level of accuracy, such that there would be little benefit to having a $\mc{S/N}$ larger than some number of $\mc{O}(1)$. In that case the objective functional, along the lines of eqn.~(\ref{AltOptLagMul}), would be
\[
\hat{\mc{A}}_{\mr{lin}}^{\mr{log}}(T)  = \int_1^T \mr{d} t \log r(t) \ - \la \left[ \int_1^T \mr{d} t \, r(t)^2 - \hat{\mc{N}}^2 \right] \ \Rightarrow \ {\delta \hat{\mc{A}}_{\mr{lin}}^{\mr{log}}(T) \over \delta \ep} = \int_1^T \mr{d} t \, \eta(t) \left[ {1 \over r(t)} - 2\,\la\, r(t) \right] = 0 \ ,\nn
\]
which implies $r(t)= 1/ \sqrt{2 \la}=$ constant. Similarly, if we also did not discount large signal to noise ratios and simply optimized the area on a linear plot, we would find
\[
\hat{\mc{A}}_{\mr{lin}}^{\mr{lin}}(T)  = \int_1^T \mr{d} t\,  r(t) \ - \la \left[ \int_1^T \mr{d} t \, r(t)^2 - \hat{\mc{N}}^2 \right] \ \Rightarrow \ {\delta \hat{\mc{A}}_{\mr{lin}}^{\mr{lin}}(T) \over \delta \ep} = \int_1^T \mr{d} t \, \eta(t) \left[ 1 - 2\,\la\, r(t) \right] = 0 \ ,\nn
\]
such that $r(t)= 1/ (2 \la)$, which is again constant. In both cases the noise term $\hat{\mc{N}}^2$ diverges linearly with $T$, which means that an appropriate regularization is pertinent. If we simply cut off the memory trace at $t=T$, such that $r(t)$ becomes a step function, we would find $\mc{S/N}(t) \sim \sqrt{N/T}$ for $1 \le t \le T$ and zero otherwise\footnote{This amounts to adding the last $T$ plasticity events with equal weights to compute the synaptic efficacies, and when embedded in a recurrent network (typically using a covariance learning rule) is essentially a time-translation invariant version of the Hopfield prescription \citep{Hopfield1982}, for which the number of storable memories scales linearly with $N$. Notice that in the case of the Hopfield network $N$ is also the number of neurons. Even though in a fully connected network of $N$ neurons there would be of order $N^2$ synapses, they would not all be statistically independent because different neurons receive basically the same input (any two neurons share $N-2$ inputs). A set of statistically independent synapses would be those on the dendritic tree of any particular neuron (for independent inputs), and therefore no larger than the total number of neurons.}.
Alternatively, if we use an exponential regulator $r(t) = \exp(-{t-1 \over 2 \tau})$, as proposed in \citet{Mezard1986},
eqn.~(\ref{SNRKernel}) leads to $\mc{S/N}(t) \sim \sqrt{N/\tau} \exp(-{t-1 \over 2 \tau})$. In both cases choosing the cutoff timescale ($T$ or $\tau$) to be $\mc{O}(N)$ leads to a memory lifetime that is also $\mc{O}(N)$, but at the cost of reducing the (initial) signal to noise ratio to $\mc{O}(1)$. Note also that while such a model would achieve a slightly better scaling of the memory lifetime (by a factor of $\log N$), the range of the synaptic weight (which is tied to the noise term) would have to grow polynomially as $\mc{O}(\sqrt{N})$, as in the proposal of \citet{Parisi1986}, rather than logarithmically. This means that a discretized synapse would require a huge number of discrete levels.

Finally, we might decide that for a particular application we really care about having a very high signal to noise ratio, even at the expense of a shorter memory lifetime, and thus discount only the latter by a logarithm, which leads to
\[
\hat{\mc{A}}_{\mr{log}}^{\mr{lin}}(T)  = \int_1^T {\mr{d} t \over t}\, r(t) \ - \la \left[ \int_1^T \mr{d} t \, r(t)^2 - \hat{\mc{N}}^2 \right] \ \Rightarrow \ {\delta \hat{\mc{A}}_{\mr{log}}^{\mr{lin}}(T) \over \delta \ep} = \int_1^T \mr{d} t \, \eta(t) \left[ {1 \over t} - 2\,\la\, r(t) \right] = 0 \ .\nn
\]
In this case we find an inverse proportionality of the memory trace to time, $r(t) = 1/ (2 \la t)$, which is approximately realized by the cascade and multistage models of \citet{FusiDrewAbbott2005,Roxin2013}. The noise integral $\hat{\mc{N}}^2$ is convergent as $T \to \infty$, such that no further regularization is needed (and correspondingly the required range of the synaptic variables is $\mc{O}(1)$ only).
The resulting signal to noise ratio behaves as $\mc{S/N}(t) \sim \sqrt{N}/t$, which means that both its initial value and the memory lifetime are $\mc{O}(\sqrt{N})$, i.e.~the lifetime is significantly reduced.

In summary, we could certainly write down other objective (utility) functionals to optimize, which would lead to different plasticity-rigidity tradeoffs.
Our solution has the benefit of exhibiting polynomial scaling with the optimal exponents (up to logarithmic corrections) for both the initial signal to noise ratio and the memory lifetime, with a smooth power-law behavior in between, and importantly requires only a small dynamical range of $\mc{O}(\sqrt{\log N})$ for the synaptic weight.
By modifying the objective, as discussed above, or deforming the $t^{-1/2}$ solution (e.g.~consider the power-law regulator with finite $\ve$), one could either increase the initial signal to noise ratio 
from $\mc{O}(\sqrt{N/\log N})$ to $\mc{O}(\sqrt{N})$ at the expense of a shorter memory lifetime, or improve the memory lifetime from $\mc{O}(N/\log N)$ to $\mc{O}(N)$ while paying a price in terms of the initial memory strength.

\subsection{Variance and spatial correlations of the dynamical variables}
\label{varianceandspatialcorr}

We can use the continuum diffusion model of Section \ref{diffusion} to obtain some insights on correlations of the dynamical variables. Recall that in the (spatial) continuum limit of our model, each variable corresponds to an interval on the one-dimensional space along which the synaptic inputs diffuse (here we are essentially undoing the coarse graining that turns a single partial differential equation for $u(x',t)$ into a set of coupled differential equation for the $u_i(t)$ variables). The impulse response of the model on an infinite line is described by the Green's function of eqn.~(\ref{Greenfunc}), 
and restricting to non-negative $x'$ by introducing a boundary at the origin merely modifies this by a factor of two (the Neumann boundary condition $\partial u/\partial x' = 0$ imposing vanishing flux at $x'=0$ is automatically respected for inputs at the origin due to the symmetry of the setup). 

Introducing a second boundary with a Dirichlet boundary condition (namely $u = 0$) at a particular value of $x'$ to limit the space to a finite interval is somewhat tedious, but feasible using the method of images. For our purposes, however, it is simpler to stick with a semi-infinite line, and model the essential effect of the second boundary by introducing an exponential cutoff term with a long timescale $T$. In other words, we can use the Green's function for a one-dimensional diffusion equation with an added decay term, which reads
\[ 
G_u(x',t) = {1 \over \sqrt{\pi D t}}\,  e^{-{x'^2 \over 4 D t}}\, e^{-{t \over T}}\ . \nn
\]
The exponential cutoff ensures that the diffusion process cannot proceed for much longer than a time $T$, which would correspond to the time it takes to reach the hypothetical second boundary. While $u(x',t)$ is not strictly zero there (as would be the case for a true boundary with Dirichlet condition), it becomes exponentially small for larger $x'$.

For a series of inputs (plasticity events) at integer-spaced times $t'$ we can write in analogy to eqn.~(\ref{KernelAnsatz}) 
\[
u(x',t) = \sum_{t' < t} \Delta w(t')\, G_u(x',t-t') \ , \nn
\]
where the value of $u(x',t)$ in the vicinity of $x'=0$ corresponds to $w(t)$, and the decay of the Green's function close to the origin to the kernel  
$r(t-t')$. For $\langle \Delta w \rangle = 0$ the expectation value of this sum vanishes, and with $\langle \Delta w(t)\, \Delta w(s)  \rangle = \delta_{t,s}$ the spatiotemporal covariance is equal to
\[
\left\langle u(x'_1,t_1) u(x'_2,t_2) \right\rangle = \sum_{t' < {\rm min}(t_1,t_2)} G_u(x'_1,t_1-t')\, G_u(x'_2,t_2-t') \ . \nn
\]
We are particularly interested in the correlation at equal times $t_1=t_2$, and as above we approximate the sum by an integral, which leads to the (stationary) spatial covariance 
\[\label{BesselCov}
\left\langle u(x'_1) u(x'_2) \right\rangle \sim \int_0^\infty {\rm d} t \,  {1 \over \pi D t}\,  e^{-{{x'_1}^2 + {x'_2}^2 \over 4 D t}}\, e^{-{2 t \over T}} = {2 \over \pi D}\,  K_0\left(\sqrt{2 ({x'_1}^2 + {x'_2}^2) \over D T} \right) \ ,
\]
expressed as a modified Bessel function of the second kind. 

\begin{figure}
	\begin{minipage}{\textwidth}
		\centering
		\includegraphics[width=3in]{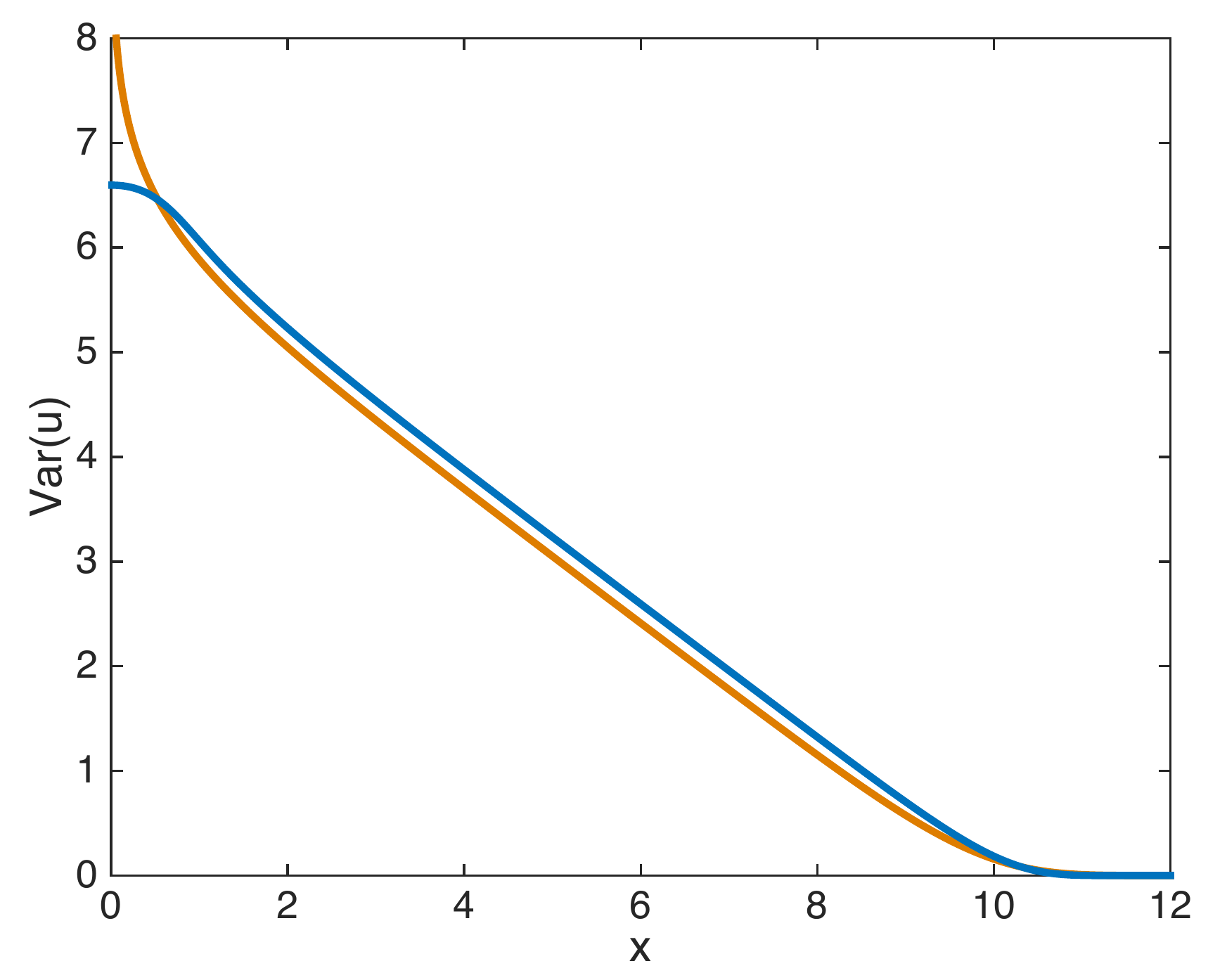}
	\end{minipage}
	\caption[]{Plots of the approximate variances $\langle u^2 \rangle$ that follow from eqns.~(\ref{BesselCov}) and (\ref{GammaCov}) with $x'_1 = x'_2 = e^{\beta x}-1$ vs~$x$, in orange and blue, respectively. Here we set the parameters (which are related to the constants $\alpha$, $m$ and $n$ of the discretized model) to $\beta=1$, $D=1$ and $T=10^9$.}
	\label{LinVar}
\end{figure}

Unsurprisingly, this implies that there are long-range correlations, since the Bessel function becomes (exponentially) small only for arguments much larger than one, which would require at least one of $x'_1$ or $x'_2$ to lie beyond the second (softly imposed) boundary.
Because the right hand side of eqn.~(\ref{BesselCov}) depends only on the combination ${x'_1}^2 + {x'_2}^2$, plotting it for the case $x'_1 = x'_2 = x'$ (which gives us the variance) is sufficient to also read off the spatial covariance. 
We use the relation $x' = e^{\beta x}-1$ to rewrite this in terms the spatial variable $x$ that is linearly related to the index $i$ of the variables $u_i$ in the discretized model (such that each variable is associated with an $x$ interval of equal size; see Section \ref{diffusion}), and plot the variance vs~$x$ in Fig.~\ref{LinVar}.

This plot diverges for $x \to 0$, because we have naively taken the lower limit of integration to be zero in eqn.~(\ref{BesselCov}).
It is easy to resolve this issue by introducing a finite lower limit of the integral, and in a further approximation we can also replace the soft exponential cutoff by a hard upper limit of $T/2$. This leads to the following expression involving incomplete Gamma functions
\[ \label{GammaCov}
\left\langle u(x'_1) u(x'_2) \right\rangle \sim \int_{1/2}^{T/2} {\rm d} t \,  {1 \over \pi D t}\,  e^{-{{x'_1}^2 + {x'_2}^2 \over 4 D t}} = {1 \over \pi  D}\,\Gamma\left(0,{{x'_1}^2 + {x'_2}^2 \over 2 D T}\right)-{1 \over \pi  D}\, \Gamma\left(0,{{x'_1}^2 + {x'_2}^2 \over 2 D}\right) \ ,
\]
which is finite at the origin. In fact its value at $x'_1=x'_2=0$, which corresponds to the variance of the synaptic efficacy, is proportional to $\log{T}$. Since in the discrete linear chain model we have $T = \mathcal{O}(n^{2m})$, and the standard deviation of the synaptic weight distribution closely approximates the memory noise (multiplied by a factor of $\sqrt{N}$), this is in agreement with the $1/\sqrt{m}$ scaling behavior of the initial signal to noise ratio shown in Fig.~\ref{scaling}C.

After the change of variables from $x'$ to $x$ we can again plot the variance that follows from eqn.~(\ref{GammaCov}); see Fig.~\ref{LinVar}. The crucial observation, both from this graph and the plot of eqn.~(\ref{BesselCov}), is that the variance decreases almost linearly over a wide range of $x$ values (as long as we are not close to one of the boundaries). Coarse graining this picture to obtain the spatially discrete linear chain model leads to variances of the variables $u_i$ that are proportional to $\log{T}$ and fall off linearly along the chain (i.e.~with the index $i$). This is consistent with the simulations shown in Fig.~\ref{discrete}, and explains why the long timescale variables require only small dynamical ranges.

\subsection{Parameter variations, potentiation/depression imbalance, and model deformations}
\label{robustness}

\begin{figure}
	\centerline{\includegraphics[width=6.5in]{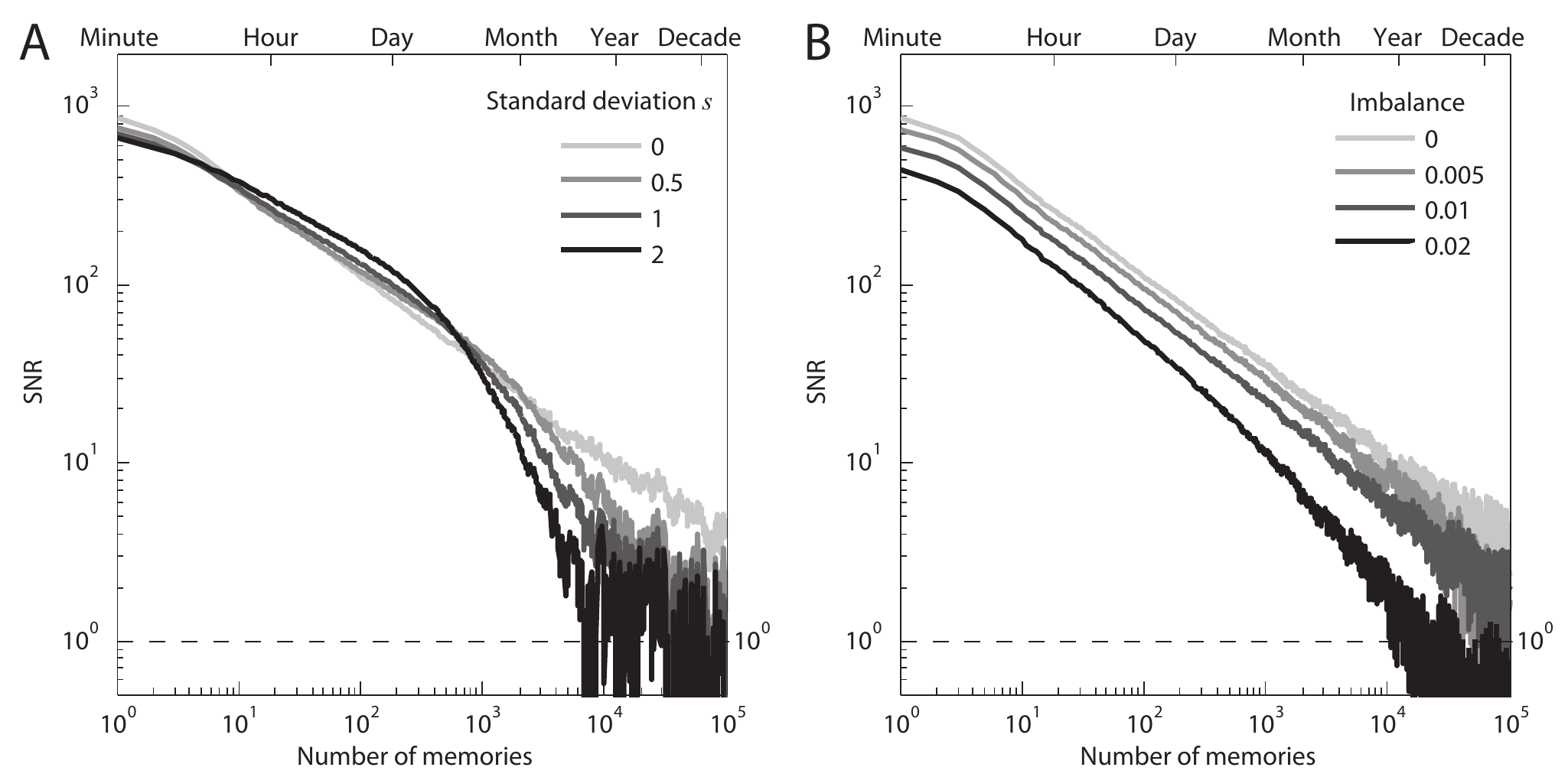}}
	\caption[]{Robustness: Effects of parameter variations and potentiation/depression imbalance. A. SNR vs number of memories when the parameters of the model are perturbed. The coupling constants between different dynamical variables (i.e.~the cross-section of the tubes connecting different beakers) are multiplied by stochastic variables drawn independently from a log-normal distribution. 
		The mean of these variables is one and their standard deviation $s$ is indicated in the figure for the different curves. Darker curves correspond to larger perturbations. The SNR of the perturbed models deviates from the $1/\sqrt{t}$ behavior. However, the difference in memory performance is still quite modest even for standard deviations that are rather large. Notice that for the curve corresponding to the smallest non-zero perturbation, the standard deviation is already as large as 50\% of the mean value of the perturbed parameter. B. SNR vs number of memories when the rates of potentiation and depression are increasingly imbalanced. The power law forgetting curves are very similar, but they are shifted downwards significantly, even for fairly small imbalances. The memory performance is rather sensitive to an uncompensated potentiation/depression imbalance.
	}
	\label{robust}
\end{figure}

As illustrated in Fig.~\ref{model}, to achieve the $1/\sqrt{t}$ decay we have tuned the parameters of the model and assumed that the synaptic modifications are balanced. Given the heterogeneity and variability in biological systems, it is important to test whether the model is robust enough to operate also when the parameters are not finely tuned. We show in Fig.~\ref{robust}A that the SNR of the perturbed model clearly deviates from the SNR of the unperturbed model. However, the deviation increases slowly and smoothly with the amplitude of the perturbations. The SNR still decays approximately as $1/\sqrt{t}$, but the time at which the exponential breakdown appears becomes progressively shorter as the perturbations grow. 

It is important to stress that for long timescales there are still synapses that retain the tracked memory. In other words, the memory signal is still significantly different from zero in a subpopulation of synapses which happen to be well tuned. When reading out all synapses, this signal is too small compared to the noise. However, a smart selection mechanism, as the one suggested in \citet{Roxin2013}, would enable the neural circuit to read out the memory even when the SNR of the entire synaptic population is too small.

The memory performance degrades differently when the synaptic modifications are imbalanced (see Fig.~\ref{robust}B). The decay remains almost unaltered, but the SNR curves are shifted downwards. Although fine-tuning does not seem to be needed, the memory system is clearly sensitive to imbalances in the statistics of potentiation and depression. 

Note that the required balance condition refers to the effective rates of potentiation and depression events that actually occur (or more precisely that would occur if there were no limits on the dynamical range), not to the relative rates that might be imposed by a certain learning rule depending on the neural activity. If for example, given a certain learning rule and statistics of neural activations, potentiation events were called for more frequently than depressions, a homeostatic mechanism internal to the synapse could compensate for this by scaling down the relative magnitude of potentiation steps (or the probability of actually executing it when demanded by the learning rule).
As long as the statistics of neural inputs do not fluctuate too strongly, such a mechanism can achieve potentiation/depression balance using only information locally available to the synapse.

The essential effect of such a balance condition is to keep the steady state distributions of the synaptic variables centered, and prevent it from becoming too concentrated near one of the boundaries of their dynamical range, since desirable modifications that would take one of the variables beyond the limits of its dynamical range are not possible, and therefore lead to errors that degrade the memory trace.

While the model is robust to detuning the parameters of its dynamical equations (see Fig.~\ref{model} or more generally eqn.~(\ref{GraphLap})), there are certainly ways of deforming it that will drastically reduce the memory performance, by effectively introducing new terms in the equations.
As in any memory model, there is an almost conserved quantity, which in our model is a weighted sum $\sum_k C_k u_k$ analogous to the total amount of liquid in all beakers. This quantity would track the sum of all past inputs (i.e.~plasticity events adding or removing liquid), except for the leakage term that connects the longest timescale beaker(s) to the reservoir, which causes it to slowly decay towards zero on a timescale $T$. Because of this term, the total amount of liquid is not precisely conserved (i.e.~equal to the sum of past inputs), but only on timescales shorter than $T$ (corresponding to a weighted sum of past inputs with an exponential discount factor). 

Recall that this leakage takes the form of an unbalanced term on the right hand side of the last one of the dynamical equations (which otherwise consist of balanced difference terms, representing nearest neighbor interactions). This term is obtained by setting $u_{m+1}=0$ in the case of the linear chain in eqn.~(\ref{introduce_n}). It serves an important purpose, which is to keep the variance of the dynamical variables finite, i.e.~it ensures that the system can settle into a steady state. However, if there were any additional leakage terms of this form with a coefficient larger than $\mathcal{O}(1/T)$ in the dynamical equations of any of the $u_k$ variables, it would introduce a new decay timescale inversely proportional to that coefficient, and this would limit the memory lifetime such a synapse could achieve. 

More generally, whatever the biophysical mechanisms may be by which a synapse stores information for a time $T$, they clearly have to respect the existence of some quantity that is at least approximately conserved on timescales of order $T$.

\subsection{More general readout schemes and synaptic non-linearities}
\label{readoutschemes}

We have assumed up to now that the synaptic efficacy is simply equal to $u_1$, and therefore, like all the variables $u_k$, has a bounded dynamical range. However, we could consider more general expressions for the synaptic efficacy, which may in principle depend on all of the internal variables $u_k$, and need not be linear. 

Such a general readout scheme would allow for an even larger class of models. Of course, reading out internal variables $u_k$ for $k>1$ would contradict the notion that these variables are hidden, i.e.~internal to the synapse, and therefore do not communicate with the outside world (the neural network) directly\footnote{Note in particular that it was precisely the idea of having only a single variable that serves as both the input and the output of the synapse that led to models with bidirectional interactions between the variables. Information enters via $u_1$, is distributed across longer timescale variables, and those in turn have to back-react on $u_1$, because that is where the information has to eventually be read out again.}. For this reason, we will not pursue such general output schemes further here. 

We will, however, briefly discuss a more restricted class of readouts that depend only on $u_1$, albeit in a non-linear fashion. In this case the synaptic efficacy could take the form of a sigmoidal function of $u_1$, possibly restricting the already limited dynamical range of $u_1$ even further. In the limit of infinite (central) slope this output degenerates to a binary readout that takes values of say $\pm 1$, in which case we can write the synaptic efficacy as $\sign(u_1)$. Such a model would thus represent a binary synapse, but with complex internal dynamics determining the binary value of its efficacy, and with inputs still acting directly on $u_1$. (Other mappings of $u_1$ to larger discrete sets are certainly conceivable as readouts, leading to synapses with more than two possible values of the efficacy.)

\begin{figure}
	\begin{minipage}{\textwidth}
		\centering
		\includegraphics[width=3in]{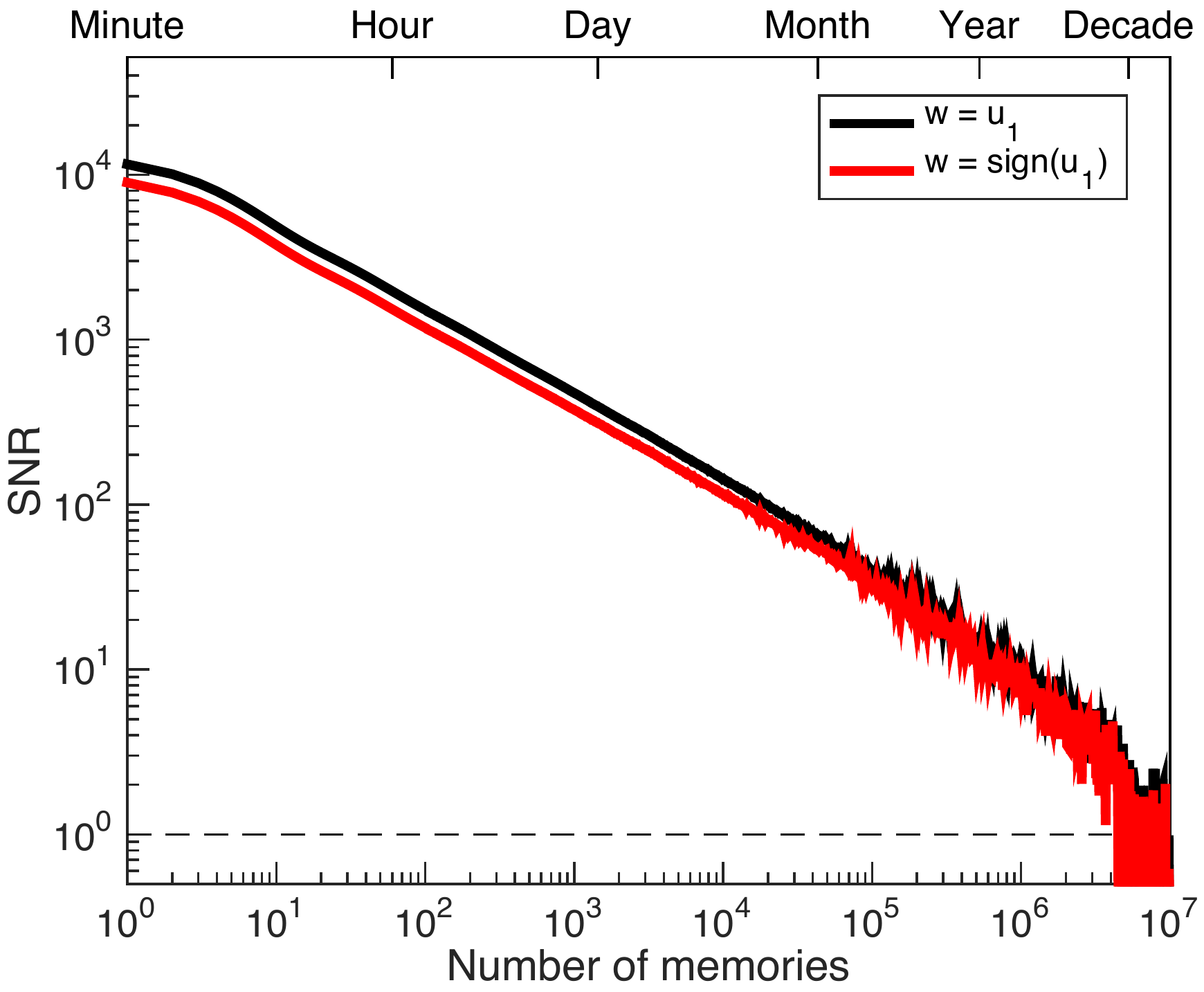}
	\end{minipage}
	\caption[]{Plot of the SNR in the fully discretized model vs~$t$ for two different readouts: $u_1$ in black and $\sign(u_1)$ in red. The parameters are the same as for the black curve in Fig.~\ref{scaling}A, with $m=10$, $N=5.4 \times 10^9$ and $40$ levels for each variable.}
	\label{SNRSign}
\end{figure}

Interestingly, reading out $\sign(u_1)$ instead of $u_1$ only has a small effect on the signal to noise ratio. Its absolute value is slightly lower,  reflecting the fact that we are now reading out at most one bit of information per synapse (compared to potentially several bits if the synaptic efficacy is equal to $u_1$, which may be continuous or discretized into multiple levels depending on which version of the model we consider), but the functional form of the signal to noise ratio and the resulting scaling behavior appear to be unchanged; see Fig.~\ref{SNRSign}.
This is not unlike taking the sign of the weight of a simple unbounded synapse after adding the contributions from all memories (i.e.~offline learning), which is known to lead to an extensive memory capacity \citep{Sompolinsky1986}, except that here the learning happens online and using bounded variables.

\subsection{Autocorrelation function} 
\label{autocorrsection}

Here we compute the autocorrelation function that follows from the simple ansatz in eqn.~(\ref{KernelAnsatz}), which proposes a synaptic weight that is a linear combination of past plasticity events multiplied by an age-dependent decay term
\[ 
w(t) = \sum_{t' < t} \Delta w(t')\, r(t-t') \ . \nn
\]
The variance of the weight of such a synapse (again assuming $\langle \Delta w(t) \rangle = 0$) is given by
\[
\sigma^2(t) = \langle w(t)^2 \rangle = \sum_{t'<t}\, \sum_{t''<t} \langle \Delta w(t')\, \Delta w(t'') \rangle\,  r(t-t')\, r(t-t'')  =  \sum_{t'<t} r^2(t-t') \ , \nn
\]
since the expectation value of a product of two synaptic modifications (of unit magnitude) leads to a Kronecker delta $\delta_{t',t''}$. Similarly, the covariance of the values of the synaptic weights at two different times is given by
\[
C(t,t+\Delta t) = \langle w(t)\, w(t+\Delta t) \rangle &=& \sum_{t'<t}\, \sum_{t''<t+\Delta t} \langle \Delta w(t')\, \Delta w(t'') \rangle\,  r(t-t')\, r(t+\Delta t-t'')   \nn \\ &=&  \sum_{t'<{\rm min}(t,t+\Delta t)} r(t-t')\, r(t+\Delta t - t') \ . \nn
\]
Note that the expression that appears inside the sum over $t'$ above depends on $t$ and $t'$ only in the combination $t-t'$, which is the age of the synaptic modification (and thus we can rewrite these formulae accordingly, summing over the age of memories).  If the synaptic weight distribution reaches a steady state, as is the case if the kernel $r(t-t')$ decays sufficiently fast (such that the sums converge), the variance $\sigma^2$ will be constant in time, and the covariance will depend only on the delay $\Delta t$. 

In case the kernel function does not quite decay fast enough it has to be regularized appropriately. A simple way to achieve this is by stipulating that the synaptic weight depends only on modifications $\Delta w$ of ages up to some (large) time constant $T$, thus cutting off the sums above after at most $T$ steps. In other words, we effectively alter the kernel function such that it vanishes for all but the $T$ most recent modifications (but is otherwise unchanged). A similar effect could be achieved in a different manner, by multiplying the original kernel function by an exponential with a long timescale $T$, leading to a soft cutoff (as discussed in Suppl.~Info.~\ref{optimization}).

The autocorrelation function $A(\Delta t)$ is then given by the ratio $C(\Delta t) / \sigma^2$. For example, if the kernel under consideration was $r(t) = t^{-1}$, we would find a variance $\sigma^2 = \sum_{t=1}^\infty t^{-2} = \pi^2/6$, and a covariance $C(\Delta t) = \sum_{t=1}^\infty t^{-1}(t+\Delta t)^{-1} = H_{\Delta t}/\Delta t$ where $H_k$ represents the $k$th harmonic number. Since these infinite sums converge, no regularization is needed in this case, though it could be incorporated if desired, leading to slightly more complicated expressions again involving harmonic numbers.

For more general kernels, however, possibly with cutoff, we won't be able to perform the required sums in closed form, and it will be useful to resort to an integral approximation. By analytic continuation of the decay function to real arguments, and replacing the sums by integrals, we obtain the following approximate autocorrelation function $\tilde{A}(\Delta t)$
\[ 
\tilde{A}(\Delta t) = {\int_1^{T-\Delta t} r(t)\, r(t+\Delta t)\, {\rm d}t \over \int_1^T r^2(t')\, {\rm d}t'} \ . \nn
\]
Here we have incorporated the cutoff $T$ in the upper limit of the integrals (assuming positive $\Delta t$), but it is easy to remove this cutoff by extending the range to infinity ($T \to \infty$) if this doesn't cause the integrals to diverge. The numerical accuracy of this expression could be improved by a continuity correction in the limits of the integrals, but we will not pursue this further here\footnote{Typically for monotonically decreasing decay functions most of the error in the approximation is accrued close to the lower limit, which one should correspondingly pick somewhat smaller than one to reduce this discrepancy. The expressions for the autocorrelation given in this section can be generalized rather easily to arbitrary lower limits of the integrals.}.

For the optimal kernel $r(t) = t^{-1/2}$ with cutoff $T$ the above integral expression leads to
\[
\tilde{A}_T(\Delta t) = {2 \over \log(T)}\, \log \left(\frac{\sqrt{T}+\sqrt{T-\Delta t}}{1+\sqrt{1+\Delta t}}\right) \nn \ ,
\]
for $\Delta t \le T-1$. First taking the limit of large cutoff $T$ and then also taking the delay $\Delta t$ large (though still much smaller than the cutoff) this reduces to
\[ \label{AutoTLimit}
\tilde{A}_T(\Delta t)\ \xrightarrow{T \to \infty,\, \Delta t \to \infty}\ 1 - \frac{\log(\Delta t/4)}{\log(T)} + \mathcal{O}\left({\Delta t^{-1/2} \over \log(T)},\, \frac{\Delta t}{T \log (T)}\right) \ .
\]
Thus if we plot the autocorrelation function versus $\log(\Delta t)$ it will be well approximated by a straight line with a slope determined by the cutoff time in this regime. There are, however, significant deviations from this behavior for small $\Delta t$ and for $\Delta t$ close to the cutoff.

Another kernel that we are interested in is the more general power law $r(t) = t^{-(1+\ve)/2}$, in particular for small $\ve>0$. In this case the integrals converge and we do not have to impose a cutoff (though it could be added without much difficulty). We find
\[
\tilde{A}_\ve(\Delta t) =
\frac{\ve\, \pi ^{3/2}\, 2^{\ve }\, \Delta t^{-\ve } \csc (\pi  \ve )}{\Gamma\left(1-\frac{\ve }{2}\right) \Gamma\left(\frac{1+\ve}{2}\right)}-\frac{2\, \ve\,  (1+\Delta t)^{(1-\ve)/2} \, _2F_1\left(1,1-\ve ;\frac{3-\ve }{2};-\frac{1}{\Delta t}\right)}{\Delta t \, (1 - \ve)} \ , \nn
\]
which involves gamma and hypergeometric functions. This expression represents the autocorrelation function for an arbitrary power-law kernel (decaying faster than $t^{-1/2}$). However, we can take similar limits as above, first $\ve \to 0$ and then large $\Delta t$, which leads to
\[
\tilde{A}_\ve(\Delta t)\ \xrightarrow{\ve \to 0,\, \Delta t \to \infty}\ 1 - \ve \log(\Delta t /4) + \mathcal{O}\left(\ve\, \Delta t^{-1/2},\, \ve^2 \log^2(\Delta t)\right) \ . \nn
\]
Again the autocorrelation function approximates a straight line in this regime when plotted against $\log(\Delta t)$, here with slope minus $\ve$, and if we identify $\ve$ with $1/\log(T)$ this is in fact the same limiting expression as in eqn.~(\ref{AutoTLimit}) for the inverse square root kernel with cutoff $T$. In other words, both small positive $\ve$ and large cutoff $T$ can act as approximately equivalent regularizations of the inverse square root kernel\footnote{It should be noted, however, that for finite $\ve$ and $T$ the autocorrelation function $\tilde{A}_\ve(\Delta t)$ looks quite different from $\tilde{A}_T(\Delta t)$ for sufficiently large $\Delta t$, since it does not have to go to zero at a finite cutoff, and exhibits a long tail with positive curvature.}.

This also gives us an idea of how closely we could possibly expect an observed synaptic decay function to approximate the idealized inverse square root kernel. If the observed decay is described by a power law with exponent minus $(1+\ve)/2$, this is sufficient for the synapse to achieve a memory lifetime $T \sim \exp(1/\ve)$, i.e.~it does not actually have to come extremely close to the $t^{-1/2}$ case to exhibit a substantial memory capacity.

\subsection{Estimating autocorrelation functions from finite length experiments}
\label{autocorrest}

In order to measure the autocorrelation function experimentally by observing the response of a synapse to a random sequence of plasticity protocols, it is necessary to first calibrate the relative rate of potentiation and depression protocols. Since one does not know the relative rate of (desired) potentiation and depression events the synapse in question would experience in its natural environment, and furthermore cannot be sure that such natural plasticity events are similar in magnitude to those induced by the chosen experimental protocols, using a random sequence of protocols of sufficient strengths to observe an effect and with a predetermined fraction of potentiation events would likely drive the synaptic efficacy to the edge of its dynamical range rather quickly. The synaptic weight may become effectively stuck in its most depressed state (if the potentiation protocol is weaker or less frequent than the synapse is accustomed to) or in its strongest state (if the potentiation protocol is too frequent or effective, and vice versa for the depression protocol), which would prevent us from measuring the relevant autocorrelation function.

Thus the experimenter has to ensure that the relative frequency of potentiation and depression protocols is chosen such that the synaptic efficacy is not systematically pushed up against one of its boundary values. The offending drift term due to a potentiation/depression imbalance (relative to the natural input conditions for the synapse in question) has to be kept small enough to prevent this from happening throughout the measurement period. 
In other words, it has to be smaller in magnitude than the dynamical range divided by the length of the experiment, which in turn implies that the calibration period required to achieve this has to be of a similar duration as the experiment itself (assuming that homeostatic mechanisms inside the synapse cannot adapt to changing input statistics on timescales as short as the length of the experiment). 
Also, the strength of the plasticity protocols has to be chosen just large enough for a single one of them to effect an observable change in the synaptic weight (at least with some finite probability), but if at all possible not so large that the magnitude of this change would be comparable to the full dynamical range of the synapse.

In addition to this calibration procedure, one has to consider carefully how accurately one can hope to estimate the true autocorrelation function from a finite amount of data.
The computations in Section \ref{autocorrsection} assumed for simplicity that the mean synaptic weight vanishes (i.e.~$\langle w \rangle = 0$, as was our convention above), but when measuring synaptic efficacies in physical units we do not know this mean value a priori. Of course it can easily be estimated from the measured data, but using this leads to a biased estimator (especially so for short experiments) of the autocorrelation function
\[ \label{autoest}
A_{\rm est}(\Delta t) = {\overline{(w(t) - \overline{w})(w(t+\Delta t) - \overline{w})} \over \overline{(w(t) - \overline{w})^2}} \ .
\]
Here overbars indicate temporal averages that are approximated by sample averages over the available data points (of which there are fewer for larger $\Delta t$). 

\begin{figure}
	\begin{minipage}{\textwidth}
		\centering
		\includegraphics[width=2.9in]{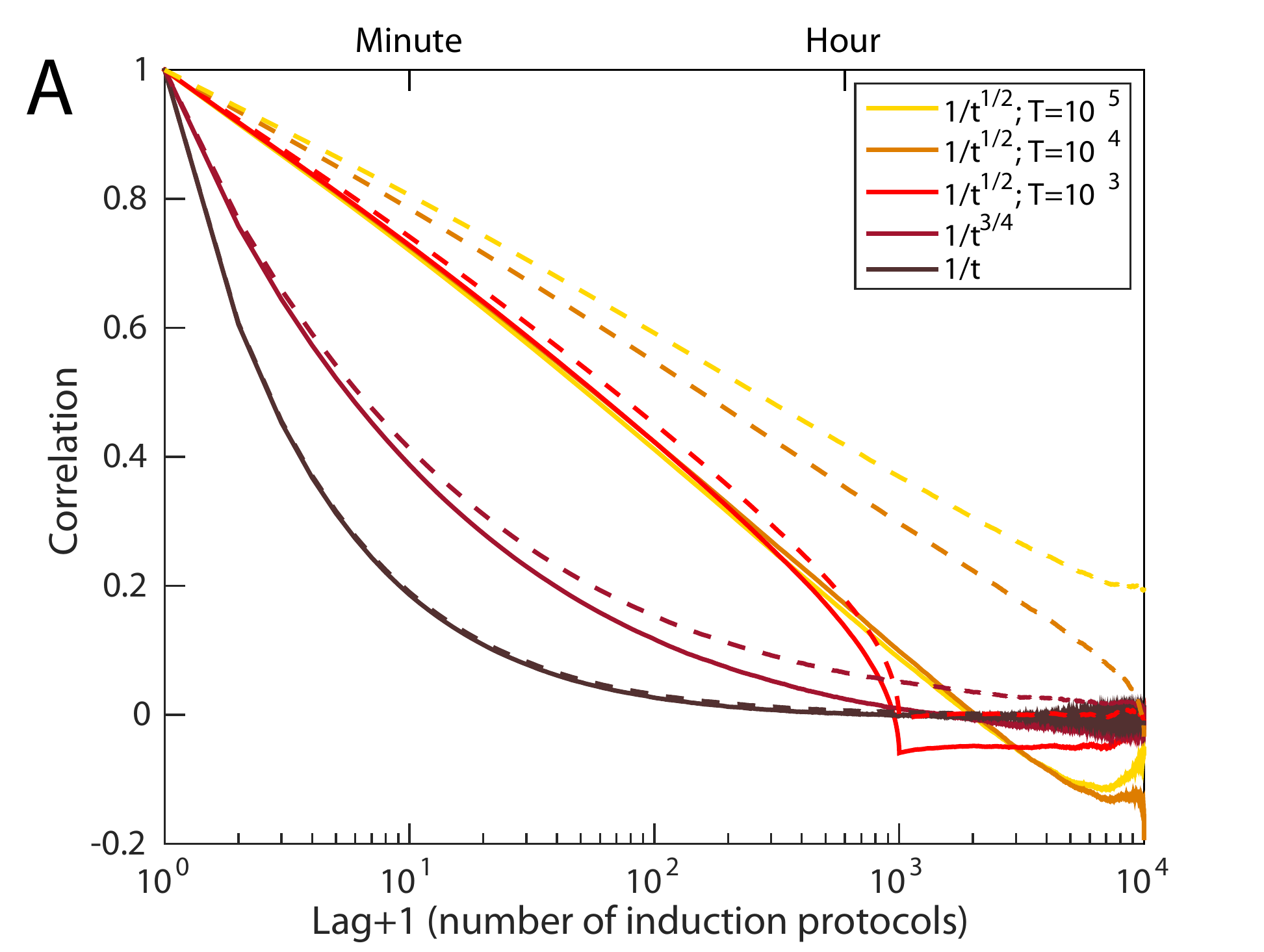}
		\includegraphics[width=2.9in]{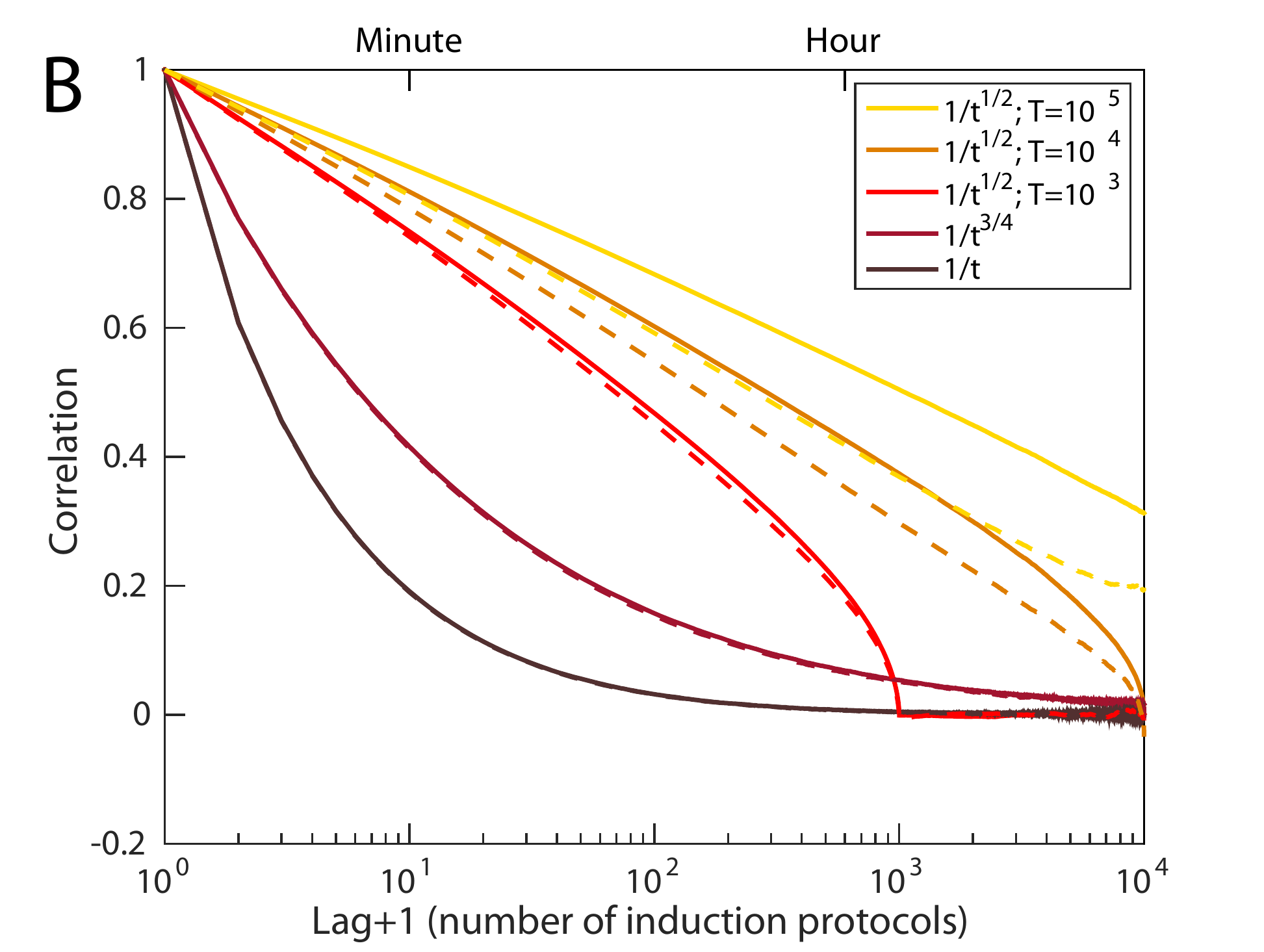}
	\end{minipage}
	\caption[]{Log-linear plots of the autocorrelation function estimated from simulated data using eqn.~(\ref{KernelAnsatz}) with different decay kernels vs $\Delta t +1$. The five models shown are the same as in Fig.~\ref{autocorr}, and for comparison the dashed lines in both panels reproduce the autocorrelation functions plotted there. A. The estimator of eqn.~(\ref{autoest}) that approximates the mean synaptic weight using the data (solid lines) introduces a significant bias, especially for slowly decaying memory traces. The curves for the three models with inverse square root decay function (and different cutoff times $T=10^3$, $10^4$ and $10^5$) lie almost on top of each other, and exhibit a rather steep slope. This estimator does not correctly capture the broad autocorrelation functions of models with slow decays and long cutoff times, but would still be sufficient to distinguish the proposed model from significantly faster power law-decays. 
		B. Autocorrelation functions estimated from simulated experiments with $10^7$ data points (solid lines), instead of $10^4$ (dashed lines and panel A). In this case the difference between using prior knowledge about the mean efficacy or estimating it from data is negligible (not shown), but the autocorrelation functions for the $1/\sqrt{t}$ models computed from such an unrealistically long experiment are significantly improved compared to those estimated from less data, and are consistent with the slopes that follow from eqn.~(\ref{AutoTLimit}).
	}
	\label{AutoEstPlot}
\end{figure}

In Fig.~\ref{autocorr} we have simulated such an idealized experiment using the simple model of eqn.~(\ref{KernelAnsatz}) with different memory decay functions, and computed the autocorrelation function assuming the true mean of the synaptic efficacy is known. If, on the other hand, we had to estimate the mean synaptic weight from the data before plotting the autocorrelation function according to eqn.~(\ref{autoest}), the result would look rather different (and highly distorted), as shown in Fig.~\ref{AutoEstPlot}A.  The curves for different cutoff times lie almost on top of each other, since the finite length of the experiment essentially imposes a lower bound on the magnitude of the slope of the (log-linear plot of the) autocorrelation function. In other words, the autocorrelation function estimated in this way would not be very flat, even if this was the case for the true autocorrelation function (say for a power-law decay with exponent close to minus one half), which could give the false impression of a faster memory decay.

For reference, we show in Fig.~\ref{AutoEstPlot}B the same graphs for an an unrealistically long experiment (with $10^7$ data points instead of $10^4$), in which case the bias becomes small, and both estimators (with known mean synaptic weight or without) approach the theoretical autocorrelation functions computed in Section \ref{autocorrsection}.

For the simple kernel models described by eqn.~(\ref{KernelAnsatz}) there are no hard limits of the dynamical range, but the variance of the synaptic efficacy is nevertheless bounded for the chosen decay functions. The limited dynamical range of real synapses may help alleviate this 
bias problem, since during the calibration procedure the experimenter is likely to encounter the extreme values the synaptic efficacy can take (by pushing it against those boundaries) and can use this additional information to establish a better estimate of the mean synaptic weight than one could obtain from the measurement phase alone.

\subsection{Comparison with other models and optimal complexity}
\label{comparison}

\begin{figure}
	\begin{minipage}{\textwidth}
		\centering
		\includegraphics[width=6.5in]{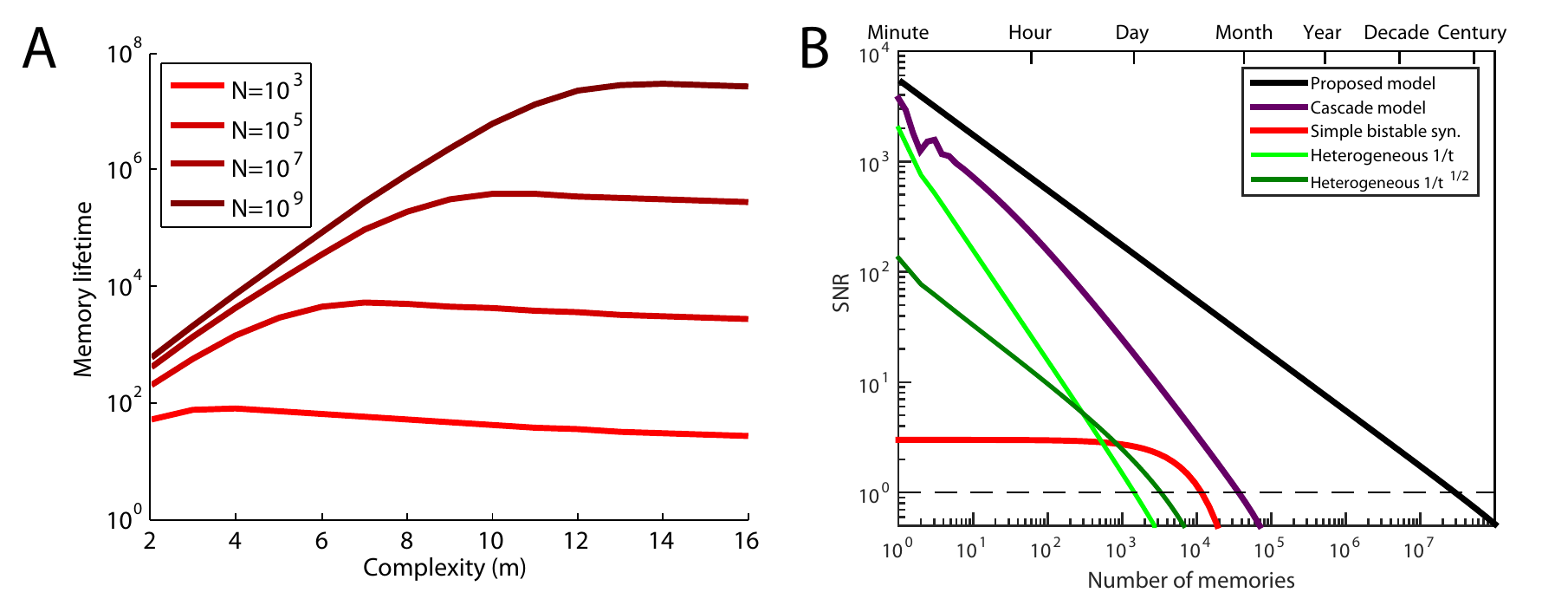}
	\end{minipage}
	\caption[]{Choosing the optimal complexity and comparing  the memory performance of different synaptic models. A. Memory lifetime as a function of synaptic complexity $m$ (i.e.~the number of variables) in the case of the proposed model. Different curves correspond to different numbers of synapses $N$. For each curve there is an optimal complexity. As the number of synapses increases, the peak shifts to the right and the optimal complexity increases. B. Comparison between five different synaptic models: our proposed model, the cascade model, a simple bistable model, and two heterogeneous population models. The five curves are the SNRs for the different models as a function of the number of memories (or time, represented on the top axis, as in the SNR plots in the main text).}
	\label{comparisonfig}
\end{figure}

Previous synaptic models are characterized by different scaling properties of the initial signal to noise ratio and the memory lifetime, the two quantities that we use to summarize the memory performance of synaptic models (see the Discussion and Table~\ref{comparisontable}): for the best complex models, like the one proposed in \citet{FusiDrewAbbott2005}, both these quantities scale like $\sqrt{N}$, where $N$ is the total number of synapses. In the model that we propose in this manuscript, the initial SNR still scales essentially like $\sqrt{N}$, but the memory lifetime is greatly extended, as it scales almost linearly with $N$. In this section we compare systematically the memory performance of different synaptic models. First, it is important to notice that all synaptic models that we consider are characterized by a certain number of parameters which determine the complexity of the synapse. To perform a fair comparison, it is important to choose the optimal parameters for each model. In the case of the cascade model described in \citet{FusiDrewAbbott2005}, the number of levels $\tilde{m}$ of the cascade is a measure of the complexity of the synapses. In the case of our model, it is the number of variables $m$. For simplicity we ignore the number of levels in the discrete version of our model. The memory performance depends on both the complexity of the synapse and the total number of synapses $N$. For a given number of synapses, there is an optimal complexity. For all of these models, small memory systems perform better with simpler synapses and complexity is required only for memory systems with large $N$. 
For example, in Figure \ref{comparisonfig}A we show the memory lifetime as a function of the complexity $m$ for our model for different numbers of synapses $N$. We use the following empirical expression to approximate the SNR
\[
\mc{S/N}(t) \simeq 0.8\, \sqrt{N \over t} {e^{-t/T} \over \sqrt{\log T }} \ , \nn
\]
where $T \simeq 6 \times 4^m$ is the longest timescale of the model.
The dependence on the parameters $m$ and $N$ was obtained from the continuum model of Sections~\ref{diffusion} and \ref{varianceandspatialcorr}, and the numerical constants were determined by fitting this expression to the SNR of simulations of the fully discretized model. We plot a different curve for each value of $N$ (namely $10^3,10^5,10^7$ and $10^9$). All curves grow with complexity, reach a maximum and then slowly decay. From the expression of the SNR  it is clear why the memory lifetime is a non-monotonic function of $m$: $T$ appears both in the exponential and in the logarithm in the denominator. When the complexity increases, the range over which the $1/\sqrt{t}$ decay prevails is extended, as $T$ increases exponentially with $m$. However, the initial SNR slowly decreases, as expressed by the logarithm in the denominator. If the SNR is well described by a power law for most of the time throughout which $SNR>1$, the exponential in the SNR expression can be expanded, and the memory lifetime $t^*$ can be approximated by
\[
t^* \simeq { 0.64\, T \over 1.28 + (T \log T)/N} \ . \nn
\]
This clearly shows that $t^*$ increases linearly with $T$, and hence exponentially with $m$, as long as $T\log T$ is negligible compared to $N$.

To compare different models, we need to choose the optimal complexity for each model. When juxtaposing our model and the cascade, we fixed the number of synapses $N=10^9$, and used the optimal parameters $m=14$ and $\tilde{m}=16$ that maximize the memory lifetime. The resulting SNRs are plotted in Figure~\ref{comparisonfig}B. For reference, we also show the SNRs of the simple synaptic model of \citet{AmitFusi1994}, in which we choose the learning rate that maximizes the memory lifetime, and of two heterogeneous models in which memories are stored in multiple populations of simple bistable synapses that are characterized by different learning rates \citep[see][and Section \ref{independent}]{Roxin2013}. The sizes of the synaptic populations have been chosen to produce two different power-law decays of the SNR: for one curve the decay is $1/t$, as in the cascade model, while for the other it is $1/\sqrt{t}$, as in the proposed model. Even though the heterogeneous models can also harness processes that operate on a wide range of timescales, these processes are not interacting (i.e.~the variables are independent). Our proposed model has an initial SNR that is only slightly larger than that of the cascade model, and orders of magnitude larger than in the simple synaptic model. However, as already discussed in the main text, the memory lifetime is several orders of magnitude larger than in any previous model of bounded synapses. This improvement would be strongly reduced for smaller memory systems (i.e.~for lower $N$). This means that for smaller brains it may be wasteful or even counterproductive to have synapses that are too complex. To take advantage of complexity, it is important to have a sufficient number of memory resources (in our case the number of synapses).

\vspace{0.5cm}
\begin{table}[h]
	\begin{center}
		\begin{tabular}{| l | c | c | c |}
			\hline
			& {\bf Time Dependence of $\mathcal{S}$} & {\bf Initial} $\mathcal{S/N}$ & {\bf Memory Lifetime} \\ \hline
			Unbounded            & const.          & $\leq\mathcal{O}(\sqrt{N})$ & $\mathcal{O}(N)$ \\  \hline
			Bistable (fast)     & $\sim e^{-Lt}$   & $\mathcal{O}(\sqrt{N})$ & $\mathcal{O}(\log(N))$ \\  \hline
			Bistable (slow)     & $\sim e^{-Lt}$   & $\mathcal{O}(1)$ & $\mathcal{O}(\sqrt{N})$\\  \hline	
			Heterogeneous        & $\sim 1/t$      & $\mathcal{O}\left({\sqrt{N} \over \log{N}}\right)$ & $\mathcal{O}\left({\sqrt{N} \over \log{N}}\right)$ \\ \hline			
			Cascade              & $\sim 1/t$      & $\mathcal{O}\left({\sqrt{N} \over \log{N}}\right)$ & $\mathcal{O}\left({\sqrt{N} \over \log{N}}\right)$ \\ \hline
			Multistage           & $\sim 1/t$      & $\mathcal{O}\left(\sqrt{{N} \over \log{N}}\right)$ & $\mathcal{O}\left(\sqrt{{N} \over \log{N}}\right)$ \\ \hline	
			Proposed Model       & $\sim 1/\sqrt{t}$ & $\mathcal{O}\left(\sqrt{{N} \over \log{N}}\right)$ & $\mathcal{O}\left({N \over \log{N}}\right)$ \\  \hline
		\end{tabular}
	\end{center}
	\caption{Approximate scaling properties of different synaptic models. ``Unbounded'' refers to models in which the synaptic variables can vary in an unlimited range, as in the Hopfield model \citep{Hopfield1982}, or more generally to models in which the dynamical range of each synapse is at least of order $\sqrt{N}$. In the case of the Hopfield model, there is no steady state, so the initial signal to noise ratio is large (as given in the table) really only for the first few memories. As more memories are stored, the noise increases, and the signal to noise ratio decreases as $1/\sqrt{t}$, where $t$ is the total number of stored memories.
		Bistable synapses have two stable synaptic values and the transitions between them are stochastic \citep{AmitFusi1994}. Fast synapses exhibit a large learning rate $L$ (i.e.~a transition probability of $\mathcal{O}(1)$), whereas slow synapses are characterized by the slowest possible learning rate (i.e.~the smallest transition probability that keeps the initial signal to noise ratio above threshold, which is $L=\mathcal{O}(1/\sqrt{N})$). In the heterogeneous model \citep{Roxin2013} the synapses have different learning rates (see also Section~\ref{independent}). The cascade model is described in \citet{FusiDrewAbbott2005} and the multistage model in \citet{Roxin2013}. In the last row we report the scaling properties of the model that we propose in this article, which are superior to other bounded models, and differ only by logarithmic factors from the unbounded case. Although the approximate scaling of the heterogeneous model is the same as for the cascade, the latter performs significantly better \citep[see][Fig.~8]{FusiDrewAbbott2005}. It is important to remember that two models with the same scaling behavior may not work equally well, as the coefficients in front of the factors reported in the table might be quite different. However, it is unlikely that a model with a better scaling behavior would perform worse, as $N$ is assumed to be very large.}
	\label{comparisontable}
\end{table}

\subsection{Models with independent variables}
\label{independent}

To illustrate why interactions between dynamical variables are beneficial for memory performance, let us briefly consider what would happen if the variables were completely decoupled from each other. We could assign different timescales to different (populations of) variables, which we can think of as independent, simple synapses. Incoming memories could then be stored in a distributed fashion in multiple synapses with a wide range of decay rates, such that at any point in time at least some of them retain a reasonably large memory signal (up to some maximal memory lifetime). A simple version of such heterogeneous models has been described in \citet{Roxin2013}.

The synaptic variables could be inherently bounded, such as e.g.~binary variables, or we could assume a priori unbounded synapses.
Either way, such simple synaptic models characterized by a single timescale $\tau$ correspond (at least in a mean field sense) to an exponential memory decay function $r(t) \sim \exp(-t/\tau)$, which is essentially flat up to a memory age of order $\tau$. In the latter case, even though the synaptic weights are a priori unbounded, we would still like to achieve a steady state distribution of synaptic weights that has a small variance, i.e.~requires us to distinguish only a small number of different values at any point during online learning. This is not possible for simple unbounded synapses, however, since the standard deviation of the weight distribution grows as $\sqrt{\tau}$. In this context, interactions are useful for achieving long timescales while maintaining small dynamical ranges for all variables\footnote{Of course one can always trivially reduce this dynamical range by simply reducing the step size for each plasticity event, but again this would not improve the situation, because what really matters is the number of distinct values that the synaptic efficacy is likely to take during learning, i.e.~the number of levels we would have to introduce when discretizing this variable in order to well approximate the dynamics in the unbounded case.}.

For binary variables with switching probability of order $1/\tau$, on the other hand, the expected memory signal will be $\mathcal{S}_\tau(t) \simeq (1/\tau) \exp(-t/\tau)$. In this case the variance is naturally bounded, with the corresponding memory noise being approximately constant in time and independent of $\tau$. Thus when considering the signal to noise ratio for an agnostic readout that takes into account all synapses equally, we can simply average the signals, while dividing by the noise merely contributes a constant factor (such that $\mathcal{S}/\mathcal{N} = \mathcal{O}(\sqrt{N})$ for a homogeneous population of size $N$ and fixed $\tau$).

If we have a distribution of timescales $\rho(\tau)$ within a population of such synapses, the total memory signal we expect can be computed by integrating over this distribution (adding the contributions from all timescales present). For example, for a power-law distribution $\rho(\tau) \sim \tau^{-\eta}$ with $\eta > 0$ (which would have to be cut off at a smallest and/or largest timescale, and normalized appropriately) we would find a total memory signal
\[ \label{IntSigDistr}
\mathcal{S}_\rho(t) \equiv \int \rm{d}\tau \, \rho(\tau)\,\mathcal{S}_\tau(t)  \sim \int_0^T \rm{d}\tau \, {e^{-t/\tau} \over \tau^{1+\eta}} = t^{-\eta}\, \Gamma \Big(\eta,\frac{t}{T}\Big) \xrightarrow{T \to \infty} \, t^{-\eta}\, \Gamma(\eta) \ ,
\]
which (approximately) exhibits a power-law decay at times $t$ smaller than some large cutoff $T$ (but larger than some shortest timescale, which we could have incorporated above by writing a finite lower cutoff for the integral). 

A commonly used distribution of timescales is $\rho(\tau) \sim 1/\tau$, since this corresponds to a uniform distribution on a logarithmic scale $(\rm{d} \tau / \tau = \rm{d}\log\tau)$. In this case, the incomplete Gamma function in eqn.~(\ref{IntSigDistr}) simplifies such that $\mathcal{S}_\rho(t) \sim (1/t)\exp(-t/T)$, i.e.~the signal exhibits a $1/t$ decay with a soft cutoff at the longest timescale $T$. 
If, on the other hand, we wanted to achieve the slower $1/\sqrt{t}$ memory decay with independent binary variables, we would need a distribution $\rho(\tau) \sim 1/\sqrt{\tau}$ that puts more emphasis on longer timescales. With this distribution eqn.~(\ref{IntSigDistr}) leads to 
$\mathcal{S}_\rho(t) \sim \sqrt{\pi/t}\, \text{erfc}\big(\sqrt{t/T}\big)$, i.e.~the desired power law, again with an exponential cutoff at times of order $T$. However, skewing the distribution in favor a longer timescales leads to a rather inefficient system (in terms of the number of variables needed to achieve a given initial signal to noise ratio) as we shall see below.

When building concrete models, we usually do not consider continuous distributions of timescales, but instead a finite number of populations with appropriately chosen timescales, i.e.~we approximate the desired decay function by a superposition of a finite number of exponentials.  Again, it is natural to choose these timescales equally spaced on a logarithmic scale. We can parameterize them e.g.~as $n^{2(i-1)}$ for $i = 1,2, \ldots m$ and for some real-valued $n >1$, which allows us to efficiently cover a large range of timescales, in this case from 1 to $T = n^{2(m-1)}$, using only a small number of timescales $m = \mathcal{O}(\log T)$. 

If these populations are of equal sizes, as in the most common heterogeneous binary model, the memory signal will be given by
\[ \label{SumInv}
\mathcal{S}(t) \sim \sum_{i=1}^m\, n^{-2(i-1)} \exp(-t \, n^{-2(i-1)}) \ ,
\]
which for $1<t<T$ and $n$ not much larger than one well approximates the $1/t$ power-law memory decay (with the quality of the approximation deteriorating as $n$ grows, since successive timescales become more widely spaced).
Similar mechanisms also generate the effective $1/t$ memory decay of the cascade \citep{FusiDrewAbbott2005} and multistage memory consolidation models \citep{Roxin2013}, even though these models do contain interactions between different timescales. 

However, if we allow the populations of increasing timescales to become progressively larger, we can slow down the memory decay. In particular, if the population size grows as $n^{(i-1)}$, the memory signal
\[ \label{SumInvSqrt}
\mathcal{S}(t) \sim \sum_{i=1}^m\, n^{-(i-1)} \exp(-t \, n^{-2(i-1)}) \ ,
\]
will approximate a $1/\sqrt{t}$ decay under the same conditions as in the previous paragraph. The crucial difference, however, lies in the total number of variables $N$ needed to achieve a given initial signal to noise ratio, which is dominated by the variables in the first population (with $i=1$; the other populations do contribute, but the sums in eqns.~(\ref{SumInv}) and (\ref{SumInvSqrt}) converge to numbers of $\mathcal{O}(1)$ even at $t=0$). For each such variable we only needed $\mathcal{O}(\log T)$ others to achieve the $1/ t$ decay, and consequently the initial signal to noise ratio in that case was $\mathcal{S}_0/\mathcal{N}_0 = \mathcal{O}(\sqrt{N}/\log T)$. For the $1/\sqrt{t}$ decay however, each variable in the fastest population corresponds to $\mathcal{O}(\sqrt{T})$ others with slower timescales, and as a result the initial signal to noise ratio drops to $\mathcal{S}_0/\mathcal{N}_0 = \mathcal{O}(\sqrt{N/T})$. 
This implies that we would need $N \gg T$ variables to achieve a reasonable initial signal to noise ratio, and in particular that this construction cannot extend the memory lifetime beyond $\mathcal{O}(\sqrt{N})$.

More generally, even though independent variables can be used to shape almost arbitrary monotonically decreasing memory functions, slowing down the decay requires investing a larger number of them in longer timescale populations, which detracts from the initial signal to noise ratio, and ultimately is not helpful in extending the overall memory lifetime. As we have shown, however, introducing properly tuned interactions between variables of different timescales can overcome these limitations.

\subsection{Relation to Markov chains and number of states}
\label{markov}

Given the quantization of the dynamical variables described in Section~\ref{discretelinearchain}, our model of a complex synapse could be rewritten as (and is mathematically equivalent to) a simple Markov chain, with the input term biasing the transition matrix depending on the memory to be stored at a given time step. A state of this Markov chain would be a joint assignment of all the quantized variables, and thus if every variable had $L$ levels the total number of states would be $L^m$, which can be a huge number. In particular, we have seen that it is sufficient to have $L = \mathcal{O}(\sqrt{\log N})$, and given that $m \sim 2 c \log{N}$ for some constant $c$ the number of states would be of order $(\log N)^{c \log N} = N^{c \log(\log N)}$, i.e.~superlinear in $N$. Due to this large number of states, our model is consistent with the general bounds on memory performance of Markov chains derived in \citet{Lahiri2013}.

The corresponding transition matrix would be an enormous $L^m$ by $L^m$ matrix, but writing out the dynamics in this language would completely hide the underlying structure of the interactions. Even though the two points of view are ultimately equivalent, the system is much simpler to describe in terms of interacting variables, rather than listing all their possible states and transitions between them, and we believe the former description is also more relevant in terms of informing possible biophysical implementations.

\subsection{Chemical reaction interpretation}\label{chemical}

The basic linear chain model we devised is essentially a particular spatial discretization of a one-dimensional diffusion process, and diffusion equations are ubiquitous in nature. For this reason there are many conceivable ways of implementing the required dynamics in terms of physical variables. The $u_i$ variables may in reality be an effective, coarse-grained description emerging from complex microscopic dynamics, but here we would like to give one simple example of a possible implementation of our model in which the $u_i$ can be identified directly with microscopic variables, namely (appropriately normalized changes in) concentrations of chemicals inside a synapse. This naive implementation is not meant to capture the complexity of the biochemical processes that are involved in synaptic memory consolidation. It is merely a simple example meant to demonstrate one possible mechanism that could instantiate the proposed dynamics.

This interpretation of the model will take the form of a chain of equilibrium chemical reactions such as
\[
A_1 + X_1\ \ \rightleftarrows \ \ A_2 + Y_2\ ; \qquad  A_3 + X_3\ \ \rightleftarrows \ \ A_4 + Y_4  \ ; \qquad A_5 + X_5\ \ \rightleftarrows \ \ A_6 + Y_6 \ \qquad \ldots \nn \quad \\
A_2 + X_2\ \ \rightleftarrows \ \ A_3 + Y_3\ ; \qquad  A_4 + X_4\ \ \rightleftarrows \ \ A_5 + Y_5 \quad \qquad \ldots \qquad \ . \qquad \qquad  \nn
\]
One can imagine that the $A_i$ are certain species of biomolecules contained inside a synapse that participate in equilibrium (hence bidirectional) reactions with other molecules of species $X_i$ and $Y_i$. These other molecules may be small, ubiquitous, and do not necessarily have to be confined inside the synapse. Each of the $X_i$ and $Y_i$ symbols may in fact represent a set of several (possibly distinct) molecules, or even the empty set (e.g.~if the $X_i$ are groups that can bind to a large biomolecule $A_i$ to form $A_{i+1}$ there may be no other reaction products $Y_{i+1}$ at all). 

For each reaction there will be a chemical equilibrium condition, such as e.g.~$A_1^* X_1^*/(A_2^* Y_2^*) =$ const.~for the first reaction, where (in an abuse of notation) we have used the same symbols that label a chemical species to also denote their concentration and it is understood that if any of the $X_i$ or $Y_i$ include multiple molecules (possibly of different species) we have to multiply their concentrations accordingly. Here and in what follows, stars indicate equilibrium (or more generally steady state) quantities. 

If we were dealing with an elementary reaction whose dynamics can be described by collision theory, we could write more explicit equations such as $r_{12}^+ = k_{12}^+ A_1 X_1$ and $r_{12}^- = k_{12}^- A_2 Y_2$ for the rates $r_{12}^\pm$ of the forward and backwards reactions, where $k_{12}^+$ and $k_{12}^-$ are constants. In equilibrium, both of these would be equal to the steady state reaction rate $r_{12}^* \equiv k_{12}^+ A_1^* X_1^* = k_{12}^- A_2^* Y_2^*$, reproducing the above equilibrium condition.

Since we have assumed that the $X_i$ and $Y_i$ are so common that their concentrations are effectively unchanged by small perturbations of the equilibrium, introducing e.g.~a small excess amount $\Delta A_1$ changes the forward rate of the first reaction by $\Delta r_{12}^+ \simeq k_{12}^+\, \Delta A_1 X_1 = r_{12}^*\,  \Delta A_1^* / A_1^*$ to first order. This will then lead to the production of more $A_2$, which in turn will increase the rates of the reactions in which it participates by $\Delta r_{12}^-$ and $\Delta r_{23}^+$ (depending in a similar fashion on $\Delta A_2$), and in this manner the perturbation denoted by $\Delta A_i$ will spread along the chain of reactions.

In fact, under the assumption of simple rate equations (collision theory), small perturbations of this system of equilibrium chemical reactions
lead to differential equations equivalent to those describing random walks (or spatially discretized heat diffusion), such as
\[
{\mathrm{d}\Delta A_2 \over  \mathrm{d} t} &=&   \Delta r_{12}^+ - \Delta r_{12}^- - \Delta r_{23}^+ +\Delta r_{23}^- \nn \\
&=& r_{12}^*\left({\Delta A_1 \over A_1^*}-{\Delta A_2 \over A_2^*}\right) - r_{23}^*\left({\Delta A_2 \over A_2^*}-{\Delta A_3 \over A_3^*}\right)  \ ,\nn
\]
and similarly for the other $\Delta A_i$. Comparing this to eqn.~(\ref{introduce_n}) we see that our dynamical variables $u_i$ can be identified with renormalized deviations from the equilibrium concentrations: $u_i \sim \Delta A_i / A_i^*$. The crucial quantity that needs to be tracked to determine the synaptic efficacy would thus be $\Delta A_1 / A_1^*$.
Furthermore, from Fig.~\ref{model} or eqn.~(\ref{GraphLap}) we can read off the relevant parameters: Steady state concentrations $A_i^*$ acts as heat capacities (or cross-sectional areas of beakers) and equilibrium reaction rates $r_{i,i+1}^*$ play the role of thermal conductivities (or tube sizes).
The appropriate tuning of these parameter would thus suggest exponentially decreasing reaction rates along the chain, and exponentially increasing equilibrium concentrations.

We can easily generalize this to a chain of equilibrium chemical reactions with arbitrary stoichiometric coefficients
\[
\lambda_1 A_1 + X_1\ \ \rightleftarrows \ \ \rho_2 A_2 + Y_2 \ ; \qquad \lambda_3 A_3 + X_3\ \ \rightleftarrows \ \ \rho_4 A_4 + Y_4 \ ; \qquad \lambda_5 A_5 + X_5\ \ \rightleftarrows \ \ \rho_6 A_6 + Y_6 \ \qquad  \quad \ \ \nn \\ 
\lambda_2 A_2 + X_2\ \ \rightleftarrows \ \ \rho_3 A_3 + Y_3 \ ; \qquad \lambda_4 A_4 + X_4\ \ \rightleftarrows \ \ \rho_5 A_5 + Y_5 \ \qquad \ \ \ldots  \qquad , \qquad \qquad \ \ \nn
\]
which, again assuming simple rate equations and expanding them to first order in small perturbations, merely changes the identification of the dynamical variables to
\[
u_i \sim {\prod_{j=2}^i  \rho_j \over \prod_{k=1}^{i-1}  \lambda_k}\, {\Delta A_i \over A_i^*} \ . \nn
\]

It is clear that one can further extend this type of model implementation to complex networks of equilibrium reactions analogous to the general graph models discussed in Section \ref{ramification}. If we are willing to forgo the one to one identification of our dynamical variables $u_i$ with individual microscopic quantities (such as concentrations) we can even imagine networks of non-equilibrium reactions that implement the required dynamics with the $u_i$ interpreted as effective variables (akin to diffusion on directed graphs, which makes sense as long as every node can be reached starting from every other node).

While assuming collision theory likely is overly simplistic, the above results really only rely on the fact that a small perturbation to chemical equilibrium leads to changes in reaction rates that are proportional to the excess concentrations $\Delta A_i$ to a first approximation, which is not implausible. 

Also, while declaring the concentrations of $X_i$ and $Y_i$ constant is convenient and leads to very simple equations, we could consider the case in which they do change significantly, and therefore influence the reactions rates and the resulting dynamics of the $\Delta A_i$. In that case, however, it would be important that these $X_i$ and $Y_i$ are confined and their numbers preserved within the synapse, since e.g.~an influx of such molecules from external sources could change the dynamics in important ways. This is in contrast to the case we have considered, where only the $A_i$ molecules have to be confined and preserved inside the synapse, up to perhaps a slow decay or leakage of these molecules over time intervals of the order of the longest implemented timescale. Since molecular turnover is not arbitrarily slow, however, the simple interpretation presented in this section does not seem suitable for very long-term memory, which likely requires multi-stable mechanisms.

\subsection{Real time versus event-triggered dynamics}
\label{realtimevsevent}

In the discrete time simulations of Figs.~\ref{scaling} and \ref{discrete} we have for simplicity assumed one new memory to be stored for each time step, but in reality there is no reason to think that new memories necessarily arrive at a constant rate. Generalizing to arbitrary, randomized timings of (desired) plasticity events does not pose a fundamental problem from a modeling point of view, either in a continuous time framework or in discrete time simulations with only some time steps coinciding with non-zero inputs. However, the question arises whether the internal dynamics of the model variables should be running even when there are no inputs, or whether their interactions should be triggered only by (desired) potentiation and depression events, with the dynamical variables otherwise frozen. Both scenarios are conceivable. 

We can consider describing the input statistics by a distribution of inter-plasticity event intervals. As long as this distribution is sufficiently concentrated around its peak (i.e.~has no long tails), such that the variance of the synaptic efficacy remains small enough to avoid hitting the boundaries of its dynamical range too frequently, we wouldn't expect the model behavior to be substantially different from the case discussed above\footnote{More generally, if we allow plasticity events of varying sizes, we could consider the joint distributions of inter-plasticity event intervals and event sizes, or perhaps even several such joint distributions that might be different for potentiation and depression, and possibly depending on the nature of the previous plasticity event.
	Again it would be important for the synaptic weights to not spend too much time at the edge of their dynamical range, which they might hit now either due to bursts of plasticity events of the same sign at an unusually high rate, or due to plasticity events of exceptionally large magnitude. 
}. 

If the internal dynamics is event-triggered with a certain mean rate of inputs, the model dynamics should be equivalent to the case of constant  (and equal) rate of arrival of memories, on timescales much longer than the inverse of that rate. On the other hand, if the internal dynamics is always running in physical time (regardless of inputs), we would have to adjust the inverse timescale $\al$ in eqns.~(\ref{introduce_n}) or (\ref{evolgen}) appropriately to achieve the same effective behavior on long timescales.

There are significant differences between these models, however, when the distribution of inter-memory input intervals is broad, e.g.~when bursts of many experiences to be stored are followed by long periods of silence (e.g.~sensory deprivation). An event-triggered mechanism could handle such variability in time rather easily (since physical time plays no role in it, as nothing happens when there is no input, and thus only the number of memories to be stored matters). On the other hand, an always-on internal dynamics would run into trouble given a small dynamical range, as mentioned above (or equivalently, would require a larger dynamical range to achieve the same memory performance).

Thus, input-triggered dynamics appears to have an advantage in terms of flexibility (it can easily adapt to changing input conditions), but physical time dynamics appears to be much simpler to implement (e.g.~as in Section \ref{chemical}, where one possible way of identifying the $u_i$ with physical variables is suggested), without the additional layer of complication required to build appropriate triggering mechanisms.
One could even imagine hybrid schemes in which only shorter timescale variables have event-triggered dynamics.

A number of secondary, and more subtle issues arise when we consider the case of discretized variables $u_i$ with stochastic transition dynamics. One would like to keep the additional variance (noise) due to the stochasticity of the internal dynamics small, which may again favor event-triggered dynamics. Taking this even further, and noting that the transition probabilities for longer timescale variables become progressively smaller (i.e.~transitions become rarer), one could then devise a scheme in which these variables are not updated every time a new input arrives, but even more infrequently, say on timescales on which their cumulative transition probability is of order one half.


\subsection{Memory retrieval}
\label{memretrievalsupp}

In order to study memory retrieval we have to consider a particular neural circuit and specify its architecture. Here we chose a feedforward, perceptron-like architecture, and a fully connected recurrent neural network. In both cases the stored memories were random and uncorrelated. More specifically, in the feedforward case we simulated one neuron receiving $N$ inputs. The memories were stored by imposing a random input pattern on the inputs and its associated desired output on the readout neuron. The neural activities of both the input and the output were $\pm 1$, chosen randomly with equal probability. Each memory was stored by modifying the synapses according to a simple covariance rule, similar to the prescription used in the Hopfield network \citep{Hopfield1982}.
For the memory stored at time $t'$
\[
\Delta w_{\frak{j}}(t') = \xi_\frak{j}(t')\,  \chi(t') \ , \nn
\]
where $\xi_\frak{j}(t')$ is the activity imposed on the presynaptic neuron $\frak{j}$ and $\chi(t')$ is the desired output. This $\Delta w_\frak{j}(t')$ determines how our complex synapses are updated (when it is positive, the synapse is potentiated, when it is negative, the synapse is depressed). Each memory was stored only once.

After all memories had been stored, we tested whether they could be retrieved by choosing one specific pattern $\xi_\frak{j}(t')$ and using it as the input. The activity of the output neuron was determined by computing the sign of the weighted sum of the inputs, i.e.~$\sign \big(\sum_{\frak{j}=1}^{N} w_\frak{j}\, \xi_\frak{j}(t')\big)$.
A memory was counted as correctly retrieved when this output matched the desired output $\chi(t)$ that was stored during memorization.

\begin{figure}
	\begin{minipage}{\textwidth}
		\centering
		\vspace{-0.2in}
		\includegraphics[width = 1 \textwidth]{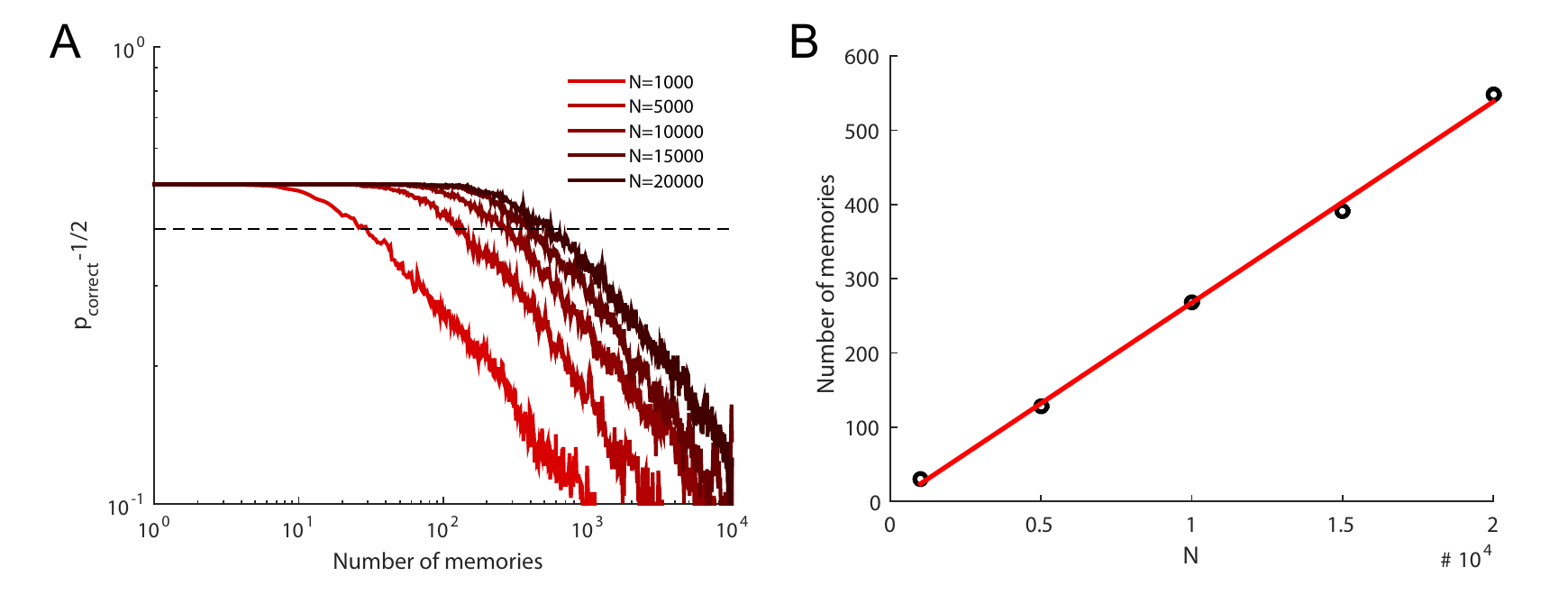}
	\end{minipage}
		\caption[]{A. Probability of correct retrieval as a function of the memory age for a feedforward network. The baseline corresponding to chance level ($p_\mathrm{correct}=1/2$) has been subtracted. Different colors correspond to different numbers $N$ of synapses, which for this architecture equals the number of inputs. The dashed line corresponding to $p_\mathrm{correct}=0.9$ is the arbitrary threshold for retrievability (i.e.~a memory of a certain age is called retrievable if the feedforward network generates the correct output with probability larger than $0.9$). B. Linear scaling of the number of retrievable memories as a function of the number of synapses $N$. The number of retrievable memories is determined by finding the intersection of the dashed line with the $p_\mathrm{correct}$ curves of panel A. The black circles are the results of simulations and the line is a linear fit (the slope is $0.027$). We used $m=8$ dynamic variables discretized with 40 levels each. The parameters have not been optimized to maximize the memory capacity (e.g.~the optimal $m$ would change depending on $N$).
		}
	\label{memretrfracvsage}
\end{figure}

\paragraph{Scaling properties (feedforward network)}

The ideal observer approach predicts that the number of storable memories should scale linearly with the number $N$ of synapses. In the neural circuit that we considered, the number of synapses is equal to the number of inputs. To test the prediction of the ideal observer approach, we progressively increased $N$ and determined the maximum number of memories that could be retrieved correctly. To estimate the number of retrievable memories, we ran multiple simulations to determine the probability $p_\mathrm{correct}$ that a memory is retrievable when followed by some number of subsequently stored memories (this number is basically the age of the memory that we are tracking, see Fig.~\ref{memretrfracvsage}A). If $p_\mathrm{correct}=1$ we have perfect retrieval, whereas $p_\mathrm{correct}=1/2$ corresponds to chance level, as the probability of guessing the correct output is $1/2$. The number of retrievable memories is estimated by determining the largest age for which $p_\mathrm{correct}>0.9$. This number increases linearly with $N$ (see Fig.~\ref{memretrfracvsage}B), as predicted by the ideal observer approach. 

The memory signal of the ideal observer approach is (the normalized expectation value of) the overlap $\chi(t)  \sum_{\frak{j}=1}^{N} w_\frak{j}\, \xi_\frak{j}(t)$ and if we assume that the distribution of this quantity is Gaussian, the probability $p_\mathrm{correct}$ of it being positive (i.e.~retrieval being correct) is
\[
p_\mathrm{correct}={1 \over 2}-{1 \over 2}\, \erf\left({1 \over \sqrt{2}}\, \mc{S/N}\right) \ , \nn
\]
where $\mc{S/N}$ is the ideal observer signal to noise ratio. Thresholding $p_\mathrm{correct}$ is thus equivalent to thresholding $\mc{S/N}$, and we have shown that the capacity defined by the latter prescription grows (almost) linearly with $N$.

\paragraph{Generalization and signal to noise ratio}

\begin{figure}
	\begin{minipage}{\textwidth}
		\centering
		\vspace{-0.2in}
		\includegraphics[width = 1 \textwidth]{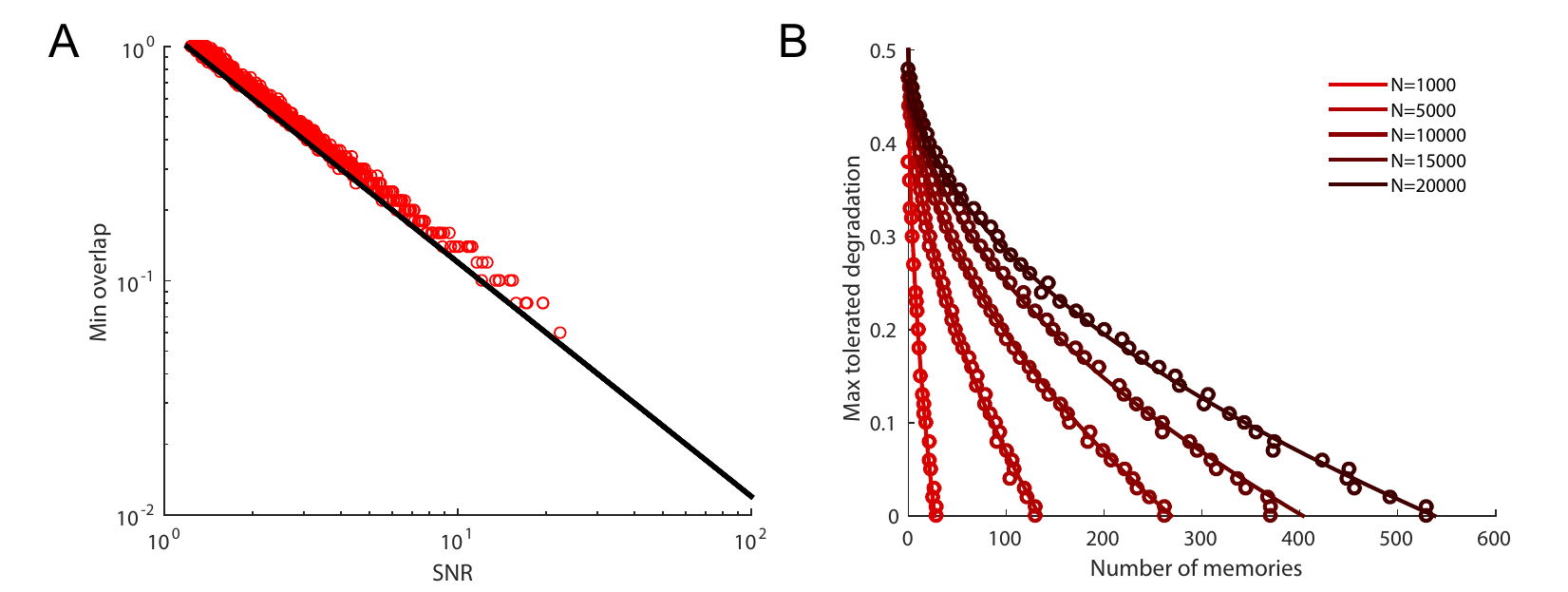}
	\end{minipage}
		\caption[]{A. Generalization ability of a feedforward network storing  random uncorrelated memories. A. Relation between the generalization ability and the ideal observer signal to noise ratio. The generalization ability is expressed on the vertical axis as the minimal overlap between the stored memory and the cue used for memory retrieval that leads to successful retrieval ($1-2\, \epsilon$, as discussed in the text). The black line is the prediction of the theory and the red dots are the results of simulations. Both the overlap and the SNR are on logarithmic scales and the line on the plot expresses the predicted power law  ($1-2\, \epsilon \propto 1/\text{SNR}$). B. Generalization ability as a function of the memory age, or equivalently, the number of memories that are stored after the tracked memory. The generalization ability is expressed as the maximum degradation $\epsilon$ that can be tolerated, i.e.~that still leads to successful retrieval. Different curves correspond to different numbers of synapses $N$ (which in this case of a feedforward network is also the number of inputs). The strong initial SNR allows for very large tolerated degradation ($\epsilon = 0.5$ corresponds to a completely uncorrelated memory). As the SNR decays the maximum tolerated degradation also decreases, and becomes zero when the memories can no longer be retrieved correctly even when the cues used for memory retrieval are not degraded. The circles are the results of the simulations and the lines are the theoretical prediction that follows from eqn.~(\ref{DegrPCorr}), i.e~maximum $\epsilon = 1/2- \gamma \sqrt{t/N}$, where $\gamma$ was fitted to the curves ($\gamma=3.05$), and $t$ is the memory age. The agreement between theory and simulation results is remarkably good. All other parameters are the same as in Fig.~\ref{memretrfracvsage}.
		}
	\label{generalization}
\end{figure}

In the simple neural circuit that we considered it is easy to estimate the generalization ability of the network, which is related to the strength of the ideal observer memory signal. Indeed, consider the feedforward network we used to validate the scaling properties. To assess the ability to generalize one can degrade the quality of the input cues, and determine the maximum degradation that is tolerated by the neural circuit, i.e.~that still produces the correct response. We decided to degrade the inputs by flipping the sign of their components with probability $\epsilon$. This form of degradation reduces the ideal observer signal by a factor $1-2\, \epsilon$, leading to:
\[
\label{DegrPCorr}
p_\mathrm{correct}={1 \over 2}-{1 \over 2}\, \erf \left({1-2\, \epsilon\over \sqrt{2}}\, \mc{S/N}\right) \ .
\]
If we demand that $p_\mathrm{correct}=0.9$ as above, the argument of the error function must be equal to some constant (that has to be determined by inverting the error function). The minimal tolerated similarity $1-2\, \epsilon$ between the stored input and the cue use for retrieval will therefore be inversely proportional to the SNR. This prediction is verified by the simulations in Fig.~\ref{generalization}.

\paragraph{Recurrent network and attractor dynamics}

\begin{figure}
	\begin{minipage}{\textwidth}
		\centering
		\vspace{-1.4in}
		\includegraphics[width = 0.6 \textwidth]{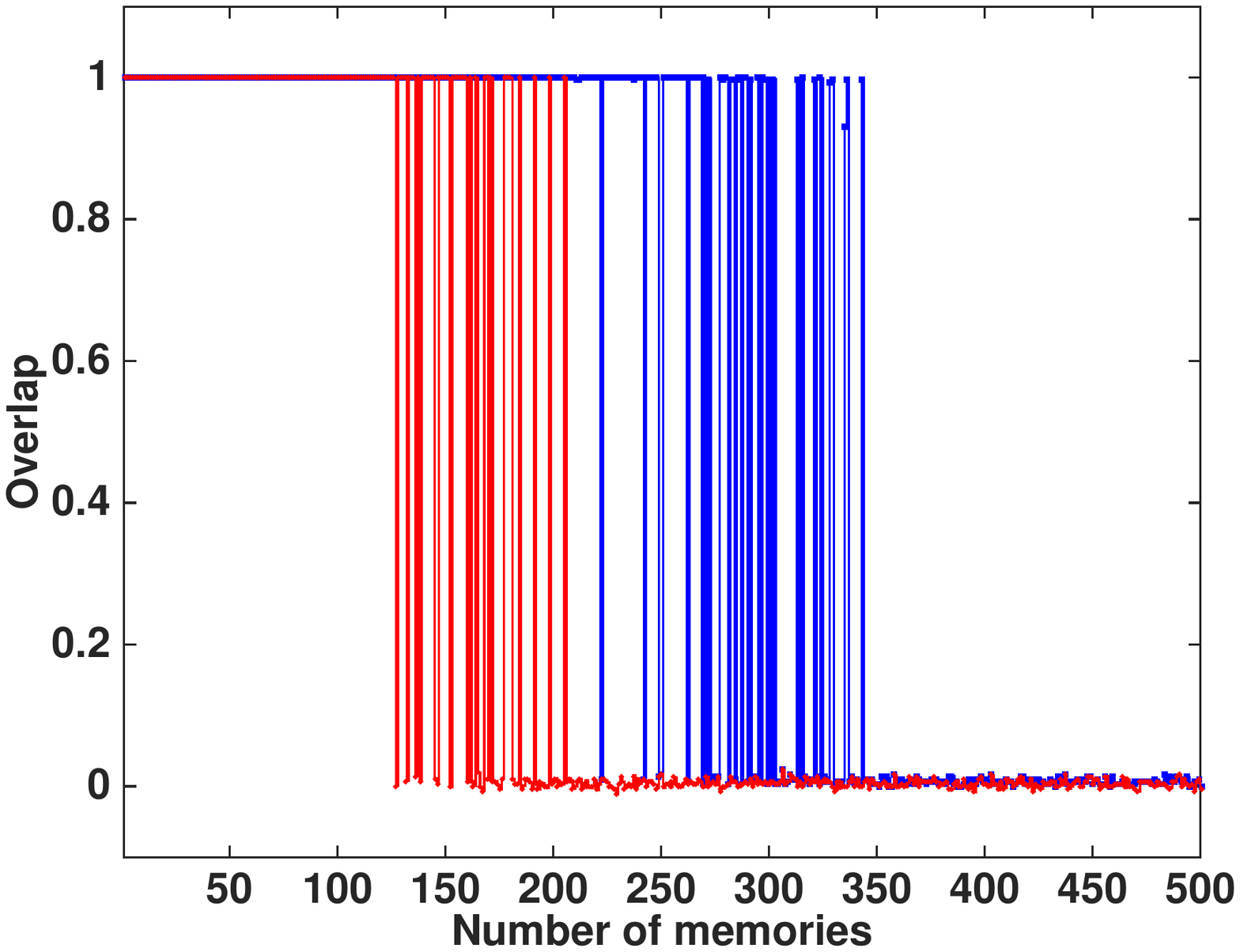}
		\vspace{-1.1in}
	\end{minipage}
		\caption[]{Overlap between the retrieved state and the stored memory patterns versus age of the memory in a simulation of a fully connected Hopfield network of binary neurons. The recall procedure is deterministic and asynchronous. For every memory we cycle through (and update the activity of) all neurons ten times in a random order, after which the resulting state typically is either very close to the stored memory (an overlap of one indicates perfect retrieval) or very far from it (an overlap of zero is chance level). The blue line corresponds to using an uncorrupted cue for retrieval ($\epsilon = 0$), while the red line corresponds to $\epsilon=1/4$. In both cases recent memories are recalled flawlessly, while for very old ones retrieval doesn't converge to an appropriate attractor state (even though the quality of one-step feedforward retrieval may still be high), with large fluctuations between those two extremes in an intermediate age range.
The number of neurons was $30000$, and for each synapse we used $m=4$ variables discretized with $30$ levels each.
		}
	\label{HopfieldRecall}
\end{figure}

We also simulated a Hopfield network  \citep{Hopfield1982} 
with complex synapses to study recall in the recurrent case. Here we apply the usual Hebbian learning rule $\Delta w_{\frak i j}(t') = \xi_\frak{i}(t')\, \xi_\frak{j}(t')$ to a fully connected, symmetric network of $N+1$ binary neurons (the total number of synapses\footnote{As noted in the discussion, even though the total number of synapses is of order $\mathcal{O}(N^2)$, the number of synapses relevant for the ideal observer analysis is $N$, since this is the maximum number that can receive independent inputs. Alternatively, we can think of computing the signal to noise ratio for the local circuit consisting of afferent neurons connected to a particular one, in which case the relevant number of synaptic weights is again $N$.} being $N(N+1)/2$). Retrieval is now a multistep procedure that iteratively and asynchronously updates randomly chosen neurons $\frak{i}$ by setting their activity to $\sign \big(\sum_{\frak{j \neq i}} w_\frak{i j}\, \hat{\xi}_\frak{j}\big)$, where $\hat{\xi}_\frak{j}$ is the current state of the neural ensemble that was obtained starting from some cue related to one of the stored patterns $\xi_\frak{j}$. The cue may again be corrupted by randomly flipping a fraction $\epsilon$ of its components. Recently stored memories corresponds attractor states under this recall dynamics (see Fig.~\ref{HopfieldRecall}).

\subsection{Sparse Representations}
\label{sparsereps}

Representing memories by sparse patterns of neural activations has long been known to carry certain computational benefits \citep{TsodyksFeigelman1988, AmitFusi1994}. 
In particular, it allows us to store a larger number of patterns.
It is important to appreciate however, that storing a larger number of items by decreasing the coding level does not necessarily imply an increased total information content, since a sparse pattern contains less information about the memory it represents (compared to the maximal amount of information that can be stored in a dense pattern of the same size).

In the main text, we have considered binary valued patterns of synaptic modifications with both values ($\pm 1$) equally likely and independently distributed, which implies a dense representation. In this case the information density of the stored patterns is maximal ($N$ bits for a binary pattern of size $N$).
Even though we have limited the discussion to the case of dense coding above, we can easily combine the advantages of sparse coding with the beneficial properties of our complex synaptic model if we are interested in increasing the number of stored patterns (at the expense of the amount of information stored per pattern).

\paragraph{Neural architectures and retrieving information}

In order to discuss sparse coding we will have to choose a particular architecture. Here we consider the usual perceptron-style setup with $N$ presynaptic neurons (indexed by $\frak{i}$) afferent to a single (postsynaptic) neuron that performs a binary classification of patterns. In order to perform such a classification correctly (i.e.~to reconstruct one binary feature of the input pattern), the readout will have to retrieve one bit per pattern from the set of $N$ afferent synapses, where this information has to be stored\footnote{This is most easily seen for (hetero)associative memory, in particular if the correct outputs are a priori independent of the inputs. In this case the recall cue alone (without the learned weights) can provide no information about the correct outputs.}.

We can generalize this architecture further by considering a number of parallel, independent readouts, receiving inputs from the same set of $N$ presynaptic neurons, but through separate sets of $N$ synapses for each of them. 
While a single binary readout can reconstruct at most one bit about each of the stored patterns of presynaptic activity, some number $R$ of parallel readouts can reconstruct up to $R$ bits of information about any one of the stored patterns using a total of $N\times R$ synapses. In particular, if we consider $N$ parallel readouts, the architecture becomes a feedforward mapping between two layers of $N$ neurons each (and using $N^2$ synapses in total)\footnote{Of course, if the task is to reconstruct the original input patterns one to one (as in autoassociative memory), some part of the correct pattern will have to be provided in the form of a corrupted cue for every recall trial, and in this case the cue does contain relevant information.}.

There are now two distinct notions of capacity we can consider, the number of patterns that can be recalled at a given time (with high fidelity), and the total amount of information about them that can be read out from the synaptic weights.
For independent memories this total information capacity is equal to the sum over all stored patterns of the retrievable information about each of them. Note that the information that is actually retrievable by a particular readout may be smaller than the total amount of information stored in the synaptic weights. 

In the case of dense coding there are $N$ bits to be reconstructed for every pattern, for which we require $N$ independent readouts, and if all of them can be retrieved successfully the total information capacity will be $N$ times the number of patterns that can be recalled. Since the number of patterns scales almost linearly with $N$ this leads to a total of almost $\mathcal{O}(N^2)$ bits stored in $N^2$ synapses (plus possibly some residual information about older patterns that cannot be recalled with high probability anymore). Because each of our synapses has only a small number $L$ of distinguishable states of its efficacy (the states of the internal variables not being accessible to the readout), we cannot hope to increase the total information capacity much further, since in this case $\log_2 L$ bits per synapse is a strict upper limit for the information that can be read out. In our model $L$ is relatively small and grows only very slowly with the number of synapses (as $\sqrt{\log N}$).
For the case of a complex synapse with binary weights discussed in Section~\ref{readoutschemes} there are only two states
and the upper limit is one bit per synapse. In both of these cases, the information capacity is close to its upper bound.

We can, however, increase the number of patterns stored by reducing the coding level, while keeping the total information content approximately constant.
If we assume for simplicity that there is a sharp transition between recent memories that can be reconstructed perfectly, and older ones of which hardly any information can be retrieved, the two notions of capacity will be proportional for any fixed coding level $f$, with the constant of proportionality being the information contained in (or equivalently the Shannon entropy of) a single pattern, which is $-N f \log_2(f) - N(1-f)\log_2(1-f)$. When $f$ is small, the information per pattern scales approximately as $Nf$ and is reduced by a factor $f$ with respect to the dense case. As we will see, sparseness may allow us to store a number of patterns that is a factor $\sim 1/f$ larger, leading to an information capacity that is approximately the same as in the dense case and again close to its upper bound.

Note that for simple synapses such as the bistable synapses studied in \cite{AmitFusi1994}, the information capacity can be very different for the dense and the sparse case. Indeed, for bistable synapses the maximum number of retrievable dense patterns scales like $\sqrt{N}$, not (almost) $N$ as in the proposed model, so the information capacity is far from its upper bound. In contrast, for sparse representations, the number of retrievable patterns scales like $N^2$, saturating the information capacity. Nevertheless, synaptic complexity remains important for at least two reasons: the first one is that the initial signal to noise ratio is much larger for complex than for simple models, and the second one is that the $N^2$ scaling requires a very small $f$ which would not be compatible with the observed $f$ (see below for a more extensive discussion).

\paragraph{Signal to noise ratio for sparse coding}

We can perform a signal to noise ratio analysis similar to that of Section~\ref{optimaldecays} for the case of sparse representations.
To this end, we have to generalize the definition of the memory signal of eqn.~(\ref{DefSignal}) slightly, in a way that makes explicit the binary decision the readout neuron faces when retrieving a previously stored pattern. We can write the signal as
\[\label{GenSignal}
\mc{S}_{t'}(t) \equiv {1 \over 2 N} \Big\langle  \sum_{\frak{i}=1}^N \big( w_\frak{i} (t)|_{\chi(t')=1}\, v_\frak{i}(t') - w_\frak{i} (t)|_{\chi(t')=0}\, v_\frak{i}(t') \big) \Big\rangle \ , 
\]
where the binary variable $\chi$ denotes the response of the postsynaptic neuron (imposed at the time the pattern in question was stored) and $\vec{v}(t')$ is a readout vector appropriate for retrieving the memory stored at time $t'$. This can be viewed as an extension of the signal to noise ratio defined in \cite{AmitFusi1994,BenDayanRubinFusi2007}, where populations of neurons were studied and $\vec{v}$ was just the pattern of pre-synaptic activity imposed on the network to trigger memory retrieval. 

For the noise (squared) we have
\[
\mc{N}^2_{t'}(t) \equiv  {1\over 2}\, \mathrm{Var}\left( {1\over N}  \sum_{\frak{i}=1}^N \big( w_\frak{i} (t)|_{\chi(t')=1}\, v_\frak{i}(t') \right) + {1\over 2}\, \mathrm{Var}\left( {1\over N}  \sum_{\frak{i}=1}^N \big( w_\frak{i} (t)|_{\chi(t')=0}\, v_\frak{i}(t') \right)  \ . \nn
\]
These definitions allow us to consider the signal to noise ratio beyond the ideal observer approach (which would imply $\vec{v}(t') = \Delta \vec{w}(t')$) and beyond the case of densely coded outputs (where $\chi=0$ and $\chi=1$ are a priori equally likely) that we have discussed in the main text.
Here and in what follows, we write equations for a single postsynaptic neuron (dropping the postsynaptic index), since the generalization to several independent outputs is trivial.

\paragraph{Sparseness with covariance learning rule}

In order to study the effects of sparse coding on the signal to noise ratio we also need to specify a learning rule (that turns neural activity patterns into patterns of desirable synaptic modifications $\Delta \vec{w}$).
A simple learning rule we will consider is the Hebbian-type covariance rule \citep{TsodyksFeigelman1988,StantonSejnowski1989} discussed below eqn.~(\ref{KernelAnsatz}), which in the case of a single postsynaptic neuron reduces to 
\[
\Delta w_\frak{i}(t)  \propto (\xi_\frak{i}(t) - a)(\chi(t) - b) \ , \nn
\]
where instead of a single coding level $f$ we distinguish between the pre- and postsynaptic coding levels $a \equiv \langle\xi_\frak{i}\rangle$ and $b \equiv \langle\chi\rangle$, assuming for simplicity that the binary variables $\xi_\frak{i}$ and $\chi$ take values 0 or 1.
We will take the readout vector to be of the form $v_\frak{i}(t) = \xi_\frak{i}(t) - c$ for some constant $c$, i.e.~it depends solely on the appropriate pattern of presynaptic activities. 

Assuming spatially and temporally uncorrelated input patterns $\vec{\xi}(t)$ and associated output labels $\chi(t)$, expectation values can easily be computed simply by multiplying the probabilities of the independent pre- and postsynaptic activations (i.e.~$a$ or $1-a$ for the pre- and $b$ or $1-b$ for the postsynaptic side) and summing over all combinations\footnote{Note that we are not averaging over the labels $\chi$ for the memory being recalled since the output is explicitly fixed to the two possible values in the two terms of eqn.~(\ref{GenSignal}).}. 

This leads to the signal to noise ratio (considering the memory stored at $t'=0$ without loss of generality)\footnote{The normalizations of $\Delta \vec{w}$ and $\vec{v}$ don't matter here, since they cancel in the signal to noise ratio. Also, the two terms in the expression for the (squared) noise turn out to be equal to each other.}
\[ \label{sparseSNRCov}
\mc{S/N}(t) = \mathcal{C}(a,b,c)\, \sqrt{N r^2(t)\over \sum_{t' < t, t' \neq 0} r^2(t-t')}  \quad \mathrm{where} \quad \mathcal{C} = {\sqrt{a\, (1-a)} \over 2\sqrt{b\, (1-b)(a - 2 a c + c^2)} } \ . 
\]
Apart from the coefficient $\mathcal{C}$ depending on the parameters $a$, $b$ and $c$, this is the same signal to noise ratio as in eqn.~(\ref{SNRKernel}). In particular, the dependence on $N$ and the kernel $r(t)$ is the same, and thus all the same considerations as above apply. The optimal time dependence is still $r(t) \sim t^{-1/2}$, and therefore the memory lifetime will scale as the square of the initial signal to noise ratio, or equivalently the square of the overall coefficient of the signal to noise ratio, which (for fixed $N$) can be made larger by making neural representations sparser.

The optimal value of $c$ that maximizes this coefficient is given by $c=a$, in which case we find $\mathcal{C} = 1/(2\sqrt{b\, (1-b)})$. In other words, the best readout is one in which the readout vector $\vec{v}$ is proportional to $\Delta \vec{w}$. This case is closely related to the ideal observer approach discussed in the main text, and in fact if in addition the postsynaptic representations are dense, i.e.~$b=1/2$, the more general definition of the signal to noise ratio in eqn.~(\ref{GenSignal}) reduces to the previously discussed case of eqn.~(\ref{DefSignal}), and we obtain exactly the same result as above.

On the other hand, if we consider a readout proportional to the presynaptic activity (i.e.~$c=0$, which might be considered more realistic than the above if we view the readout neuron as a linear classifier with fixed threshold computing the overlap $\vec{w}.\vec{\xi}$) the numerical coefficient in front of the signal to noise ratio simplifies to $\mathcal{C} = \sqrt{1-a}/(2\sqrt{b\, (1-b)})$.  
Unlike case of the optimal readout (with $c=a$), there remains a dependence on the presynaptic coding level.
We will distinguish two scenarios: one in which the presynaptic coding level remains constant as $b$ changes, and one in which $a=b$. 

\begin{figure}
	\begin{minipage}{\textwidth}
		\centering
		\vspace{-1.4in}
		\includegraphics[width = 0.6 \textwidth]{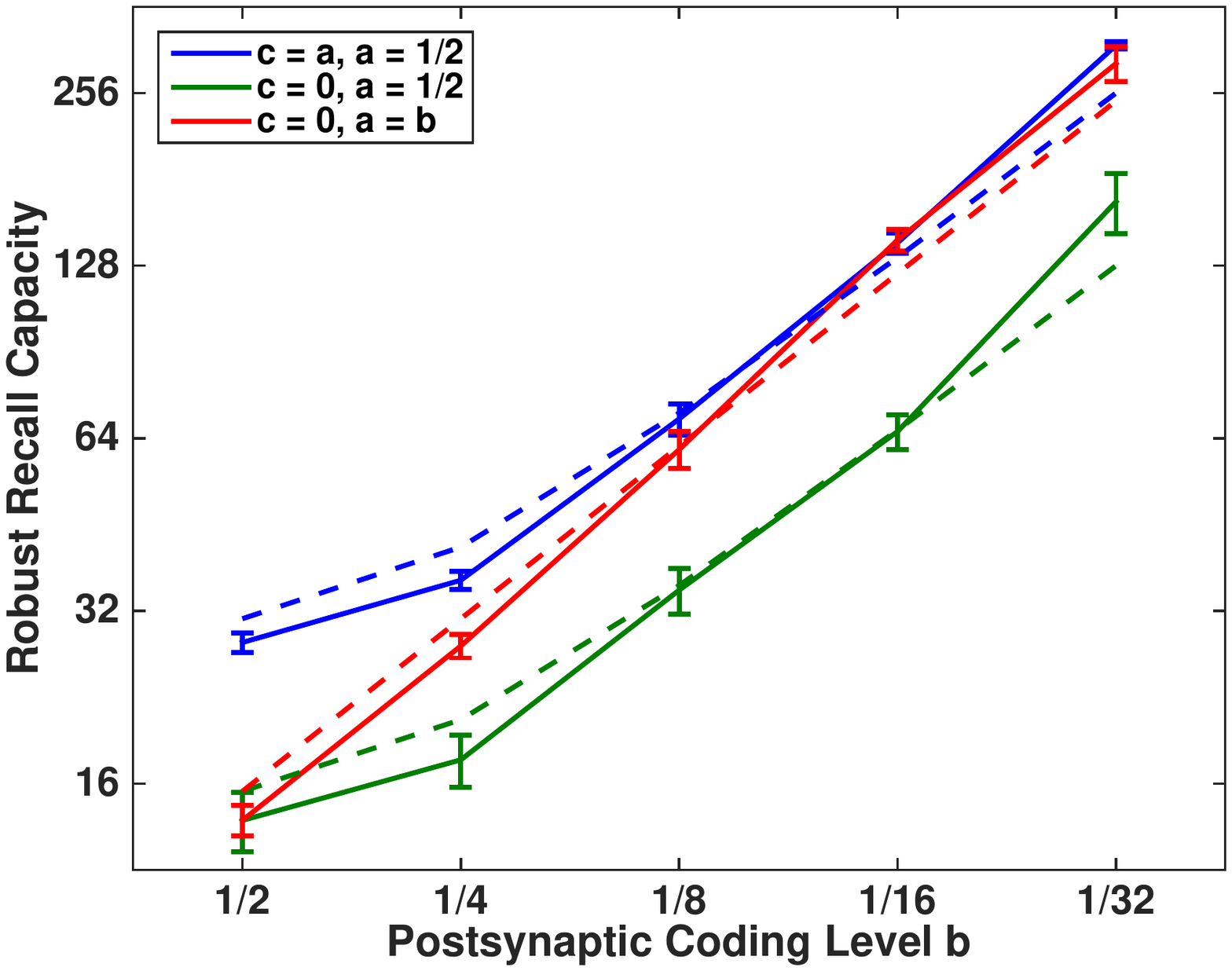}
		\vspace{-1.1in}
	\end{minipage}
		\caption[]{Doubly logarithmic plots of the robust recall capacity versus postsynaptic coding level $b$ for the case of the covariance learning rule. Data points are simulations of the fully discretized version of our synaptic model with $m=5$ variables and $N=2000$ synapses. We call a pattern robustly recalled if the probability of a readout neuron classifying it correctly is larger than $0.99$ (we estimate this probability by sampling $500$ independent readouts). This very stringent criterion (which leads to numerically rather small capacities) is necessary to study the scaling behavior, since the chance level for recall grows as we decrease $b$, and the threshold must be chosen high enough such that even for the sparsest patterns considered it is still well above chance level. For each data point we numerically optimize the bias term fixing the location of the decision boundary, which unlike the case of dense coding is no longer zero. 
			We normalize the learning rule such that $\langle\Delta w_\frak{i}\rangle = 1$ for all $\frak{i}$, while the spacing between adjacent levels of the discretized variables remains unity.
			The three curves (from top to bottom) correspond to the optimal readout ($c=a$, shown here with in blue with constant $a=1/2$, though the numerical results for $a=b$ are not significantly different), the simple readout ($c=0$) with $a=b$ (red), and the simple readout with $a=1/2$ (green). The dashed lines indicate the corresponding predictions of eqn.~(\ref{sparseSNRCov}), where one parameter common to the three curves (the overall magnitude) has been adjusted to match the data. This leads to a good fit, even though the calculation leading to eqn.~(\ref{sparseSNRCov}) is agnostic about the implementation of the internal synaptic dynamics and the discretization of variables.
		}
	\label{SparseCovariance}
\end{figure}

In all of these cases the coefficient in front of the signal noise ratio grows asymptotically as $1/\sqrt{b}$ as the output coding level is reduced, and therefore (since the optimal time dependence is still $1/\sqrt{t}$) the memory lifetime (the number of patterns that can be recalled at a given time) grows asymptotically as $\mathcal{C}^2  \propto 1/b$, as shown in Fig.~\ref{SparseCovariance}.

\paragraph{Sparseness with Amit-Fusi learning rule}

We can perform a similar analysis for the perhaps more biologically plausible Amit-Fusi learning rule \citep{AmitFusi1994},
which potentiates a synapse by a certain amount if both the pre- and postsynaptic neurons are active, and depresses it by a fraction $f/(1-f)$ of that amount when only the presynaptic neuron is activated, where $f$ is the coding level\footnote{Here we take the pre- and postsynaptic coding levels to be equal (both $f$) for simplicity, as in \cite{AmitFusi1994}.
	In that paper a whole family of learning rules was studied for simple binary synapses. Here we consider just one particularly straightforward rule, according to which potentiation events occur with probability $f^2$ and depression events with probability $f(1-f)$, and apply it to our model of a complex synapse.}. When the presynaptic neuron is quiescent, no plasticity events occur. Note that this presynaptically gated learning rule is balanced, in the sense that the expectation value of the weight change vanishes. 
We again use a readout vector that is linearly related to the presynaptic activity, i.e.~$v_\frak{i}(t) = \xi_\frak{i}(t) - c$, with $0 \le c \le 1$.

In this case the signal to noise ratio of eqn.~(\ref{GenSignal}) is 
\[ \label{sparseSNRAF}
\mc{S/N}(t) = \mathcal{\hat{C}}\, \sqrt{N r^2(t)\over \sum_{t' < t, t' \neq 0} r^2(t-t')}  \ \ \ \mathrm{with} \ \ \mathcal{\hat{C}} = \frac{1-c}{2\sqrt{(1-f)^2 \left((c-f)^2+f\right)+N f (1-f)(c-f)^2}} \ , \nn
\]
which again differs from the result for dense coding only by a coefficient (here called $\hat{\mathcal{C}}$). The computation of the noise term leading to this result is slightly more involved than for the covariance learning rule, since even though $\langle \Delta w_\frak{j}(t') \rangle = 0$ for the stored patterns (averaging over the labels $\chi$), they are not spatially uncorrelated, i.e.~for a given pattern $\langle \Delta w_\frak{i}(t')\, \Delta w_\frak{j}(t') \rangle$ doesn't vanish for $\frak{i} \neq \frak{j}$.
This is easy to see, since according to the learning rule all nonzero components of patterns for which the output neuron was active are positive (whereas they are all negative if the output neuron was quiescent). These correlations introduce a dependence on $N$ in the coefficient $\hat{\mathcal{C}}$, which multiplies the signal to noise ratio and therefore rescales the number of patterns that can be recalled ($\propto \hat{\mathcal{C}}^{2}$)

\begin{figure}
	\begin{minipage}{\textwidth}
		\centering
		\includegraphics[width = 0.49 \textwidth]{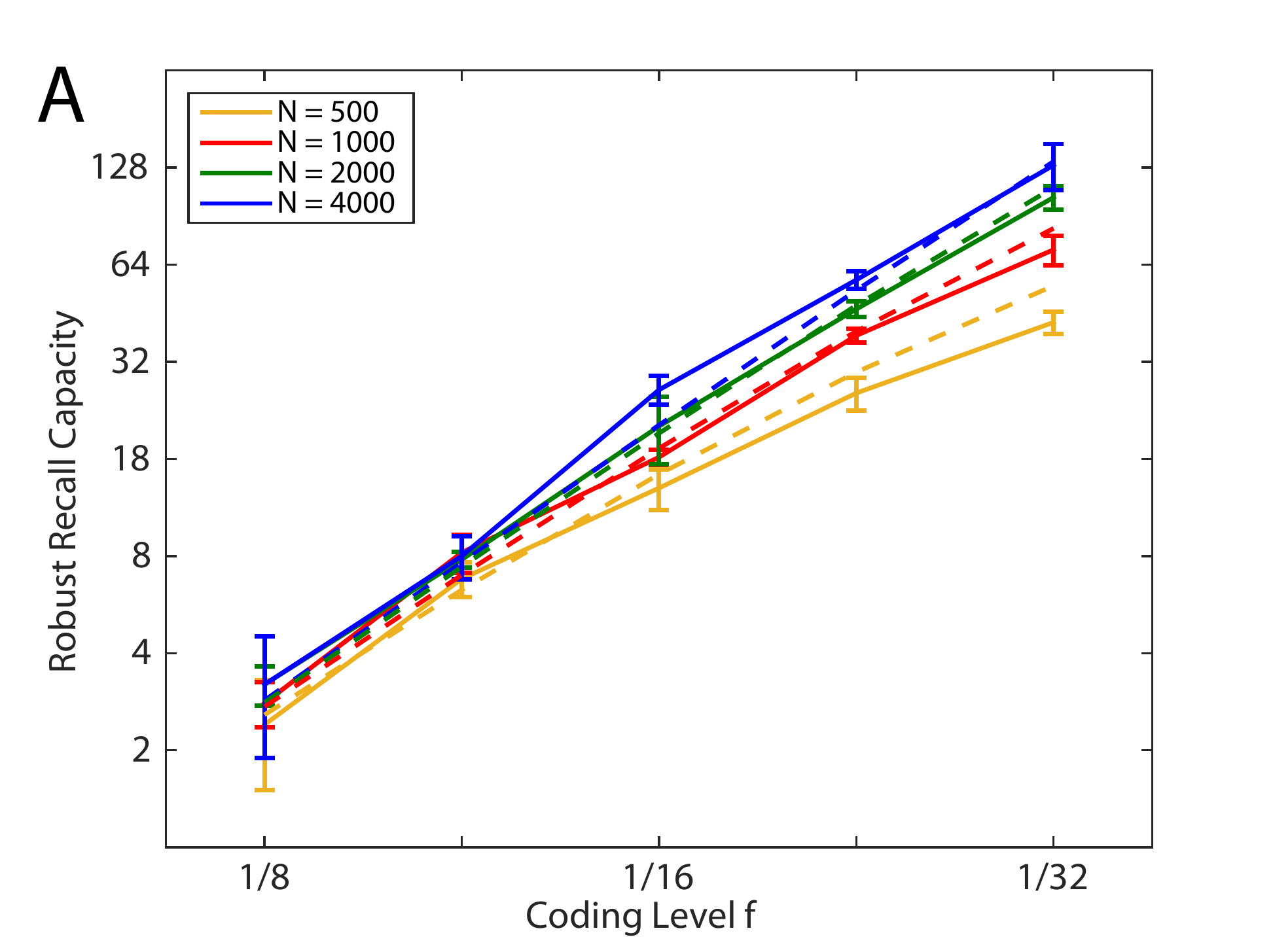}
		\includegraphics[width = 0.49 \textwidth]{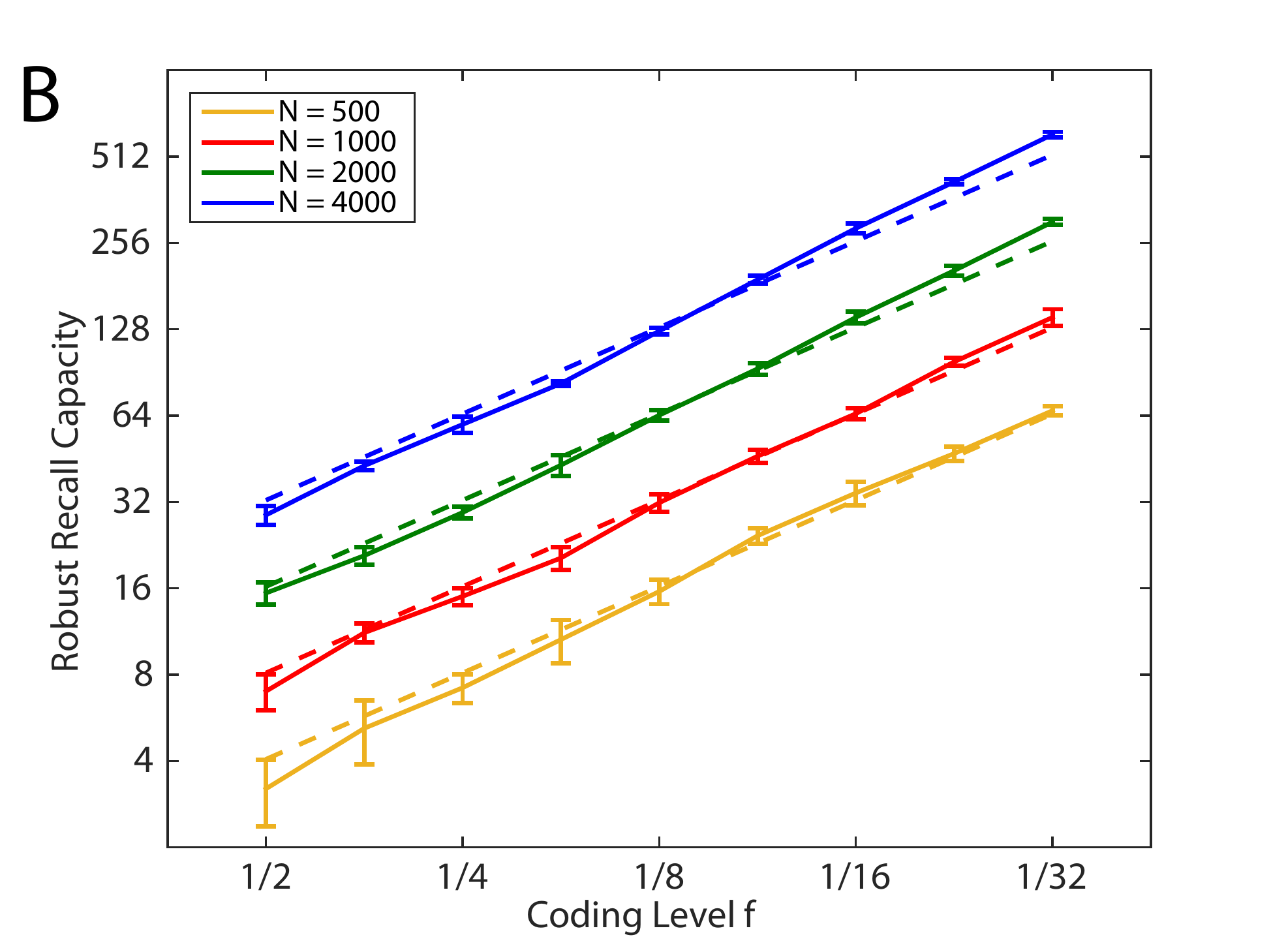}
	\end{minipage}
		\caption[]{Doubly logarithmic plots of the robust recall capacity versus coding level $f$ for the case of the Amit-Fusi learning rule. The parameters and recall procedure are otherwise as for Fig.~\ref{SparseCovariance}, except that we vary the number of presynaptic neurons, from $N=500$ (yellow) to $N = 4000$ (blue).
			A: Simple readout with presynaptic activity ($c=0$). Spatial correlations of the patterns of synaptic modifications severely limit the capacity for moderate coding levels in this case, and only for very small $f$ (when $N f^2 \ll 1$) does the growth of the capacity approach $1/f$. Solid lines are numerical simulations, while dashed lines indicate the predicted scaling with $\hat{C}^2$, where one parameter (the overall magnitude) has been fit for this family of curves. 
			B: For the balanced readout ($c=f$) the capacity is much larger and grows essentially linearly with $\hat{C}^2 = 1/(4f)$ (dashed lines, again a one parameter fit). 
		}
	\label{SparseAF}
\end{figure}

In the case of the simple readout (thresholding $\sum_{\frak{j}=1}^{N} w_\frak{j}\, \xi_\frak{j}(t')$, i.e.~$c=0$), this coefficient reduces to $\mathcal{\hat{C}} = 1/\sqrt{4 f\, (1-f)\, (1 + f^2\, (N-1))}$. Even though the number of patterns that can be retrieved grows as the coding level decreases, correlations severely limit the capacity unless $N f^2$ is small (see Fig.~\ref{SparseAF}).

A much better readout (almost, but for finite $N$ not precisely optimal) is again obtained by setting $c=f$. In this case we have 
$\mathcal{\hat{C}} = 1/\sqrt{4 f}$, i.e.~we expect the capacity to grow in inverse proportion to $f$. 
While asymptotically (as $f \to 0$ for fixed $N$) equivalent to the simple $c=0$ case, this readout leads to a much larger capacity when $N f^2$ is not small, because it circumvents the problem of spatial correlations of individual patterns of synaptic modifications (see Fig.~\ref{SparseAF}).
Note that this is in essence a prescription for learning a separate threshold (or bias term) for every readout neuron, in a manner that is linear in the sum of the incoming weights, while before we had assumed that all readouts have the same bias (though we numerically optimized this common value for each combination of $f$ and $N$).


The results of this section show that we can easily combine the computational benefits of complex synapses, which provide us with a close to optimal information capacity even for dense representations, and those of sparse coding, which allow us to further increase the number of patterns stored (at the cost of reducing the amount of information per pattern, such that the total information capacity remains roughly constant).

We have seen that signal to noise ratio calculation for the abstract model of a synapse with decay function $r(t)$ and sparse coding factorizes into a kernel-dependent part (which is identical to the case of dense coding and leads to all the same considerations about the optimal kernel$~\sim t^{-1/2}$ and its implementation using internal synaptic dynamics), and a part depending on the coding levels of neural representations, readout schemes, and the learning rule, which can be optimized separately.   

The above computation for the deterministic Amit-Fusi rule can be generalized to the stochastic case (e.g.~with a synaptic depression step that has the same magnitude as the potentiation step, but occurs only with probability $f/(1-f)$ if the pre- but not the postsynaptic neuron is active, i.e.~the conditions for synaptic depression are otherwise met).  
More generally, the concrete implementation of our fully discretized (Markov chain) model is of course always stochastic, even if the learning rule determining $\Delta \vec{w}$ is not. At this more detailed level of description (which we will not investigate further here), the parameters of the learning rule and neural representations can no longer be viewed as entirely independent of those of the synaptic model. Instead, they should of course be chosen to complement each other, e.g.~given a certain number and spacing of discrete levels for the synaptic efficacy, the learning rule should be normalized to optimally use the available dynamical range.

\newpage

\vspace{1cm}
\bibliographystyle{neuron}
\bibliography{mem}


\pagebreak

\end{document}